\documentclass[a4paper,11pt]{article}
\pdfoutput=1 % if you are submitting a pdflatex (i.e. if you have
\usepackage{jcappub} % for details on the use of the package, please
\usepackage[T1]{fontenc}
\usepackage[utf8]{inputenc}
\usepackage{float}
\usepackage{enumitem}
\usepackage{graphics}
\usepackage{mdframed}
\usepackage{mathtools}
\usepackage[dvipsnames]{xcolor}
\usepackage{soul}
\usepackage{amstext}
\usepackage{amsmath}
\usepackage{bm}	% Better vectors: \bm{}
\usepackage{booktabs}% Enables *rule commands for horizontal lines in tables
\usepackage{cleveref}% Automatic eq. fig., etc.
\usepackage[section]{placeins}% Places \FloatBarrier at the beginning of sections
\usepackage{pgfplots} % matplotlib plots .pgf are rendered inside latex
\usepackage{array,collcell} %1
\usepackage{siunitx} %2
\usepackage{caption}
\usepackage{subcaption}
\usepackage{makecell}
\usepackage{ctable} % for \specialrule command
\usepackage{adjustbox}
\usepackage{parskip}
\usepackage{graphicx}
\usepackage{tabularx}
\usepackage{multirow}
\usepackage{arydshln}
\usepackage{rotating}
\usepackage{tikz}
\usetikzlibrary{positioning}
\usepackage{pstricks}
\usepackage{color}
\usepackage{amsfonts}
\usepackage{mathrsfs}
\usepackage{epsfig}\usepackage{amsmath,amssymb,amsthm,graphicx,latexsym,braket}
\usepackage{dutchcal} % lowercase calligraphy
\usepackage{ulem}
\usepackage{pifont}
\renewcommand{\arraystretch}{2}

\newcolumntype{H}{>{\centering\arraybackslash}X}

\newcommand{\be}{\begin{equation}}
\newcommand{\ee}{\end{equation}}
\newcommand{\bea}{\begin{eqnarray}}
\newcommand{\eea}{\end{eqnarray}}
\newcommand{\dd}{\mathrm{d}} % For the derivatives
\usepackage{verbatim}

\title{\textsc{\fontsize{30}{65}\selectfont \sffamily \bfseries Exploring the role of SKA surveys with upcoming cosmic microwave background missions in probing primordial features}}

\author[1]{Debabrata Chandra\note{E-mail: {deb.iitdelhi@gmail.com}}}

\affiliation{Physics and Applied Mathematics Unit, Indian Statistical Institute, \\203 B.T.Road, Kolkata 700108, India}

\abstract{{\fontfamily{bch}This present article is dedicated to thoroughly exploring the competency of the synergy of the upcoming Cosmic Microwave Background~(CMB) missions and Square Kilometer Array~(SKA) surveys in detecting features in the primordial power spectrum. Features are by definition specific scale-dependent modifications to the minimal power-law power spectrum. The functional form of the features depends on the inflationary scenarios taken into consideration. The identification of any conclusive deviation from the feature-less power-law power spectrum will allow us to largely fathom out the microphysics of the primordial universe. Here, we consider three vital theoretically motivated feature models, namely, \textbf{Sharp feature signal}, \textbf{Resonance feature signal}, and \textbf{Bump feature}. To investigate these features, we associate each feature model with a specific scale-dependent function called a template. Here we explore three distinct fiducial models for each feature model and for each fiducial model we compare the sensitivity of 36 different combinations of the cosmological surveys. We implement the Fisher matrix forecast method to obtain the possible constraints on the feature model parameters for the future CMB missions, namely, \textbf{PICO}, \textbf{CORE-M5}, \textbf{LiteBIRD} and \textbf{CMB-S4} in synergy with upcoming SKA surveys, wherein we explore \textbf{SKA-Cosmic Shear} and \textbf{SKA-Intensity Mapping} surveys. Furthermore, the significance of combining \textbf{EUCLID-Galaxy} surveys with the \textbf{SKA-Intensity Mapping} survey is also explored. To consider the feasibility of propagating theoretical uncertainties of nonlinear scales in estimating the uncertainties on the feature parameters, we adopt redshift dependent upper limits of scales. To demonstrate the relative sensitivities of these future surveys towards the parameters of the feature models, we present a comparative analysis of all three feature models.}}

\DeclareUnicodeCharacter{2212}{-}
\begin{document}
\maketitle
\section{\textbf{Introduction}}\label{intro}
{\fontfamily{qpl}
The continuous improvement in the accuracy of measurements has elevated the stature of Cosmology to a precision science. The stunningly scrupulous measurements delivered by diverse probes, ranging from Cosmic Microwave Background observations~(CMB), Large Scale Structure observations~(LSS), weak gravitational lensing observations, Type-Ia supernovae~(SNe) to Lyman-alpha forest~(Ly$ \alpha $), have consolidated our current understanding regarding the Universe. The Planck's latest release~\cite{Planck:2018vyg,Planck:2018jri,Planck:2019kim}, along with other probes, suggests with strong statistical confidence that the best known phenomenological description of the Universe to-date can be summed up with the following key points:
\begin{itemize}[itemsep=-.3em]
\item[\ding{109}] The Universe is spatially flat, $\Omega_\mathrm{K} = 0.001 \pm 0.002 $~(68~\%~CL).
\item[\ding{109}] The Universe is having the dark matter density, $\omega_\mathrm{c}~(\Omega_\mathrm{c} h^2)$ = 0.120 $\pm$ 0.001~(68~\%~CL) and the baryon density, $\omega_{\mathrm{b}}~(\Omega_\mathrm{b} h^2)$ = 0.0224 $\pm$ 0.0001~(68~\%~CL).
\item[\ding{109}] The observed value for angular acoustic scale is 100$\theta_\ast$ = 1.0411 $\pm$ 0.0003~(68~\%~CL) and for optical depth is $\tau$ = 0.054 $\pm$ 0.007~(68~\%~CL).
\item[\ding{109}] The initial fluctuations are purely adiabatic and Gaussian in nature. The scalar perturbations exhibit a power-law power spectrum with scalar spectral index, $ n_\mathrm{s}= 0.965 \pm 0.004$~(68~\%~CL)~(so far the results do not indicate towards any scale dependency of the scalar spectral index) and amplitude~($A_\mathrm{s}$), $ \ln(10^{10} A_\mathrm{s}) = 3.044 \pm 0.014$~(68~\%~CL).  
\end{itemize}
It is remarkable to find that the Universe can be modelled with astonishing accuracy just by six-parameters only, from the very age of recombination all the way down to the present epoch, which is called the Standard model of Cosmology or $ \Lambda \text{CDM} $ Cosmology. The six parameters, demonstrating the $ \Lambda \text{CDM} $ cosmology, can be classified into primordial and late-time parameters. The two primordial parameters~($A_\mathrm{s}$, $n_\mathrm{s}$) characterize the behavior of primordial density perturbations, and the rest are referred to as the late-time parameters, which describe the evolution of the fluctuations after horizon re-entry. As a phenomenological prescription to describe the observed Universe with fine accuracy, $ \Lambda \text{CDM} $ cosmology has proven to be extremely successful compared to its other competing cosmological theories and models; and has firmly been recognized as the most viable cosmological model as per state-of-the-art experimental facilities. However, despite its remarkable fit to the available data, there are several concepts which are fundamental in understanding our Universe, that are not deciphered yet at any fundamental level, such as: 
\begin{itemize}[itemsep=-.3em]
\item[\ding{109}] What is the true nature of dark matter and dark energy? Do they interact with each other? Is dark energy a cosmological constant or dynamical in nature? If dark energy is dynamical in nature, then what is its exact functional form?
\item[\ding{109}] Did the Universe go through an inflationary phase? If yes, then how did it happen or what is the detailed mechanism driving inflation? 
\item[\ding{109}] What is the behaviour of gravity at all scales? Is there a more fundamental modified theory of gravity that eventually boils down to General Relativity as a special limiting case?
\item[\ding{109}] The detailed microphysics of reionization history is yet to be found. The epoch of first star formation and how it occurred are still eluding us.
\item[\ding{109}] What is the mechanism of the genesis of primordial density fluctuations which eventually evolved into cosmological structures, like stars, galaxies, etc?
\item[\ding{109}] What is the physics behind the asymmetry between matter and anti-matter?
\item[\ding{109}] Besides, there are certain implicit assumptions in the $ \Lambda \text{CDM} $ model, such as that it assumes there are three species of neutrinos, with the total mass being the smallest mass, consistent with the neutrino oscillation experiment. However, what is the actual sum of these masses? How many species of relativistic degrees of freedom are associated with recombination? Were there only three species at the time of recombination as predicted by the standard model? Or did the numbers differ, as suggested by many alternative models?
\end{itemize}
Hence, a further rigorous and in-depth investigation into these fundamental aspects of physics is presently the cardinal requirement.
 
The description of the Universe is based on the concordance principle, where inflation prepares the seeds for structure formation~\cite{Mukhanov:1981xt} and the ensuing evolution is taken care of by $ \Lambda \text{CDM} $ cosmology. The latest observations demand initial perturbations to be Gaussian, purely adiabatic, and scalar, characterized by a power-law power spectrum with a spectral index, $n_\mathrm{s}= 0.965 \pm 0.004$~(68~\%~CL), to offer a best fit to the data. These behaviours of the primordial fluctuations required to confer the best fit to the latest data are the salient predictions of the standard single-field slow-roll inflationary models with Einstein gravity. Despite making predictions that are consistent with the observations, the detailed mechanism governing inflation is still unclear. Thus far, current data have only confirmed the generic predictions of inflationary theory, which certainly refrain us from concluding about the model that is realized for our Universe. 
However, by improving the stringency of the bounds on the model parameters describing inflationary dynamics, current data have ruled out several inflationary proposals of different classes and shrunk the vast space of viable models~\cite{Vennin:2015eaa,Planck:2018jri}. Thus, to dig deeper into the microphysics of inflation, we need to conclusively probe other observables related to the dynamics of inflation, such as:
\begin{itemize}[itemsep=-.3em]
\item[\ding{109}] Primordial gravitational waves, which are one of the important predictions of inflation, and a robust confirmatory test of inflation. Like primordial density fluctuations, primordial gravitational waves are also the results of primordial quantum fluctuations. The primordial gravitational waves are the tensor perturbations of the spacetime metric.
\item[\ding{109}] Another major prediction of inflation is primordial non-Gaussianity. Inflation suggests that the statistical behaviour of primordial density perturbations is not exactly Gaussian. The simplest quantitative measurement of primordial non-Gaussianity for primordial density fluctuations is the three-point function, or Bispectrum.
\item[\ding{109}] Besides the two aforesaid observables of inflation, which probe the physics of inflation for two different energy scales. There is another probe of inflationary dynamics, that is, primordial features. Depending upon the inflationary models, it introduces different energy scales in the inflationary paradigm as primordial features in the power spectrum.
\end{itemize}
The current status of these observables as per the latest data can be summarized as follows:  
\begin{itemize}[itemsep=-.3em]
\item[\ding{109}] The latest upper limit on the tensor to scalar ratio, the possible probe of primordial gravitational wave, imparted by the observations, is $ r<0.032$ at 95\% confidence~\cite{Tristram:2021tvh}.
\item[\ding{109}] The current constraints on bispectrum amplitudes for three key shapes are $f_{\mathrm{NL}}^{\mathrm{local}}=-0.9 \pm 5.1$; $f_{\mathrm{NL}}^{\mathrm{equil}}= -26 \pm 47$, and $f_{\mathrm{NL}}^{\mathrm{ortho}}= -38 \pm 24$. No compelling evidence for non-Gaussianity has been found yet. Instead, weak constraints have been imparted only~\cite{Planck:2019kim}. 
\item[\ding{109}] Some earlier attempts to search for inflationary features in CMB data include, e.g. refs. \cite{WMAP:2003syu,Martin:2003sg,Covi:2006ci,Meerburg:2011gd,Meerburg:2013dla,Easther:2013kla,Achucarro:2014msa,Hazra:2016fkm}. In Planck's latest release, an extensive search for theoretically motivated parameterized primordial features that exhibit shifts from the standard feature-less power spectrum has been conducted. Thus far, we have not identified any statistically significant features.
\end{itemize}
In this article, we only focus our discussion on the quest for the status of inflationary features in light of upcoming CMB and SKA surveys. The inflationary feature is defined as a correction over the standard power-law form of the primordial power spectrum~\cite{Chen:2010xka,Chluba:2015bqa,Slosar:2019gvt}, which encapsulates all the meticulous details of the physics of the inflationary model. Commonly, the very cause of the genesis of primordial features is rooted in the effective model building of inflation. The inflationary models that display a striking deviation from scale invariance in the dynamics give rise to features in their power spectra. Simple single-field slow-roll inflation predicts an almost scale-invariant power spectrum, which is referred to as the power-law power spectrum. However, there are a myriad of models of the primordial universe, which predict departures from scale invariance as a consequence of differing from simple single-field slow-roll conditions. Such departures appear in the form of local or/and global scale-dependent function in the power spectrum as an additional factor together with the minimal feature-less form.    
In general, models with more complicated scenarios considering a greater number of degrees of freedom along with inflaton, and considering interaction between them, demand for additional symmetries, which are eventually broken in quantum gravity theory, results in affecting inflationary dynamics~\cite{Baumann:2009ds,Baumann:2014nda}. Despite avoiding the stringent effects of quantum gravity, the residuals of such tension may show up as diverse sub-leading violations of scale invariance in the form of features in the primordial power spectrum. 

There are several models of inflation that produce features in their power spectra caused by the departure of any time-dependent background quantity from a slow-roll evolution. Such a background quantity can be any one of, or a combination of, the following quantities, which typically appear in the inflationary model building:
\begin{itemize}[itemsep=-.3em]
\item[\ding{109}] slow-roll parameters are one class of such background parameters that parameterize the time variation of the scale factor, 
\item[\ding{109}]  sound-speed of the curvature perturbation is another class of background parameters that measures the deviation from the canonical condition,
\item[\ding{109}] parameters characterizing the mutual interaction of inflaton and other degrees of freedom are also an example of background parameters.
\end{itemize}
Furthermore, there are models which produce features without causing disruption in the slow-roll condition~\cite{Achucarro:2010da,Palma:2020ejf}. Diverse physical scenarios generate primordial features that serve as windows, enabling us to probe into the fundamental physics of the primordial universe, revealing invaluable information about the very nature of our universe. Such signals get embedded in cosmological observables; identifying such signatures will allow us to unravel the mystery of the primeval universe. The CMB maps allow us to have the 2-dimensional information of the initial perturbations. This not only curbs the level of precision achievable by a CMB measurement~(i.e., cosmic variance), but also extenuates the information of features in order to project it on a 2-dimensional surface. In contrast, the large scale structure of the Universe preserves the 3-dimensional information of primordial perturbations. In fact, for an adequately large survey volume, it is possible to push the limit of cosmic variance beyond the reach of CMB surveys. Thus, here, we primarily aim to exploit how the possible 3-dimensional information from SKA surveys would complement the information from the future CMB missions in singling out the features in primordial density perturbations by compensating for the intrinsic limitations of CMB surveys.

The main objective of this work is to search for the prospect of identifying inflationary models with non-trivial features in their power spectrum in upcoming CMB and SKA surveys. In this article, we study the synergy of upcoming CMB and SKA surveys in order to measure how the SKA surveys will improve the uncertainties on the feature parameters when combined with future CMB missions. In article~\cite{Chandra:2022utq}, the same set of feature models has been examined using the same set of fiducial values of the model parameters, for the same set of CMB experiments, as considered in this work. However, in article~\cite{Chandra:2022utq}, DESI and EUCLID galaxy surveys have been considered as future LSS surveys, whereas this present work is dedicated to studying the performance of SKA surveys in conjuction with future CMB missions. Thus, to have a comparison of how SKA surveys alter the constraints on the feature parameters compared to the analysis~\cite{Chandra:2022utq}, interested readers can refer to the same. In this present work, along with the SKA surveys, we have also studied how the combination of the SKA intensity mapping survey with the EUCLID cosmic shear and galaxy clustering surveys would perform compared to the individual surveys. This analysis would be conducted by employing the Fisher matrix forecast method and by using the mock likelihoods of the surveys; in the following sections, the detailed methodology has been provided. We use the publicly available cosmological code, \textbf{MontePython}\footnote{https://github.com/brinckmann/montepython\_public}~\cite{Audren:2012wb,Brinckmann:2018cvx}, interfaced with the Boltzmann code \textbf{CLASS}\footnote{http://class-code.net}~\cite{Blas:2011rf} to obtain the Fisher matrices in this work.   

The layout of this article is as follows: in order to begin with, we offer a succinct discussion in section~\ref{Feature:Template} on three distinct and salient feature models along with their relevant templates which have been investigated in this article. Section~\ref{COS:EXPSPEC} presents a brief introduction to the CMB and LSS experiments that have been dealt with in this investigation and discusses their associated instrumental specifications. The forecast methodology and the results obtained from this analysis are discussed in sections~\ref{forecast} and~\ref{Results}, respectively. The conclusion of this analysis is discussed in section~\ref{Con}. All the results obtained in this analysis are tabulated in appendix~\ref{tables}. In the end, appendix~\ref{pwrspec:likhd} delivers a short discussion on the method of constructing the power spectrum and the likelihood of the 21cm intensity mapping survey.
}
\section{Feature models}\label{Feature:Template}
{\fontfamily{qpl}
This section presents a succinct introduction of the features explored in this work. For a comprehensive review on primordial features, readers can refer to~\cite{Palma:2017wxu,Chen:2016vvw,Chen:2010xka,Chluba:2015bqa,Slosar:2019gvt}. Three vital classes of feature models, i.e., sharp feature signal, resonance feature signal and bump feature, have been thoroughly investigated in this article. Each feature is recognized with a scale-dependent function that encapsulates all the intricate details of the microphysics of the inflationary models in the primordial power spectrum, called a template. Both the scenarios of primordial power spectrum for scalar perturbations, the simple feature-free power-law power spectrum and the power spectrum with a feature, are delineated in equation~(\ref{Pri:Power}), 
where, $ A_\mathrm{s} $ stands for the scalar amplitude at pivot scale $ k_{*} = 0.05~\text{Mpc}^{-1}$, $ n_\mathrm{s} $ denotes the scalar spectral index, and ${\mathscr F}(k) $ describes any modification over the minimal feature-free power spectrum. 
The specific functional forms of $ {\mathscr F}(k) $ related to different feature models taken into consideration in this article are explicitly given in equation~(\ref{Pri:Template}) and elaborated afterwards.
\begin{equation}\label{Pri:Power}
\textbf{Power Spectrum}=\left\lbrace
\begin{array}{lll}
{\mathscr P}_{0}(k)=A_\mathrm{s}\left(\frac{k}{k_{*}}\right)^{\left(n_\mathrm{s}-1 \right)} & ~~~~~\textbf{Feature Free} & \\
{\mathscr P}(k)={\mathscr P}_{0}(k)\left[1+{\mathscr F}(k)\right] & ~~~~~\textbf{With Feature} & \\
\end{array}
\right.
\end{equation}
\begin{equation}\label{Pri:Template}
\textbf{Template}~[{\mathscr F}(k)=\dfrac{\Delta {\mathscr P}}{{\mathscr P}_{0}}(k)]=\left\lbrace
\begin{array}{lll}
{{C_s}} \sin \left( \frac{2 k}{{{k_s}}} + {{\phi_s}} \right) & ~~~\textbf{Sharp Feature} & \\
{{C_b}} \left( \dfrac{\pi e}{3} \right)^{3/2} \left( \frac{k}{{{k_b}}}\right)^3 e^{-\frac{\pi}{2}(\frac{k}{{{k_b}}})^2} & ~~~\textbf{Bump Feature} & \\
{{C_r}} \sin \left[ {{k_r}} \log \left( 2k \right) + {{\phi_r}} \right] & ~~~\textbf{Resonance Feature}  & 
\end{array}
\right.
\end{equation}
\subsection{Feature model I: sharp feature signal (linear oscillation)}
\label{Sec:Sharp}
There are various examples of inflationary models with distinct physical considerations, where features of this category emerge. In all these models, the key factor liable for generating sharp features is the fleeting shift of any background quantity associated with the evolution of curvature perturbation from the attractor solution at some point during evolution~\cite{Starobinsky:1992ts}. A few examples of such scenarios include single field inflationary models having non-smooth potential~\cite{Achucarro:2010da,Starobinsky:1992ts,Adams:2001vc,Ashoorioon:2006wc,Chen:2006xjb,Chen:2008wn,Bean:2008na,Miranda:2012rm,Bartolo:2013exa,Hazra:2014goa}, multi-field models with abrupt bends in the trajectory of inflaton field~\cite{Achucarro:2010da,Palma:2020ejf,Ashoorioon:2008qr,Cespedes:2012hu}, and models with sharp deviations in the sound speed of inflaton fields~\cite{Achucarro:2010da,Cespedes:2012hu,Park:2012rh,Achucarro:2012fd,Achucarro:2013cva}. These types of features introduce a linear oscillatory module in the primordial power spectrum as an extension to the simple feature-less structure. The template used in this analysis for sharp feature signal is given in equation~(\ref{Pri:Template}). The parameter ${{C_s}}$ in the template~(\ref{Pri:Template}) represents the amplitude, ${{k_s}}$ denotes the characteristic scale and ${{\phi_s}}$ stands for the phase angle~\cite{Chen:2011zf,Chen:2016vvw,Palma:2017wxu,Fergusson:2014hya,Fergusson:2014tza}.
\subsection{Feature model II: bump feature}
\label{Sec:Bump}
This class of features turns up as a sudden bump within a narrow band of scale in the power spectrum of curvature perturbations. There are a variety of physical processes causing the generation of bump-like features. For instance, production of particles in the time of inflation can yield such features in the primordial power spectrum. For bump-like features, in this work, we have examined the template shown in equation~(\ref{Pri:Template})~\cite{Barnaby:2009dd,Barnaby:2009mc,Barnaby:2010ke,Romano:2008rr,Chantavat:2010vt,Palma:2017wxu,Chen:2016vvw}. Here, the parameter $ {{C_b}} $ in the template~(\ref{Pri:Template}) represents the bump amplitude and the scale $ {{k_b}} $ selects the location in the momentum space where the bump appears. Here, the parameter~$ {{C_b}} $ in the template~(\ref{Pri:Template}) represents the bump amplitude and the scale~$ {{k_b}} $ selects the location of the bump in the momentum space~$ {{k_b}} $.
\subsection{Feature model III: resonance feature signal (logarithmic oscillation)}\label{Sec:Resonance}
Here we introduce another important class of features that we consider for our analysis, which comes into play when a background parameter oscillates around the attractor solution with a frequency that is greater in comparison to the Hubble scale~(H), allowing it to resonate with the sub-horizon quantum modes of the curvature perturbations, which in turn induces a scale-dependent oscillatory factor in the density fluctuations~\cite{Chen:2008wn}. Some of the inflationary models with this class of features include brane inflation~\cite{Bean:2008na}, axion monodromy inflation~\cite{Flauger:2009ab,McAllister:2008hb}, and natural inflation~\cite{Freese:1990rb,Wang:2002hf}. The template~\cite{Chen:2008wn,Palma:2017wxu,Chen:2016vvw} that has been used to parametrize the resonance feature signal is shown in equation~(\ref{Pri:Template}), where the parameters ${{C_r}}$, ${{k_r}}$, and ${{\phi_r}}$ connote the feature amplitude, characteristic scale of the feature, and associated phase factor, respectively.
\begin{figure}[htb]
\begin{mdframed}
\captionsetup{font=footnotesize}
\begin{minipage}[b1]{1.0\textwidth}         
\begin{subfigure}{0.48\textwidth}\centering
    \raisebox{0.22\linewidth}{\rotatebox{90}{\scalebox{0.3}{$\mathbf{\bm{\ell}(\bm{\ell}+1)C_{\bm{\ell}}^{TT}/2\bm{\pi}~[\bm{\mu} \textbf{K}^2]}$}}}\,
    \includegraphics[width=0.9\linewidth]{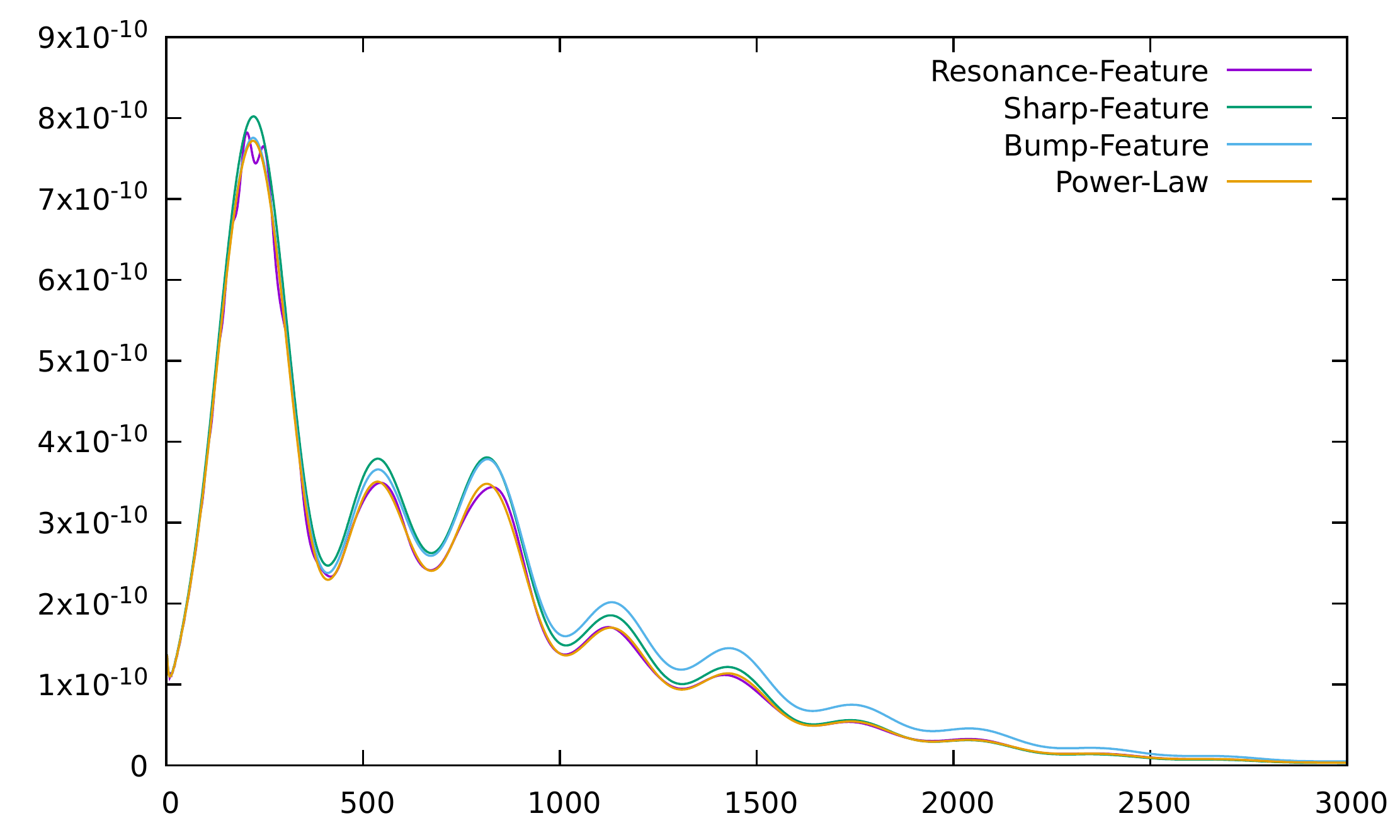}
\vspace*{-2mm}
                      ~~~~~~\scalebox{0.40}{$\bm{\ell}$} 
   
  \end{subfigure}\hfill
  \begin{subfigure}{0.48\textwidth}\centering
    \raisebox{0.2\linewidth}{\rotatebox{90}{\scalebox{0.35}{$ \mathbf{P_m^{nl}(k)~[\textbf{h}^{-3}\textbf{Mpc}^3]} $}}}\,
    \includegraphics[width=0.9\linewidth]{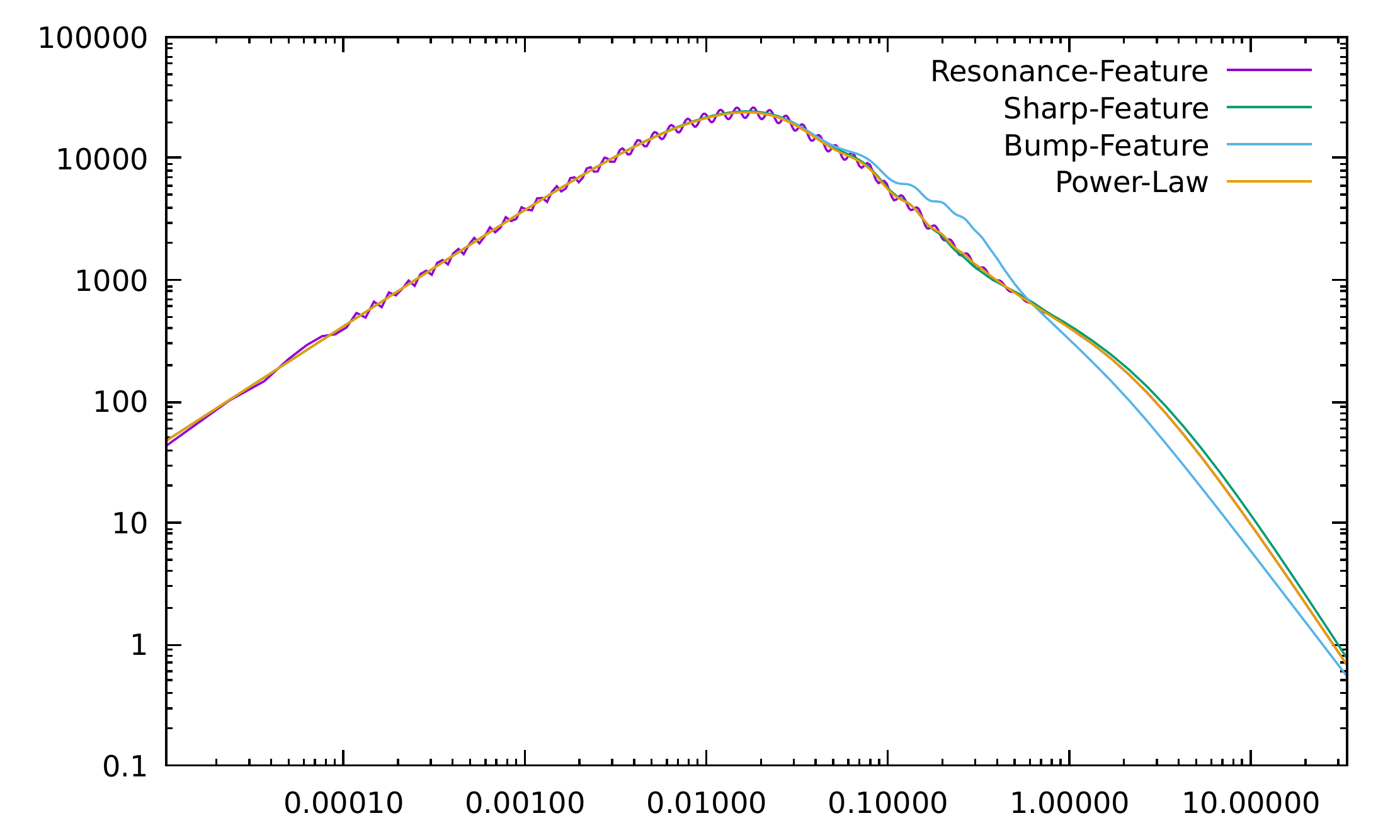}

                        ~~~~~~~~ \scalebox{0.35}{$\mathbf{\textbf{k}~[\textbf{h}~\textbf{Mpc}^{-1}]}$}
     \end{subfigure}
  
    \vspace*{3mm}

\centering

\begin{subfigure}{0.60\textwidth}\centering
    \includegraphics[width=0.9\linewidth]{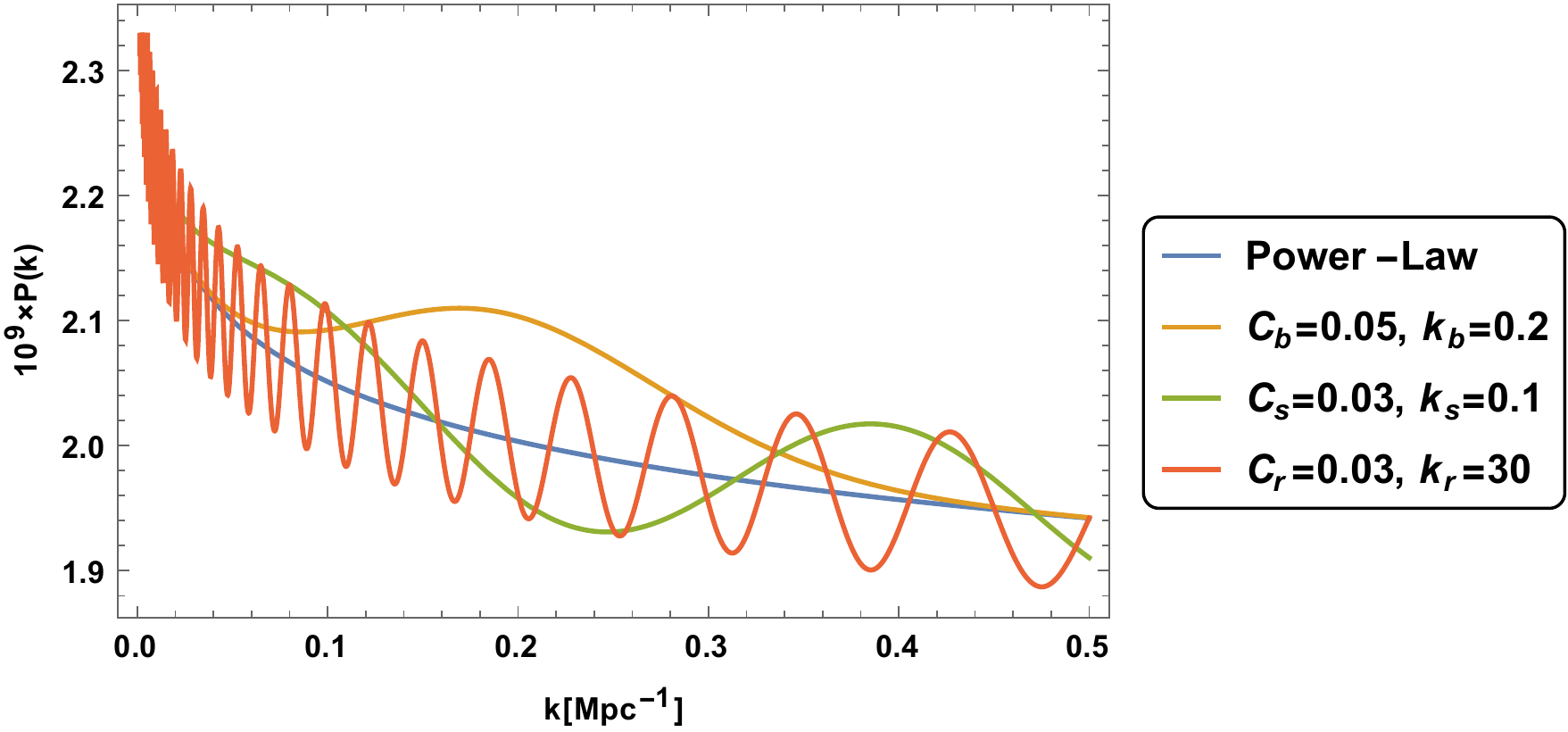}
     \end{subfigure}
\caption{{\fontfamily{bch}\textit{Top left}: Here, we plot temperature angular power spectra for four different scenarios. \textit{Top right}: In this figure, we depict the non-linear matter power spectra. \textit{Centre}: This figure shows the primordial power spectra for different cases, plotted using equations~(\ref{Pri:Power}) and~(\ref{Pri:Template}).}}\label{fig:matrices}
\end{minipage}
\end{mdframed}
\end{figure}

In this work, the Fisher forecast method has been employed to estimate the possible constraints on the feature model parameters. To execute the Fisher forecast method, the adopted fiducial values for the model parameters are separately tabulated in tables~\ref{tab:BaseFidVal} and~\ref{tab:FeaFidVal} of section~\ref{forecast}.
In figure~\ref{fig:matrices}, the \textit{Top left} plot exhibits the temperature angular power spectra associated with three feature models and the feature-free case; similarly, the \textit{Top right} plot displays the non-linear matter power spectra corresponding to three feature models and the feature-free scenario, and finally, the \textit{Centre} plot shows the primordial power spectra for three feature models and for feature-free scenario. In order to generate the plots in figure~\ref{fig:matrices}, we have used the equations~(\ref{Pri:Power}) and~(\ref{Pri:Template}). To generate the plots for temperature angular power spectra and non-linear matter power spectra, we have adopted fiducial values for the feature parameters different from those given in table~\ref{tab:FeaFidVal} just to display the behaviour of individual features distinctly; but for the forecast analysis, we have considered the fiducial values, which are shown in table~\ref{tab:FeaFidVal}.%In figure~\ref{fig:matrices}, we have exhibited the plots for the power spectra, where on the \textit{Top left}, the temperature angular power spectra has been; on the \textit{Top right}, the non-linear matter power spectra; and at the \textit{Centre}, the primordial power spectra, for the three feature models and simple power-law power spectrum, using equations~(\ref{Pri:Power}) and~(\ref{Pri:Template}). To plot the temperature angular power spectra and non-linear matter power spectra, we have adopted fiducial values different from those given in table~\ref{tab:FeaFidVal} for the feature parameters only to explicitly display the behaviour of individual features; but for the forecast, we have considered the value shown in table~\ref{tab:FeaFidVal}.
}
{\fontfamily{qpl}
\section{Cosmological experiments and their specifications}\label{COS:EXPSPEC}
The cosmological missions and surveys that are taken into consideration in this analysis, as well as their respective experimental specifications, have been the subject of discussion in this section. We take into consideration the upcoming SKA weak lensing and intensity mapping surveys in synergy with the forthcoming CMB missions to conduct the Fisher forecast analysis for the above-stated features. Furthermore, we analyze the combination of SKA intensity mapping surveys and EUCLID galaxy surveys. Below, we explicitly mention the individual missions taken under consideration for this analysis:
\begin{itemize}[itemsep=-.3em]
\item[\ding{109}] \textsf{Cosmic microwave background missions~(CMB)}: \textbf{PICO~(PC)}, \textbf{CORE-M5~(M5)}, \textbf{LiteBIRD~(LB)}, \textbf{CMB-S4~(S4)}, \textbf{Planck~(PL)}
\item[\ding{109}] \textsf{Intensity mapping surveys~(IM)}: \textbf{SKA1 band 1~(SKA1-IM1)}, \textbf{SKA1 band 2~(SKA1-IM2)}
\item[\ding{109}] \textsf{Galaxy clustering surveys~(GC)}: \textbf{EUCLID}
\item[\ding{109}] \textsf{Cosmic shear surveys~(CS)}: \textbf{SKA1}, \textbf{SKA2}, \textbf{EUCLID}
\end{itemize}
\subsection{Cosmic microwave background missions~(CMB)}\label{CMB:EXSPEC}
A short discussion on the characteristics of the CMB missions considered in this work has been presented here, as well as a brief outline of the scheme of combining two CMB missions has been provided here. Exactly the same experimental and instrumental specifications as in the article~\cite{Chandra:2022utq} have been assumed for all the above-mentioned CMB missions in this investigation of primordial features; the details of which are tabulated in the appendix \textbf{B} of ref.~\cite{Chandra:2022utq}. For all CMB experiments, we use precisely the same methodology and mock likelihood as in the article~\cite{Chandra:2022utq}.
\begin{itemize}[itemsep=-.3em]
\item[\ding{109}] $\textbf{Ground-based project}$
\begin{itemize}[itemsep=-.3em]
\item[•] $\textbf{CMB-S4}$ is a stage-4 mission~\cite{CMB-S4:2016ple,CMB-S4:2017uhf} covering a smaller fraction of the sky in contrast to satellites; however, it significantly outperforms LiteBIRD and CORE-M5 in both beam resolution and sensitivity. Hereafter, \textbf{S4} stands for \textbf{CMB-S4}.
\end{itemize}
\end{itemize}
\begin{itemize}[itemsep=-.3em]
\item[\ding{109}] $\textbf{Space-based project}$
\begin{itemize}[itemsep=-.3em]
\item[•] $\textbf{LiteBIRD}$ is a project~\cite{Matsumura:2013aja,Suzuki:2018cuy} devised to achieve optimal performance in measuring CMB B-mode. It is moderately equipped with resolution capacity but has sound sensitivity. Hereafter, \textbf{LB} stands for \textbf{LiteBIRD}. 
\item[•] $\textbf{CORE-M5}$, another anticipated proposal for CMB measurements~\cite{CORE:2017oje}. In sensitivity, it is roughly comparable to LiteBIRD, but is equipped with a much stronger beam resolution. Hereafter, \textbf{M5} stands for \textbf{CORE-M5}.
\item[•] $\textbf{PICO}$~\cite{Sutin:2018onu,Young:2018aby} has better resolving power compared to LiteBIRD for all of its channels, whereas it surpasses CORE-M5 for only a few channels. However, CMB-S4 marginally surpasses PICO in resolution. In sensitivity, PICO is at par with the CMB-S4~(only for a few channels), but surpasses CORE-M5 and LiteBIRD for all of its channels. Hereafter, \textbf{PC} stands for \textbf{PICO}.
\item[•] $\textbf{Planck}$ is an existing space-based mission and we have the results from the final release of the full Planck mission. Here, for this investigation, we have used the mock spectra created by the Montepython code by running the mock likelihood instead of using the real Planck data or the actual Planck likelihood. % where the assumed sensitivity and resolution for mock likelihood of Planck is roughly close to the full Planck mission, since it is not possible to exactly mimic the real results, because of several theoretical assumptions, such as perfect cleaning of foreground as well as  . 
Hereafter, \textbf{PL} stands for \textbf{Planck}.
\end{itemize}
\end{itemize}

Since it is recognized that the performance of the CMB-S4 mission would be best at small angular scales, whereas LiteBIRD and CORE-M5 will be devised to offer the best performance at large angular scales, a quantitative analysis becomes really compelling to figure out the potential of the synergy of a ground-based CMB mission with a space-based CMB mission in probing the feature parameters. Keeping this in mind, we consider CORE-M5 and LiteBIRD missions in synergy with the CMB-S4 mission. With the same outlook, in addition to these forthcoming CMB missions, the already existing satellite-based Planck mission has been taken into consideration in synergy with the forthcoming ground-based CMB-S4 mission, not as an individual CMB mission. Here, we consider the PICO mission only as an individual CMB mission, not in combination with the CMB-S4 mission, whereas for LiteBIRD, it has only been considered in combination with the CMB-S4 mission, not as an individual CMB mission. For the CMB-S4 and CORE-M5 missions, they have been studied as individual CMB missions and as well as in mutual combination.
\begin{itemize}[itemsep=-.3em]
\item[\ding{109}] $\textbf{Synergy Scheme}$

Here, we explicitly elaborate on the scheme of combining the space-based CMB missions, namely, Planck, LiteBIRD, and CORE-M5, with the ground-based CMB mission, CMB-S4.
%Here, we discuss the potential combinations of the CMB missions taken up in this study. The studied combinations are: \textbf{Planck+CMB-S4~(PL+S4)}, \textbf{LiteBIRD+CMB-S4~(LB+S4)}, and \textbf{CORE-M5+CMB-S4~(M5+S4)}.
\begin{itemize}[itemsep=-.3em]
\item[•] In $\textbf{Planck+CMB-S4}$ combination, only Planck data with full sky coverage is considered for low-$ \ell $~($ \ell\leq50 $), whereas for high-$ \ell $~($ \ell>50 $), the CMB-S4 data for its full sky coverage and Planck data with $17\%$ sky coverage are used~\cite{CMB-S4:2016ple}. For the Planck experiment, instead of the full Planck mission~\cite{Planck:2018vyg} in this analysis, we have considered the mock Planck likelihood. Hereafter, \textbf{PL+S4} stands for \textbf{Planck+CMB-S4}.
\item[•] In the case of $\textbf{LiteBIRD/CORE-M5+CMB-S4}$ combination, low-$ \ell $ ($ \ell\leq50 $) data is taken from LiteBIRD/CORE-M5 for its full sky coverage ($70\%$ of the sky), whereas high-$ \ell $ ($ \ell>50 $) data is taken from LiteBIRD/CORE-M5 for $30\%$ of the sky and from CMB-S4 for $ 40\% $ of the sky~\cite{CMB-S4:2016ple}. Hereafter, \textbf{LB+S4} and \textbf{M5+S4} stands for \textbf{LiteBIRD+CMB-S4} and \textbf{CORE-M5+CMB-S4}, respectively.
\end{itemize}
\end{itemize}
%\end{itemize}
\subsection{Large scale structure surveys~(LSS)}\label{LSS:EXSPEC}
\begin{itemize}[itemsep=-.3em]
\item[\ding{109}] $\textbf{EUCLID}$~\cite{EuclidTheoryWorkingGroup:2012gxx,Amendola:2016saw}, in its spectroscopic survey, will detect approximately $10^7$ galaxies within a redshift range of $ 0.7-2.0 $. EUCLID will also carry out photometric studies to measure the cosmic shear, which will roughly measure 30 galaxies per $\text {arcmin}^2$, covering a redshift range of $0-3.5$ across 15,000 $\text {deg}^2$ of the sky. In this article, we have considered both the galaxy clustering and the weak lensing measurements. For EUCLID galaxy clustering and the weak lensing, we use precisely the same methodology and mock likelihoods as in~\cite{Chandra:2022utq} and adopt the same experimental specifications as in~\cite{Chandra:2022utq}.
\item[\ding{109}] $\textbf{SKA}$ will map neutral hydrogen using the 21cm intensity mapping survey; along with that, it will also provide the galaxy clustering and cosmic shear information. All this information will allow us to trace the LSS distribution up to redshift $ z\sim20$. In this present work, both the 21cm intensity mapping and the weak lensing surveys of SKA have been taken under consideration. The scheme to construct the 21cm power spectrum and the corresponding mock likelihood for the intensity mapping experiment of SKA is discussed in appendix~\ref{pwrspec:likhd}. The experimental specifications we consider in this article for SKA intensity mapping surveys are summarized below in table~\ref{tab:SKACSExpspec}, and are adopted from the article~\cite{Sprenger:2018tdb}. For the cosmic shear experiment of SKA, we use precisely the same methodology to construct the weak lensing angular power spectrum and associated mock likelihoods as in~\cite{Chandra:2022utq}. The experimental specifications of SKA cosmic shear surveys are also tabulated in table~\ref{tab:SKACSExpspec}, which are adopted from the article~\cite{Sprenger:2018tdb}.
\begin{itemize}[itemsep=-.3em]
\item[•] $\textbf{Intensity mapping~(IM):}$
The specifications of the SKA intensity mapping surveys are illustrated in table~\ref{tab:SKACSExpspec}, where the parameters $f_{\text {sky}}$ and $z$ designate 
sky fraction and redshift. In this analysis, we allow the parameters $\Omega_{\text{HI},0}$ and ${\mathscr M}_{\text{HI}}$ of equation~(\ref{reddep:omega}) to vary around their fiducial values of $4\times10^{-4}$ and $0.6$, respectively. To perform the analysis, the parameter $ \sigma_{\text{nl}} $ in equation~(\ref{Terms}) related to the FoG effect  has also been allowed to vary around the fiducial values of $7~\text{Mpc}$. To model the HI bias~($ b_{\text {HI}}(z)$) with further accuracy, we use two nuisance parameters, ${\mathscr N}_1^{\text{IM}}$ and ${\mathscr N}_2^{\text{IM}}$ with mean value 1, which are allowed to vary in this analysis. The noise power~($ {\mathscr P}_{\text{noise}}(z) $) defined in equation~(\ref{eq:P21_PN}) is provided by
\begin{equation}
{\mathscr P}_{\text{noise}}(z) = 4\pi f_{\text{sky}} \frac{{\mathscr T}^2_{\text{sys}}}{\nu_0t_{\text{tot}}{\mathscr N}_{\text{dish}}} \frac{ r^2(z)\,(1+z)^2}{2H(z)} \ ,
\label{eq:PN}
\end{equation}
where $ {\mathscr T}_{\text{sys}}$,  $t_{\text{tot}}$, and $ N_{\text{dish}} $ represent system temperature, total observation time and number of dishes. The adopted values of $t_{\text{tot}}$ and $ N_{\text{dish}}$ are provided in table~\ref{tab:SKACSExpspec}.

By definition system temperature is as follows 
\begin{equation}
{\mathscr T}_{\text{sys}} = {\mathscr T}_{\text{inst}} + {\mathscr T}_{\text{sky}}\ ,
\label{eq:PN}
\end{equation}
where the instrument's temperature~(${\mathscr T}_{\text{inst}}$) is given in table~\ref{tab:SKACSExpspec} and the expression of the sky temperature~($ {\mathscr T}_{\text{sky}}$) is as follows 
\begin{equation}
{\mathscr T}_{\text{sky}} = 20\left(\frac{408\,\text{MHz}}{\nu}\right)^{2.75} \text{K} \ .
\end{equation}

The full width at half maximum to the rms has been applied in order to quantify the Gaussian suppressions arising in the power spectrum and that can be provided by $\text{FWHM} = 2\sqrt{2\ln2} \ \sigma$. The channel width resulting from the band separation into 64,000 channels is considered as FWHM for frequency.
\begin{align}
\sigma_{\theta} = \left(2\ln2\right)^{-0.5} \frac{\lambda_0}{2{\mathscr D}} (1+z) && \sigma_{\nu} = \left(2\ln2\right)^{-0.5} \frac{\delta \nu}{2}
\end{align}
One can evaluate the angular resolution for single dish mode using the diameter~($ {\mathscr D} $) of a single dish, which is ${\mathscr D} = 15$ $\text{m}$.
Consequently, $\sigma_{\theta}$ is as large as $0.34\,^{\circ}(1+z)$.
\item[•] $\textbf{Cosmic shear~(CS):}$
For the cosmic shear experiment of SKA surveys, we use precisely the same methodology and mock likelihood as in~\cite{Chandra:2022utq}. The specifications are summarized in table~\ref{tab:SKACSExpspec} and are adopted from the article~\cite{Sprenger:2018tdb}. We have considered in each bin the identical number of galaxies and split the whole redshift range into 10 bins. The parameter $ {\cal n}_i $ in equation~(\ref{Galaxy:Number}) denotes the number of galaxies per steradian in each bin. The distribution of galaxy number density with respect to redshift has been shown in equation~(\ref{Number:Density}), where the numerical values of the parameters $ \alpha $, $ \beta $, $ \gamma $, and $ z_0 $ for both the configurations of SKA-CS are provided in table~\ref{tab:SKACSExpspec}. The function describing the uncertainty in redshift measurements is provided in equation~(\ref{Error:Func}), where $z$ stands for true redshift and $z_\text{m}$ refers to measured redshift. The assigned fiducial value of $\sigma_{\text{lensing}}$ of equation~(\ref{Shear:Noise}), a parameter related to the innate ellipticity of the galaxies, is $0.3$.
\end{itemize}
\be
\label{Number:Density}
\frac{\dd {\cal n}_{\text{gal}}}{\dd z} = z^{\beta} e^{-\left(\frac{\alpha z}{z_\text{0}}\right)^{\gamma}}
\ee
\noindent\begin{minipage}{.5\linewidth}
\begin{equation}
\label{Galaxy:Number}
{\cal n}_i = \frac{{\cal n}_{\text{gal}}}{10} \times 3600\left(\frac{180}{\pi}\right)^2
\end{equation}
\end{minipage}%
\begin{minipage}{.5\linewidth}
\begin{equation}
\label{Shear:Noise}
{\cal N}_{\ell}^{ij} = \delta_{ij}\frac{\sigma_{\text{lensing}}^2}{{\cal n}_i}
\end{equation}
\end{minipage}
The redshift uncertainty is parameterized as follows:
\begin{equation}
\label{Error:Func}
{\mathscr E}(z,z_\text{m}) = \begin{cases}
\dfrac{1-f_{\text{spec-z}}}{\sqrt{2\pi}\sigma_{\text{ph-z}}(1+z)}\exp\left[-\dfrac{(z-z_\text{m})^2}{2\sigma^2_{\text{ph-z}}(1+z)^2}\right] + f_{\text{spec-z}}\delta(z-z_\text{m}),&z\leq z_{\text{spec-max}}\\[15pt]
\dfrac{1}{\sqrt{2\pi}\sigma_{\text{ph-z}}(1+z)}\exp\left[-\dfrac{(z-z_\text{m})^2}{2\sigma^2_{\text{ph-z}}(1+z)^2}\right],&z\leq z_{\text{ph-max}} \\[15pt]
\dfrac{1}{\sqrt{2\pi}\sigma_{\text{no-z}}(1+z)}\exp\left[-\dfrac{(z-z_\text{m})^2}{2\sigma^2_{\text{no-z}}(1+z)^2}\right],&z\geq z_{\text{ph-max}}.
\end{cases}
\end{equation}
\end{itemize}
\setlength{\tabcolsep}{1.4pt} % Default value: 6pt
\renewcommand{\arraystretch}{1.2} % Default value: 1
\newcolumntype{C}[1]{>{\Centering}m{#1}}
\renewcommand\tabularxcolumn[1]{C{#1}}
\begin{minipage}{\linewidth}
\small
\centering
\captionsetup{font=footnotesize}
\label{tab:SKACSExpspec}
\begin{tabular}{|cccccccccccc|}
\hline
\multicolumn{12}{|c|}{\textbf{SKA Intensity Mapping (IM)}}                                                                                                                                                                                                                                                                                                                                                                                                                                                                                                                                                                                                                                                \\ \hline
\multicolumn{1}{|c|}{\text{Mission}}                                                  & \multicolumn{1}{c|}{$z_{\text {min}}$} & \multicolumn{1}{c|}{$z_{\text {max}}$} & \multicolumn{1}{c|}{\begin{tabular}[c]{@{}c@{}}$\nu_{\text {min}}$\\ $[\text {MHz}]$\end{tabular}}   & \multicolumn{1}{c|}{\begin{tabular}[c]{@{}c@{}}$\nu_{\text {max}}$\\ $[\text {MHz}]$\end{tabular}}    & \multicolumn{1}{c|}{\begin{tabular}[c]{@{}c@{}}$ \delta\nu $\\ $[\text {kHz}]$\end{tabular}} & \multicolumn{1}{c|}{\begin{tabular}[c]{@{}c@{}}${\mathscr T}_{\text {inst}}$\\ $[\text {K}]$\end{tabular}} & \multicolumn{1}{c|}{\begin{tabular}[c]{@{}c@{}}$t_{\text {tot}}$\\ $[\text {h}]$\end{tabular}} & \multicolumn{1}{c|}{${\mathscr N}_{\text {dish}}$}                & \multicolumn{1}{c|}{$f_{\text {sky}}$}                  & \multicolumn{1}{c|}{$ \Omega_{\text {HI}}(z)$}                   & $ b_{\text {HI}}(z)$                                \\ \hline
\multicolumn{1}{|c|}{\begin{tabular}[c]{@{}c@{}}\text{SKA1}\\  (\text{band 1})\end{tabular}} & \multicolumn{1}{c|}{$0.45$}   & \multicolumn{1}{c|}{$2.65$}   & \multicolumn{1}{c|}{\begin{tabular}[c]{@{}c@{}}$\sim 400$ \\ ($350$)\end{tabular}}  & \multicolumn{1}{c|}{\begin{tabular}[c]{@{}c@{}}$\sim 1000$ \\ ($1050$)\end{tabular}} & \multicolumn{1}{c|}{$10.9$}                                            & \multicolumn{1}{c|}{$23$}                                            & \multicolumn{1}{c|}{\multirow{2}{*}{$10^4$}}                           & \multicolumn{1}{c|}{\multirow{2}{*}{$200$}} & \multicolumn{1}{c|}{\multirow{2}{*}{$0.58$}} & \multicolumn{1}{c|}{\multirow{2}{*}{{$\Omega_{\text{HI},0} (1+z)^{{\mathscr M}_{\text{HI}}}$}}} & \multirow{2}{*}{${\mathscr N}_1^{\text{IM}} [0.904+0.135(1+z)^{1.696 {\mathscr N}_2^{\text{IM}}}]$} \\  \cline{1-7}
\multicolumn{1}{|c|}{\begin{tabular}[c]{@{}c@{}}\text{SKA1} \\ (\text{band 2})\end{tabular}} & \multicolumn{1}{c|}{$0.05$}   & \multicolumn{1}{c|}{$0.45$}   & \multicolumn{1}{c|}{\begin{tabular}[c]{@{}c@{}}$\sim 1000$ \\ ($950$)\end{tabular}} & \multicolumn{1}{c|}{\begin{tabular}[c]{@{}c@{}}$1421$\\  ($1760$)\end{tabular}} & \multicolumn{1}{c|}{$12.7$}                                            & \multicolumn{1}{c|}{$15.5$}                                          & \multicolumn{1}{c|}{}                                                 & \multicolumn{1}{c|}{}                     & \multicolumn{1}{c|}{}                      & \multicolumn{1}{c|}{}                        &                                   \\ \hline
\multicolumn{12}{|c|}{\textbf{SKA Cosmic Shear (CS)}}                                                                                                                                                                                                                                                                                                                                                                                                                                                                                                                                                                                                                                                     \\ \hline
\multicolumn{1}{|c|}{\text{Mission}}                                                  & \multicolumn{1}{c|}{$ \alpha $}  & \multicolumn{1}{c|}{$ \beta $}   & \multicolumn{1}{c|}{$  z_{\text {ph-max}}$}                                                  & \multicolumn{1}{c|}{$f_{\text {spec-z}}$}                                                 & \multicolumn{1}{c|}{$f_{\text {sky}}$}                                        & \multicolumn{1}{c|}{$ \sigma_{\text {ph-z}} $}                                        & \multicolumn{1}{c|}{$z_0$}                                               & \multicolumn{1}{c|}{$ \gamma $}                 & \multicolumn{1}{c|}{$ \sigma_{\text {no-z}} $                              }                  & \multicolumn{1}{c|}{$ z_{\text {spec-max}}$}                & \text{Galaxy~number~density}~[${\cal n}_{\text{gal}}$] \\ \hline
\multicolumn{1}{|c|}{\text{SKA1}}                                                     & \multicolumn{1}{c|}{$ \sqrt{2} $}      & \multicolumn{1}{c|}{$2$}      & \multicolumn{1}{c|}{$2.0$}                                                  & \multicolumn{1}{c|}{$0.15$}                                                   & \multicolumn{1}{c|}{$0.1212$}                                             & \multicolumn{1}{c|}{$0.05$}                                          & \multicolumn{1}{c|}{$1.1$}                                              & \multicolumn{1}{c|}{$1.25$}                  & \multicolumn{1}{c|}{$0.3$}                & \multicolumn{1}{c|}{$0.6$}                     & $2.7$~[$\text {per arcmin}^2$] \\ \hline
\multicolumn{1}{|c|}{\text{SKA2}}                                                     & \multicolumn{1}{c|}{$ \sqrt{2} $}      & \multicolumn{1}{c|}{$2$}      & \multicolumn{1}{c|}{$2.0$}                                                  & \multicolumn{1}{c|}{$0.5$}                                                    & \multicolumn{1}{c|}{$0.7272$}                                             & \multicolumn{1}{c|}{$0.03$}                                          & \multicolumn{1}{c|}{$1.3$}                                              & \multicolumn{1}{c|}{$1.25$}                   & \multicolumn{1}{c|}{$0.3$}                & \multicolumn{1}{c|}{$2.0$}                     & $10$~[$\text {per arcmin}^2$]                               \\ \hline
\end{tabular}\par
\bigskip
\parbox{17cm}{\captionof{table}{Here, the experimental specifications adopted in this analysis for SKA intensity mapping and SKA cosmic shear surveys have been presented.}} 
\end{minipage}
}
{\fontfamily{qpl}
\section{\textbf{Forecast methodology}}\label{forecast}
In this section, we present in brief the Fisher Matrix Forecast Method~\cite{Tegmark:1996bz,Verde:2007wf,Verde:2009tu,Coe:2009xf} and the fiducial models analyzed in this work. We have expounded all the results obtained from this analysis in subsequent sections. The Fisher matrix method is an easy-to-implement and efficient technique for estimating errors on the model parameters for a given experimental set-up and for comparing the performance of different instruments or different survey strategies for a given instrument. The major utility of the Fisher matrix approach is that it enables us to estimate the measurement errors on the model parameters when marginalizing over a large number of parameters without performing the real experiment. Thus, the Fisher matrix method is the work-horse of survey design, which allows us to explore different experimental set-ups and optimize the experiment. Furthermore, it allows us to study the complementarity of different, independent and uncorrelated experiments and hence helps us to explore how different combinations of experiments are effective in improving the constraints on the model parameters as well as lifting parameter degeneracies. The Fisher matrix technique analytically estimate the measurement uncertainty in cosmological parameters by approximating the logarithm of the likelihood function $\mathscr L(\theta)$ as a multivariate Gaussian function of the cosmological parameters $\left\lbrace \theta_{i} \right\rbrace $ around a maximum at $\left\lbrace \theta_0 \right\rbrace $ (or at the fiducial model). The Fisher information matrix~(\text{F} with components $\text{F}_{ij}$) is characterized by the second order derivatives of the log-likelihood function or the effective chi-square function with respect to the model parameters evaluated at their best fit values or assumed fiducial values, as shown in equation~(\ref{Fisher Information Matrix}). The Fisher matrix encases all the information about the feasible uncertainties on the individual model parameters as well as about their mutual statistical correlations. The Fisher information matrix, when inverted, provides us with the covariance matrix. Each diagonal element of the covariance matrix offers the square of the possible 1-$\sigma$ constraint on each corresponding model parameter, as demonstrated in equation~(\ref{Fisher Information Matrix}). %The inverse of the Fisher matrix provides us with the covariance matrix, and the square root of the diagonal elements of the covariance matrix gives the bounds on the individual parameters, as demonstrated in equation~(\ref{Fisher Information Matrix}).
In this analysis, we employ the Fisher matrix method to forecast the expected errors on the model parameters for the aforesaid experiments around their respective fiducial values. The Fisher matrices must be positive-definite and invertible for the fiducial models. %For this forecast, we have assumed $\Lambda \text{CDM}$ as our baseline cosmology. The fiducial values for the baseline and the feature model parameters are listed in table~\ref{tab:BaseFidVal} and~\ref{tab:FeaFidVal}, which are adopted from the Planck results\footnote{These fiducial values are adopted from column no. 2 with a heading "Plik best fit" of table 1. of ref.~\cite{Planck:2018vyg}.}~\cite{Planck:2018vyg} and ref.~\cite{Palma:2017wxu}, respectively.
In this forecast, the $\Lambda \text{CDM}$ model has been chosen as the baseline cosmology. For both the $\Lambda \text{CDM}$ model and the three feature models, the accepted fiducial values are explicitly listed in table~\ref{tab:BaseFidVal} and~\ref{tab:FeaFidVal}, respectively. From the latest Planck results\footnote{These fiducial values are adopted from column no. 2 with a heading "Plik best fit" of table 1. of ref.~\cite{Planck:2018vyg}.}~\cite{Planck:2018vyg}, and article~\cite{Palma:2017wxu}, we have taken the fiducial values for the parameters of the $\Lambda \text{CDM}$ model, and feature models, respectively. To determine the Fisher matrices, we have used the publicly available code, \textbf{MontePython}\footnote{https://github.com/brinckmann/montepython\_public}~\cite{Audren:2012wb,Brinckmann:2018cvx} \textbf{v3.4}, which is interfaced with the Boltzmann solver code \textbf{CLASS}\footnote{http://class-code.net}~\cite{Blas:2011rf} \textbf{v2.9.4}. The required modifications in the Boltzmann solver code, namely, \text{CLASS}, have been made in order to incorporate the feature models of our consideration in the code for this investigation, and \text{MontePython} has been employed to obtain the Fisher matrices. All the Fisher matrices are computed straight from the likelihoods by producing the fiducial data sets for individual surveys and missions considered in this work and finally applying equation~(\ref{Fisher Information Matrix}); for the detailed recipe, the reader can refer to the article~\cite{Brinckmann:2018cvx}.

\setlength{\tabcolsep}{1.0pt} % Default value: 6pt
\renewcommand{\arraystretch}{1.2} % Default value: 1
\newcolumntype{C}[1]{>{\Centering}m{#1}}
\renewcommand\tabularxcolumn[1]{C{#1}}
\begin{minipage}{\linewidth} 
\small
\centering
\captionsetup{font=footnotesize}
\begin{tabular}{|ccccccc|}
\hline
\multicolumn{7}{|c|}{\textbf{$\Lambda$CDM baseline cosmology}}                                                                                                                                                                                                    \\ \hline
\multicolumn{1}{|c|}{\textbf{Parameter}}     & \multicolumn{1}{c|}{\text{CDM density}} & \multicolumn{1}{c|}{\text{Baryon density}} & \multicolumn{1}{c|}{\text{Hubble constant}} & \multicolumn{1}{c|}{\text{Optical depth}} & \multicolumn{1}{c|}{\text{Spectral index}} & \text{Amplitude} \\ \hline
\multicolumn{1}{|c|}{\textbf{Fiducial value}} & \multicolumn{1}{c|}{$\omega_\mathrm{cdm}=0.12011$}     & \multicolumn{1}{c|}{$\omega_\mathrm{b}=0.022383$}       & \multicolumn{1}{c|}{$ H_0=67.32$~\text{km/s$\cdot$Mpc}}           & \multicolumn{1}{c|}{$ \tau_\mathrm{reio}=0.0543$}        & \multicolumn{1}{c|}{$n_\mathrm{s}=0.96605$}        & $ 10^{9}\cdot A_\mathrm{s}=2.100$            \\ \hline
\end{tabular}\par
\bigskip
\parbox{14cm}{\captionof{table}{This table provides the fiducial values that have been adopted in this analysis for the six parameters describing the $\Lambda$\text{CDM} baseline cosmology.}}\label{tab:BaseFidVal}
\end{minipage}
\setlength{\tabcolsep}{1.4pt} % Default value: 6pt
\renewcommand{\arraystretch}{1.2} % Default value: 1
\newcolumntype{C}[1]{>{\Centering}m{#1}}
\renewcommand\tabularxcolumn[1]{C{#1}}
\begin{minipage}{\linewidth} 
\small
\centering
\captionsetup{font=footnotesize}

\begin{tabular}{|cccc|}
\hline
\multicolumn{4}{|c|}{\textbf{Primordial feature}}                                                                                                                             \\ \hline
\multicolumn{1}{|c|}{\textbf{Feature model}} & \multicolumn{1}{c|}{\textbf{Fiducial model I}} & \multicolumn{1}{c|}{\textbf{Fiducial model II}} & \textbf{Fiducial model III} \\ \hline
\multicolumn{1}{|c|}{Sharp feature}          & \multicolumn{1}{c|}{${{C_s}}=0.03; {{k_s}}=0.004; {{\phi_s}}=0$}  & \multicolumn{1}{c|}{${{C_s}}=0.03; {{k_s}}=0.03; {{\phi_s}}=0$}    & ${{C_s}}=0.03; {{k_s}}=0.1; {{\phi_s}}=0$      \\ \hline
\multicolumn{1}{|c|}{Bump feature}           & \multicolumn{1}{c|}{${{C_b}}=0.002; {{k_b}}=0.05$}          & \multicolumn{1}{c|}{${{C_b}}=0.002; {{k_b}}=0.1$}            & ${{C_b}}=0.002; {{k_b}}=0.2$             \\ \hline
\multicolumn{1}{|c|}{Resonance feature}      & \multicolumn{1}{c|}{${{C_r}}=0.03; {{k_r}}=5; {{\phi_r}}=0$}      & \multicolumn{1}{c|}{${{C_r}}=0.03; {{k_r}}=30; {{\phi_r}}=0$}      & ${{C_r}}=0.03; {{k_r}}=100; {{\phi_r}}=0$      \\ \hline
\end{tabular}\par
\bigskip
\parbox{14cm}{\captionof{table}{Here, we provide the fiducial values that have been adopted for the model parameters of the three primordial features considered in this analysis.}}\label{tab:FeaFidVal}
\end{minipage}

\begin{equation}\label{Fisher Information Matrix}
\textbf{Definitions}=\left\lbrace
\begin{array}{lll}
\textbf{F}_{ij}=\frac{1}{2}\dfrac{\partial^{2} {\chi^{2}}}{\partial \theta_{i}\partial \theta_{j}}= - \dfrac{\partial^{2} \ln{\mathscr L}}{\partial \theta_{i}\partial \theta_{j}} & ~~~~~\textbf{Fisher Information Matrix} & \\
\textbf{Cov}~(\theta_{i},\theta_{j})\geq [\textbf{F}^{-1}]_{ij} & ~~~~~\textbf{Covariance Matrix} & \\
\sigma(\theta_{i}) = \sqrt{[\textbf{F}^{-1}]_{ii}} & ~~~~~\textbf{1-$\sigma$ Marginalized Constraint}  & 
\end{array}
\right.
\end{equation}
We have applied upper and lower limits on the scales for galaxy clustering measurements of EUCLID (\textbf{Euclid-GC}). A \textbf{cut-off} $k_{\text{min}}=0.02~\text{Mpc}^{-1}$ has been set for the lower limit of the scales to maintain the approximation of small-angle or to exclude the scales that are greater compared to the assumed bin width. And, a redshift dependent \textbf{non-linear cut-off} $k_{\text{NL}}(z) = k_{\text{NL}}(z=0)\cdot(1+z)^{2/(2+n_\mathrm{s})}$ on the upper limit of scales has been imposed by adopting the conservative scheme, where $k_{\text{NL}}(z=0)=0.2 $\,$h\text{Mpc}^{-1}$ for \textbf{Euclid-GC}. Similarly, for the cosmic shear observations of both EUCLID (\textbf{Euclid-CS}) and SKA (\textbf{SKA-CS}), we have imposed limits on the multipole range, where the lower and the upper limits have been set to $\ell_{\text{min}}=5$  and  $\ell_{\text{max}}^i = k_{\text{NL}}(z) \cdot \bar{r}^{i}_{\text{peak}}$, respectively; where $k_{\text{NL}}(z=0)=0.5 $\,$h\text{Mpc}^{-1}$ and $\bar{r}^{i}_{\text{peak}}$ is provided by,
\begin{equation}
\bar{r}^{i}_{\text{peak}}\equiv  \frac{1}{(N-i)}\sum_{j>i} \frac{\displaystyle \int_0^{\infty}\frac{\dd r \cdot r}{r^2} {\cal K}_i(r) {\cal K}_j(r)}{\displaystyle \int_0^{\infty}\dfrac{\dd r}{r^2} {\cal K}_i(r) {\cal K}_j(r)} \ 
\end{equation}
where $N$ denotes the number of bins, ($ i $, $ j $) represents a pair of redshift bins, and the functions $ {\cal K}_i(r) $ represent the lensing kernels.
}

{\fontfamily{qpl}
\section{\textbf{Results and analysis}}\label{Results}
We now present a comparative analysis of the results obtained from the combined surveys taken into consideration by employing the Fisher method based on the fiducial models described in table~\ref{tab:FeaFidVal}. We tabulate marginalized 1-$\sigma$ uncertainties for all the parameters, including baseline and feature model parameters, separately, for each feature model and survey combination considered in this analysis in appendix~\ref{tables}. Below, we graphically illustrate and analyse the results obtained from this analysis, which allows us to gauge the relative performances of different survey combinations. These plots are obtained from the results tabulated in appendix~\ref{tables}. The readers can turn to those numerical results given in appendix~\ref{tables} at any point of their interest. Here, we restrict our discussion only to feature parameters, since investigating features is the prime focus of this current work. However, the same graphical analysis can be performed for other parameters as well.  

Three typical fiducial models~(see table~\ref{tab:FeaFidVal}) associated with three distinct values of the characteristic scale for each feature model have been considered in this analysis. In subsequent discussions we have analyzed all these distinct cases of the feature models individually. The graphical illustrations for all the cases under consideration have been depicted in the following figures~(\ref{fig:Sharp-1}-\ref{fig:Resonance-3}). The names of all the CMB experiments (individual as well as combined) that have been taken into consideration have been explicitly mentioned on the right-hand side of each plot. Along the \textbf{X-axes} we have represented six different synergies of experiments that have been taken under consideration by six distinct numbers from \textbf{1} to \textbf{6}. The survey combinations, namely, \textbf{CMB+SKA1-CS}, \textbf{CMB+SKA2-CS}, \textbf{CMB+SKA1-IM2}, \textbf{CMB+SKA1-IM1}, \textbf{CMB+SKA1-(IM1+IM2)}, and \textbf{CMB+EUCLID-(GC+CS)+SKA1-IM2}, have been represented explicitly by the numbers \textbf{1}, \textbf{2}, \textbf{3}, \textbf{4}, \textbf{5}, and \textbf{6} on the \textbf{X-axes} of all the plots~(\ref{fig:Sharp-1}-\ref{fig:Resonance-3}), respectively.  Along \textbf{Y-axes}, the marginalized 1-$\sigma$ bounds on the feature parameters are depicted. In all these plots, the variation of the 1-$\sigma$ uncertainties on the model parameters with each survey combination of the LSS surveys considered in this work with a given CMB experiment is shown along the horizontal direction, whereas the variation of the 1-$\sigma$ bounds with different CMB experiments in combination with a given LSS survey is shown along the vertical direction.

\begin{figure}[h!]
\begin{mdframed}
\captionsetup{font=footnotesize}
\center
\begin{minipage}[b1]{1.0\textwidth}
$\begin{array}{rl}
\includegraphics[width=0.5\textwidth]{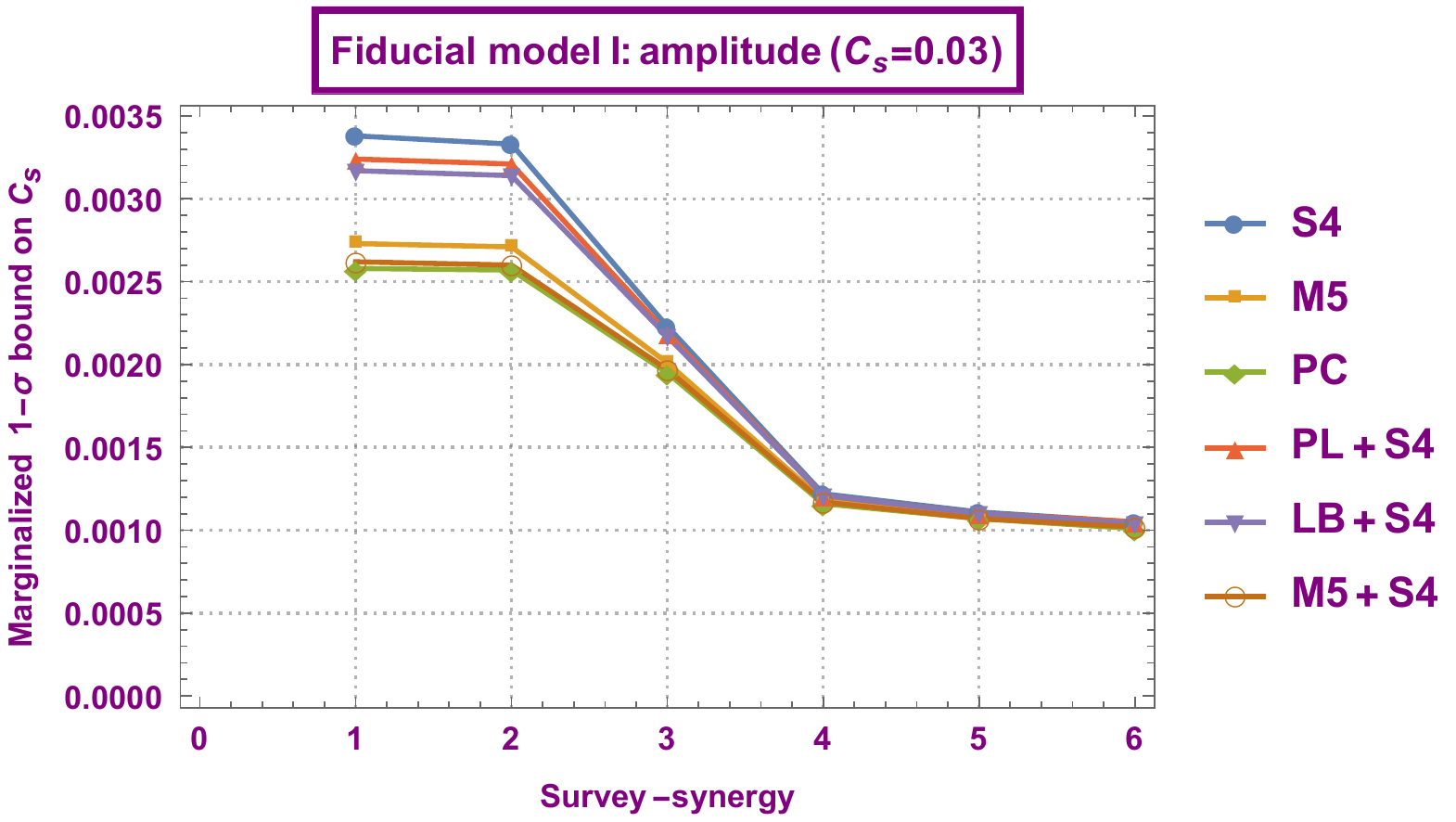} &
\includegraphics[width=0.5\textwidth]{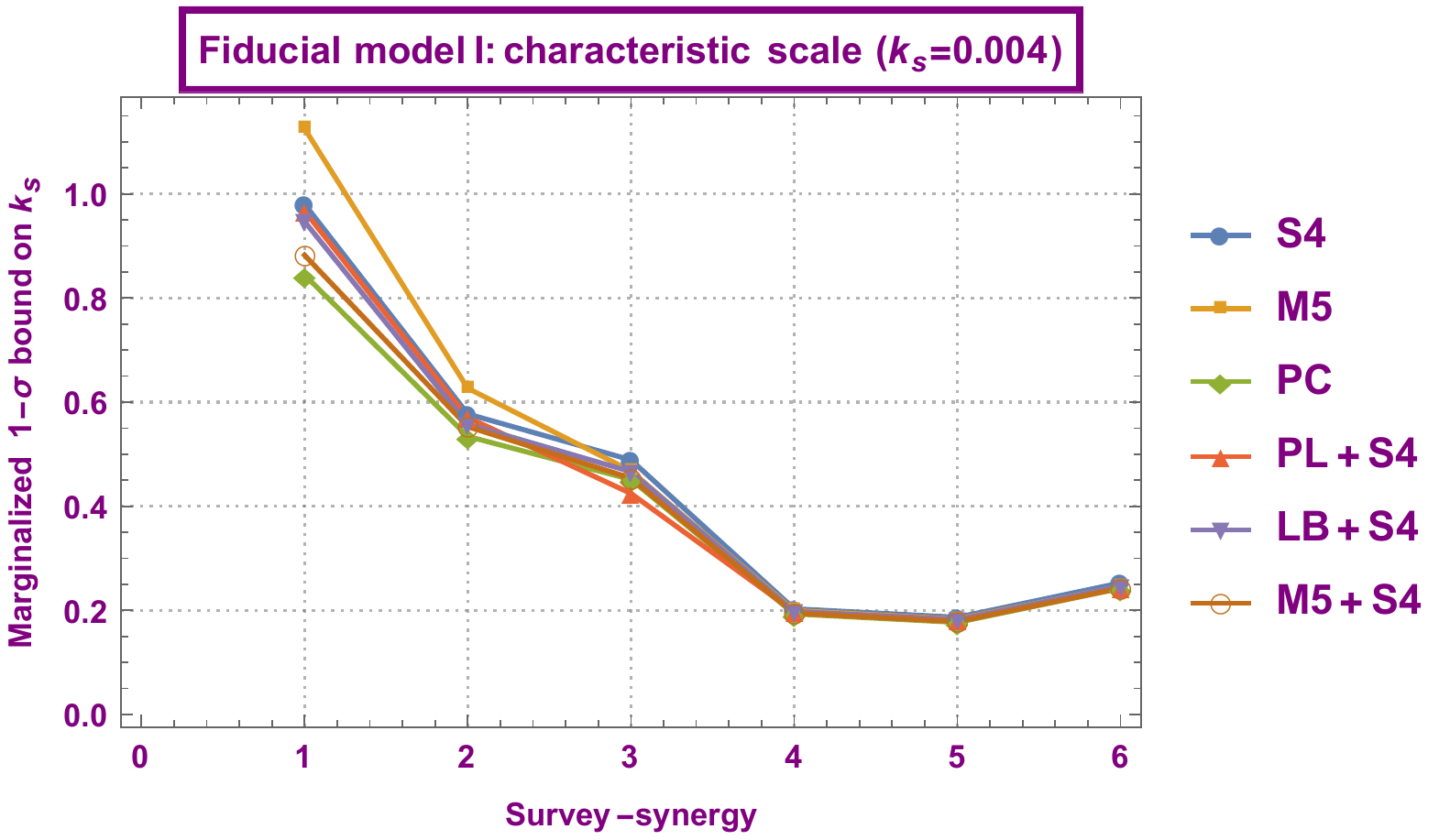}\\
\multicolumn{2}{c}{\includegraphics[width=0.5\textwidth]{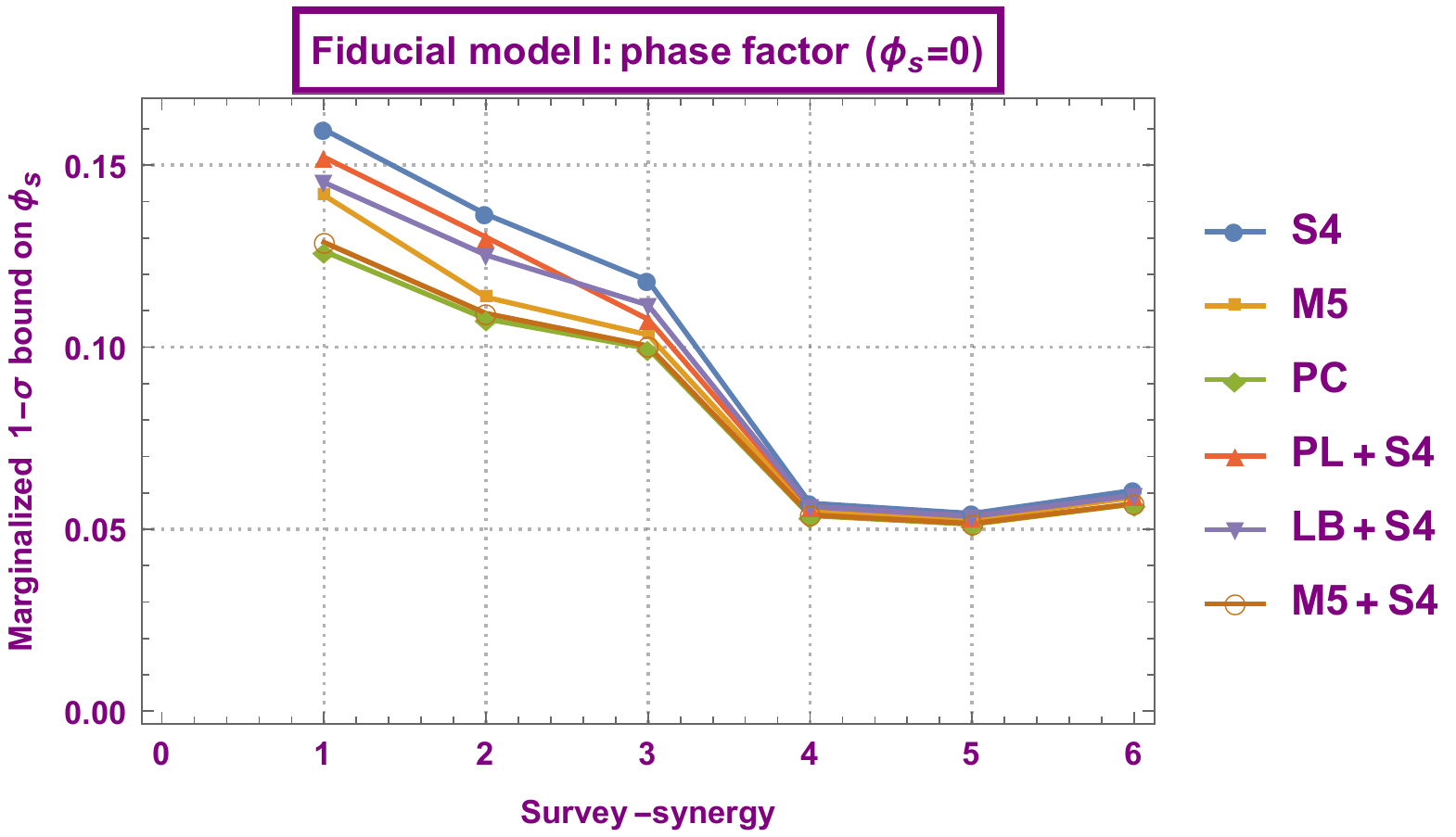}}
\end{array}$
\caption[]{\label{fig:Sharp-1}{\fontfamily{bch}
In the upper \textit{left}, upper \textit{right}, and lower \textit{middle} figures, we illustrate the marginalized 1-$\sigma$ bounds that have been obtained for the \textit{amplitude}~($ {{C_s}} $), \textit{characteristic scale}~(${{k_s}}$), and \textit{phase factor}~(${{\phi_s}}$) of the \textit{sharp feature} signal, respectively. These graphical plots exhibit the found results for the \textit{fiducial model I} of the \textit{sharp feature}, which can be found in table~\ref{table:CMB-S4-Sharp} to~\ref{table:CORE-M5+CMB-S4-Sharp} of appendix~\ref{tables}. On the right side of each plot, the CMB experiments taken into consideration are listed. The numbers \textbf{1}, \textbf{2}, \textbf{3}, \textbf{4}, \textbf{5}, and \textbf{6} along \textbf{X-axes} denote the combinations of cosmological surveys considered in this analysis, namely, \textbf{CMB+SKA1-CS}, \textbf{CMB+SKA2-CS}, \textbf{CMB+SKA1-IM2}, \textbf{CMB+SKA1-IM1}, \textbf{CMB+SKA1-(IM1+IM2)}, \textbf{CMB+EUCLID-(GC+CS)+SKA1-IM2}, respectively. Along \textbf{Y-axes}, we depict the marginalized 1-$\sigma$ errors that are achieved for the feature model parameters.\footnote{The upper \textit{right} plot exhibiting the marginalized 1-$\sigma$ bounds on the \textit{characteristic scale}~(${{k_s}}$) is scaled with a factor of $10^5$.}
}}
\end{minipage}
\hfill
\end{mdframed}
\end{figure}
\subsection{Feature model I: sharp feature}
Let us begin our analysis with the results obtained for the sharp feature signal. As previously stated, for each feature model, there are three distinct fiducial models related to three individual cases of characteristic scale of the feature. We have subsequently discussed each fiducial model step-wise for the sharp feature.
\subsubsection{Fiducial model I: ${{C_s}}=0.03$; ${{k_s}}=0.004$; ${{\phi_s}}=0$}
The obtained results for the fiducial model I of sharp feature signal are exhibited graphically in figure~\ref{fig:Sharp-1}, from which subsequent conclusions can be derived.
\begin{itemize}[itemsep=-.3em]
\item[\ding{109}] $\textbf{Oscillation amplitude~($\mathbf{{{C_s}}}$):}$
 
Here we have examined the \textit{left} plot of figure~\ref{fig:Sharp-1}, where the results for the oscillation amplitude~($ {{C_s}} $) of the sharp feature signal are shown. 
\begin{itemize}[itemsep=-.3em]
\item[•] For cosmic shear experiments of SKA~(SKA1-(CS), SKA2-(CS)), the strongest bounds on the oscillation amplitude arise from PC and from the combination of M5 and S4 data~(M5+S4), whereas S4 offers the weakest constraints on the amplitude. 
\item[•] SKA1-(CS) and SKA2-(CS) have similar constraining ability for a given CMB experiment.
\item[•] The combinations of CMB and SKA intensity mapping experiments~(CMB+SKA1-IM2, CMB+SKA1-IM1) improve the results in comparison to the survey combinations of CMB and SKA cosmic shear experiments~(CMB+SKA1-(CS), CMB+SKA2-(CS)) for all CMB experiments. 
\item[•] Between SKA1-IM1 and SKA1-IM2, the constraining capability of SKA1-IM1 is better than SKA1-IM2 for all CMB experiments. When both SKA1-IM1 and SKA1-IM2 are considered together with CMB experiments~(CMB+SKA1-(IM1+IM2)), it further improves the bounds by a tiny amount.
\item[•] For a given CMB experiment, the following three survey combinations, CMB+SKA1-(IM1), CMB+SKA1-(IM1+IM2) and CMB+EUCLID-(CS+GC)+SKA1-(IM2), provide almost similar bounds on oscillation amplitude~($ {{C_s}} $), although, among them, the best one comes from CMB+EUCLID-(CS+GC)+SKA1-(IM2) and the weakest one from CMB+SKA1-(IM1). Therefore, it shows that when we combine SKA1-IM2 with EUCLID-(GC+CS), it surpasses the constraining strength of both SKA1-IM1 and SKA1-(IM1+IM2) for all CMB experiments.
\end{itemize}

\item[\ding{109}] $\textbf{Characteristic scale~($\mathbf{{{k_s}}}$):}$

Here we have summarized the inferences drawn from the \textit{right} plot of figure~\ref{fig:Sharp-1}, where the results found for the characteristic scale~(${{k_s}}$) of sharp feature signal has been depicted.
\begin{itemize}[itemsep=-.3em]
\item[•] The finest bound on characteristic scale~(${{k_s}}$) comes from PC and the weakest one from M5 for the survey combinations of CMB and SKA1-(CS)~(CMB+SKA1-(CS)).
\item[•] In contrast to the oscillation amplitude~(${{C_s}}$), the characteristic scale~(${{k_s}}$) behaves in a different way. Here SKA2-(CS) improves the bounds on the characteristic scale~(${{k_s}}$) compared to the bounds that come from SKA1-(CS).
\item[•] The combination of CMB and SKA1-(IM2)~(CMB+SKA1-(IM2)) provides better constrains compared to the combination of CMB and SKA2-(CS)~(CMB+SKA2-(CS)) for all CMB surveys.
\item[•] Among all the SKA experiments and their possible synergy, the best constraints are coming from SKA1-(IM1) and SKA1-(IM1+IM2). In the case of SKA1-(IM1+IM2), combining  SKA1-(IM2) with SKA1-(IM1) is not showing any significant change but a tiny improvement. 
\item[•] If only SKA surveys are considered, then SKA1-(IM1+IM2) improves the results by a factor of more than 4 compared to SKA1-(CS), depending on different CMB experiments.
\item[•] The inclusion of SKA1-(IM2) data with EUCLID-(CS+GC) data offers weaker constraints for ${{k_s}}$ compared to SKA1-(IM1+IM2), unlike the case of amplitude~(${{C_s}}$). However, combining EUCLID-(CS+GC) with SKA1-(IM2) improves the bounds that come from the SKA1-(IM2)-only case for all CMB experiments.
\end{itemize}
\item[\ding{109}] $\textbf{Phase factor~($\mathbf{{{\phi_s}}}$):}$

The \textit{middle} plot of figure~\ref{fig:Sharp-1}, where the 1-$\sigma$ uncertainties on the phase angle~($ {{\phi_s}} $) of the sharp feature signal have been exhibited, is the subject of discussion here.
\begin{itemize}[itemsep=-.3em] 
\item[•] The bounds on the phase angle~($ {{\phi_s}} $) lie between S4 and PC, where the strongest and the weakest constraints arise from PC and S4, respectively, for all survey combinations. 
\item[•] SKA2-(CS) provides better constraints on the phase factor~($ {{\phi_s}} $) compared to the constraints coming from SKA1-(CS).
\item[•] SKA intensity mapping surveys are imparting better constrains on $ {{\phi_s}} $ compared to SKA weak lensing surveys, similar to the oscillation amplitude~($ {{C_s}} $) and characteristic scale~(${{k_s}}$), for all CMB experiments.
\item[•] SKA1-(IM1) tightens the bounds by nearly a factor of 2 compared to the bounds coming from SKA1-(IM2), depending on different CMB experiments.
\item[•] Combining SKA1-(IM2) with SKA1-(IM1)~(SKA1-(IM1+IM2)) shows a slight improvement in the results compared to SKA1-(IM1) for all CMB experiments.
\item[•] Combination of EUCLID-(CS+GC) and SKA1-(IM2)~(EUCLID-(CS+GC)+SKA1-(IM2)) improves the constraints arising from SKA1-(IM2). However, EUCLID-(CS+GC)+SKA1-(IM2) gives slightly weaker constraints compared to SKA1-(IM1) and SKA1-(IM1+IM2).
\end{itemize}
\end{itemize}

\begin{figure}[h!]
\begin{mdframed}
\captionsetup{font=footnotesize}
\center
\begin{minipage}[b1]{1.0\textwidth}
$\begin{array}{rl}
\includegraphics[width=0.5\textwidth]{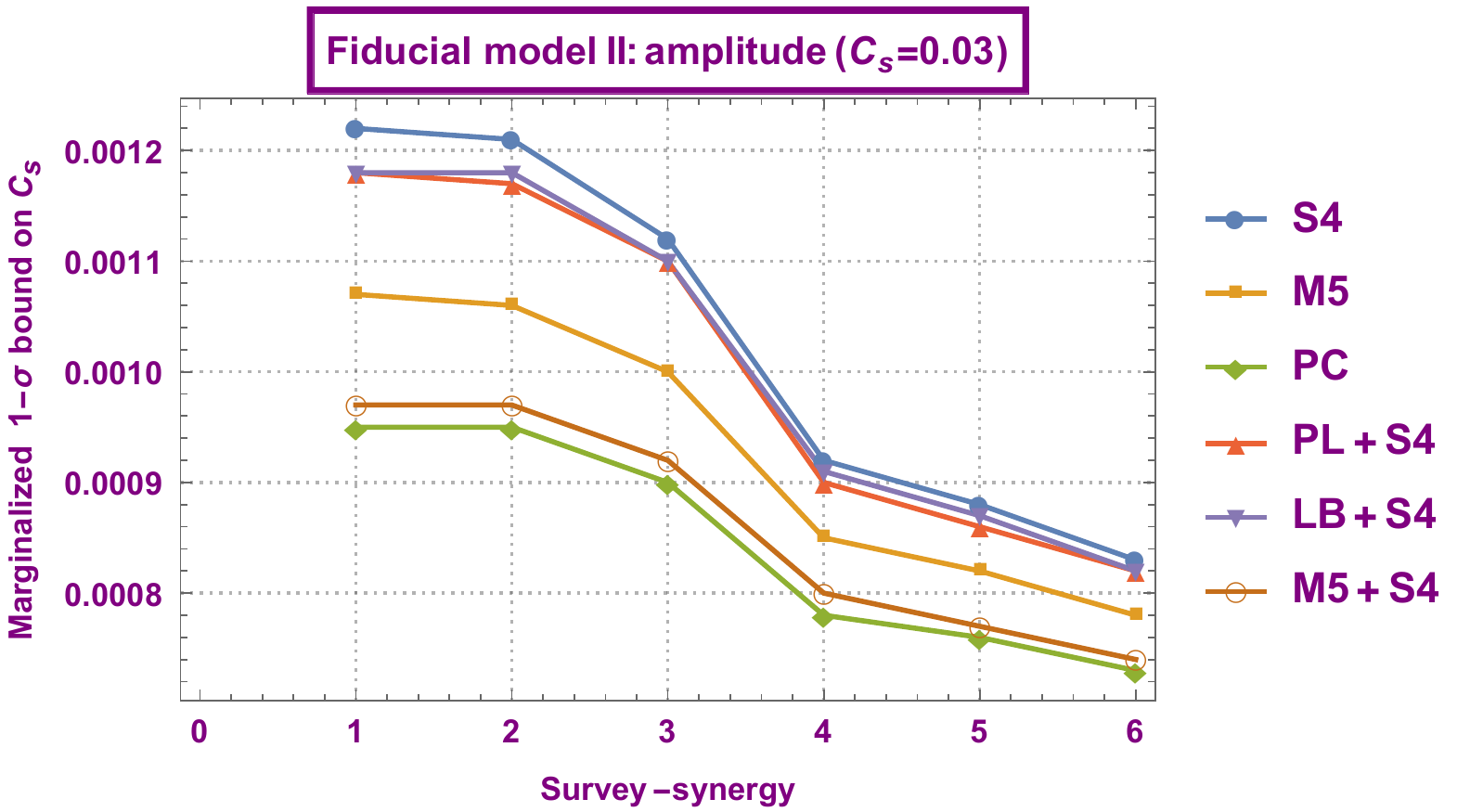} &
\includegraphics[width=0.5\textwidth]{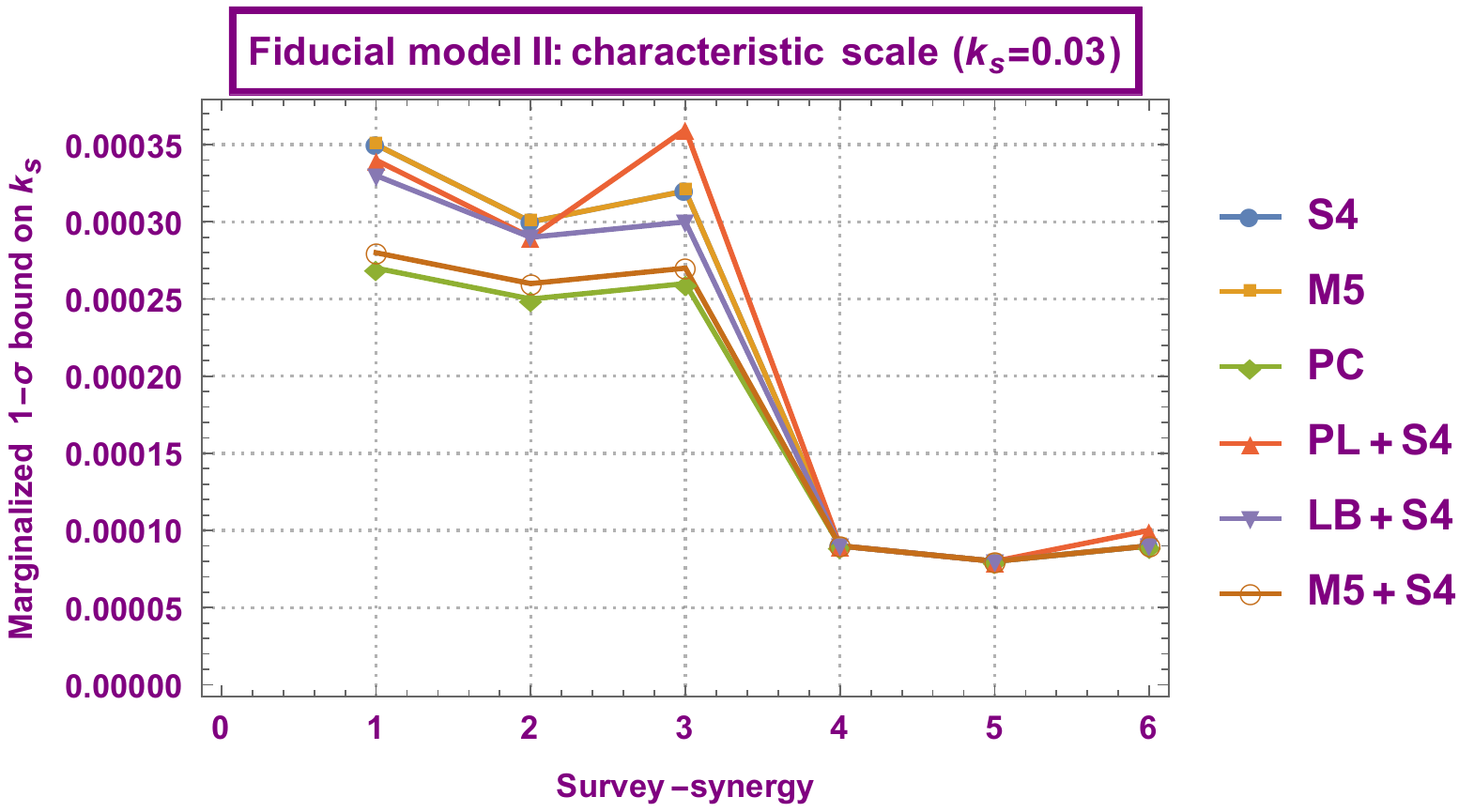}\\
\multicolumn{2}{c}{\includegraphics[width=0.5\textwidth]{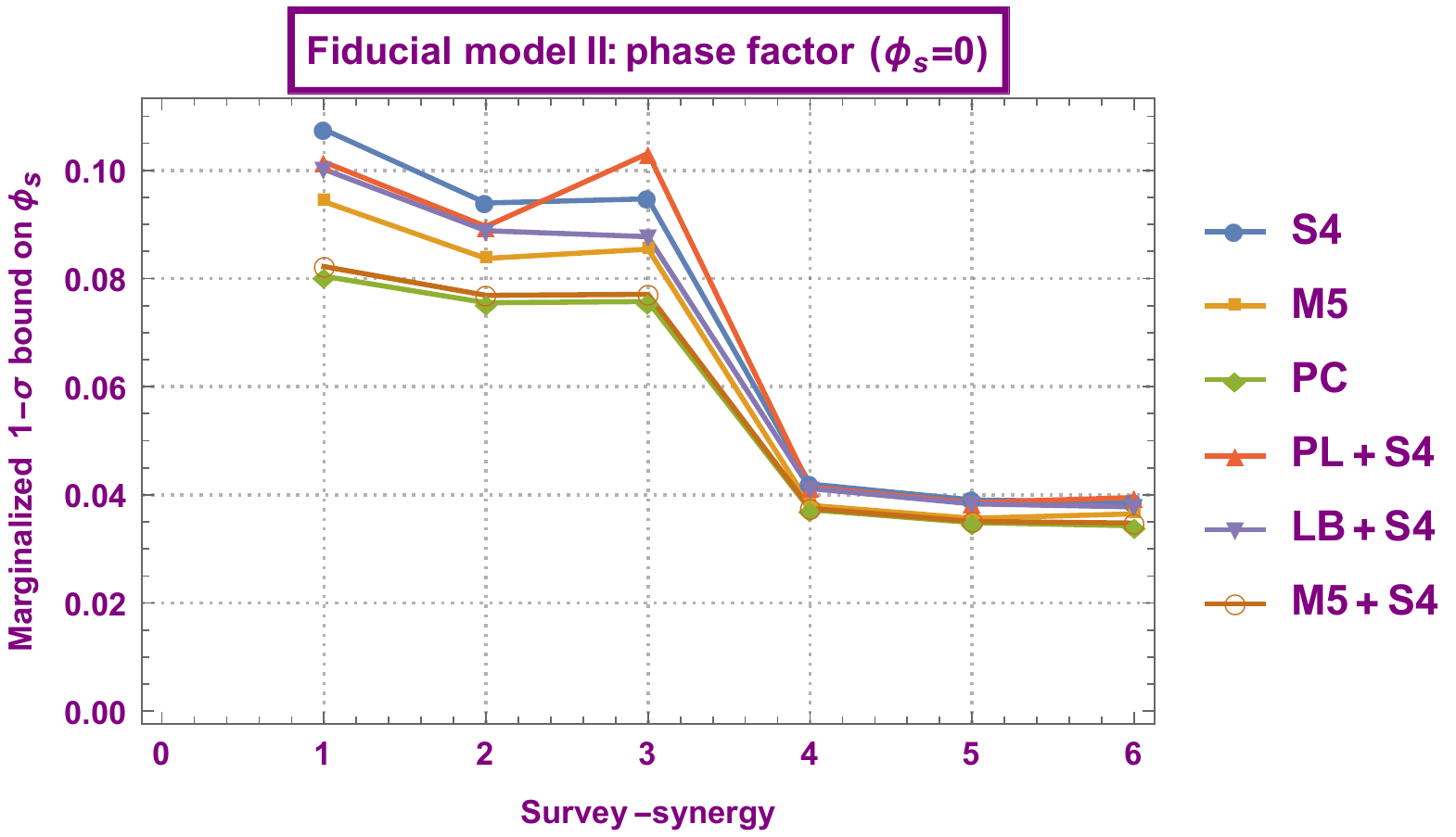}}
\end{array}$
\caption[]{\label{fig:Sharp-2}{\fontfamily{bch}
In the upper \textit{left}, upper \textit{right}, and lower \textit{middle} figures, we illustrate the marginalized 1-$\sigma$ bounds that have been obtained for the \textit{amplitude}~($ {{C_s}} $), \textit{characteristic scale}~(${{k_s}}$), and \textit{phase factor}~(${{\phi_s}}$) of the \textit{sharp feature} signal, respectively. These graphical plots exhibit the found results for the \textit{fiducial model II} of the \textit{sharp feature}, which can be found in table~\ref{table:CMB-S4-Sharp} to~\ref{table:CORE-M5+CMB-S4-Sharp} of appendix~\ref{tables}. On the right side of each plot, the CMB experiments taken into consideration are listed. The numbers \textbf{1}, \textbf{2}, \textbf{3}, \textbf{4}, \textbf{5}, and \textbf{6} along \textbf{X-axes} denote the combinations of cosmological surveys considered in this analysis, namely, \textbf{CMB+SKA1-CS}, \textbf{CMB+SKA2-CS}, \textbf{CMB+SKA1-IM2}, \textbf{CMB+SKA1-IM1}, \textbf{CMB+SKA1-(IM1+IM2)}, \textbf{CMB+EUCLID-(GC+CS)+SKA1-IM2}, respectively. Along \textbf{Y-axes}, we depict the marginalized 1-$\sigma$ errors that are achieved for the feature model parameters.
}}
\end{minipage}
\hfill
\end{mdframed}
\end{figure}

\subsubsection{Fiducial model II: ${{C_s}}=0.03$; ${{k_s}}=0.03$; ${{\phi_s}}=0$}
The obtained results for the fiducial model II of sharp feature signal are shown graphically in figure~\ref{fig:Sharp-2}, from which the subsequent conclusions can be derived.
\begin{itemize}[itemsep=-.3em]
\item[\ding{109}] $\textbf{Oscillation amplitude~($\mathbf{{{C_s}}}$):}$ 

Here we have discussed the \textit{left} plot of figure~\ref{fig:Sharp-2}, where the results for the oscillation amplitude~($ {{C_s}} $) of the sharp feature signal are shown. 
\begin{itemize}[itemsep=-.3em]
\item[•] For all survey combinations, the strongest and the weakest constraints arise from PC and S4, respectively. 
\item[•] PL+S4 reaches the sensitivity of LB+S4 for all survey combinations, and S4 too nearly meets their sensitivity except for SKA1-(CS) and SKA2-(CS).
\item[•] M5+S4 and PC show almost similar sensitivity for the oscillation amplitude~($ {{C_s}} $).
\item[•] SKA1-(CS) and SKA2-(CS) have almost similar sensitivity towards oscillation amplitude~($ {{C_s}} $) for a given CMB experiment.
\item[•] Both SKA1-IM1, SKA1-IM2 and their combination~(SKA1-(IM1+IM2)) impart stronger constraints on the oscillation amplitude~($ {{C_s}} $) than the SKA cosmic shear experiments.  
\item[•] SKA1-IM1 has a stronger constraining capacity than SKA1-IM2 for all CMB experiments.
\item[•] When SKA1-IM2 is combined with SKA1-IM1, it improves the bounds coming from SKA1-IM1 alone, and when SKA1-IM2 is combined with EUCLID-(GC+CS), it provides even tighter bounds than SKA1-(IM1+IM2).
\end{itemize}

\item[\ding{109}] $\textbf{Characteristic scale~($\mathbf{{{k_s}}}$):}$

Here we have summarized the inferences drawn from the \textit{right} plot of figure~\ref{fig:Sharp-2}, where the results found for the characteristic scale~(${{k_s}}$) of the sharp feature signal are depicted.

\begin{itemize}[itemsep=-.3em]
\item[•] For characteristic scale~(${{k_s}}$), SKA2-(CS) assigns better constraints on ${{k_s}}$ than SKA1-(CS), unlike oscillation amplitude~($ {{C_s}} $).
\item[•] The characteristic scale~(${{k_s}}$) shows a different behaviour for the fiducial value ${{k_s}}=0.03$ compared to ${{k_s}}=0.004$ for the survey combinations of CMB and SKA1-(IM2). Here, for the fiducial value ${{k_s}}=0.03$, the SKA1-(IM2) experiment, in contrast to the fiducial value ${{k_s}}=0.004$, deteriorates the results instead of improving the uncertainties compared to the uncertainties coming from SKA2-(CS). Strikingly, for PL+S4, the constraint is even weaker than SKA1-(CS), whereas for the rest of the CMB experiments, the constraints are slightly tighter than SKA1-(CS).
\item[•] The SKA1-(IM1) survey has a better constraining ability in comparison to the SKA1-(IM2) survey. Depending upon different CMB experiments, SKA1-(IM1) narrows down the errors by a factor of 2.5 to 4 in contrast to the bounds given by SKA1-(IM2).  
\item[•] SKA1-(IM1), SKA1-(IM1)+SKA1-(IM2) and EUCLID-(CS+GC)+SKA1-(IM2) exhibit roughly identical sensitivity towards the characteristic scale~(${{k_s}}$) for its fiducial value of ${{k_s}}=0.03$, and among these three, SKA1-(IM1)+SKA1-(IM2) gives the strongest bounds.
\end{itemize}

\item[\ding{109}] $\textbf{Phase factor~($\mathbf{{{\phi_s}}}$):}$

The conclusions drawn from the \textit{middle} plot of figure~\ref{fig:Sharp-2} are discussed here. In the \textit{middle} plot of figure~\ref{fig:Sharp-2}, the 1-$\sigma$ uncertainties on the phase angle~($ {{\phi_s}} $) of the sharp feature signal are summarized.

\begin{itemize}[itemsep=-.3em] 
\item[•] The bounds on the phase angle~($ {{\phi_s}} $) are between PC and S4 (except few instances of PL+S4), where the lowest uncertainties are arising from PC, for all survey combinations. 
\item[•] In regard to phase angle~($ {{\phi_s}} $), constraining capacity of SKA2-(CS) is higher compared to SKA1-(CS).
\item[•] Except for M5 and PL+S4, for the rest of the CMB experiments, SKA2-(CS) and SKA1-(IM2) show nearly equal sensitivity to the phase angle~($ {{\phi_s}} $).
\item[•] For all survey combinations, M5+S4 almost reaches the sensitivity of PC. 
\item[•] For phase angle~($ {{\phi_s}} $), SKA1-(IM1), SKA1-(IM1+IM2) and EUCLID-(CS+GC)+SKA1-(IM2) have almost equivalent constraining capacity for all the CMB experiments of our consideration.  
\item[•] Depending upon which CMB mission is being considered, the uncertainties that come from SKA1-(IM1), SKA1-(IM1+IM2), and EUCLID-(CS+GC)+SKA1-(IM2) are shrunken by roughly a factor of 2 or more, in contrast to the bounds that come from SKA1-(CS), SKA2-(CS), and SKA1-(IM2).
\end{itemize}
\end{itemize}

\begin{figure}[h!]
\begin{mdframed}
\captionsetup{font=footnotesize}
\center
\begin{minipage}[b1]{1.0\textwidth}
$\begin{array}{rl}
\includegraphics[width=0.5\textwidth]{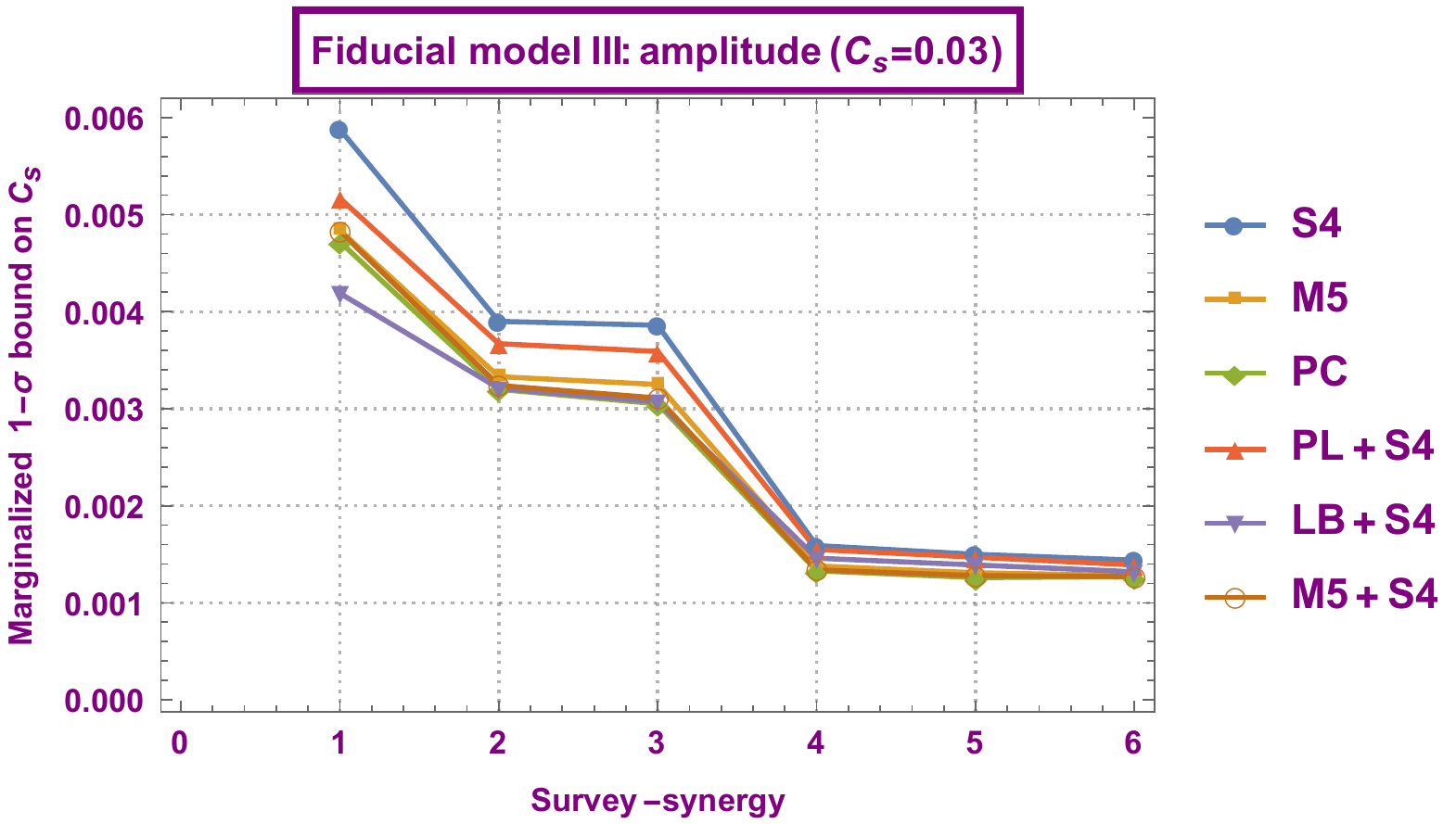} &
\includegraphics[width=0.5\textwidth]{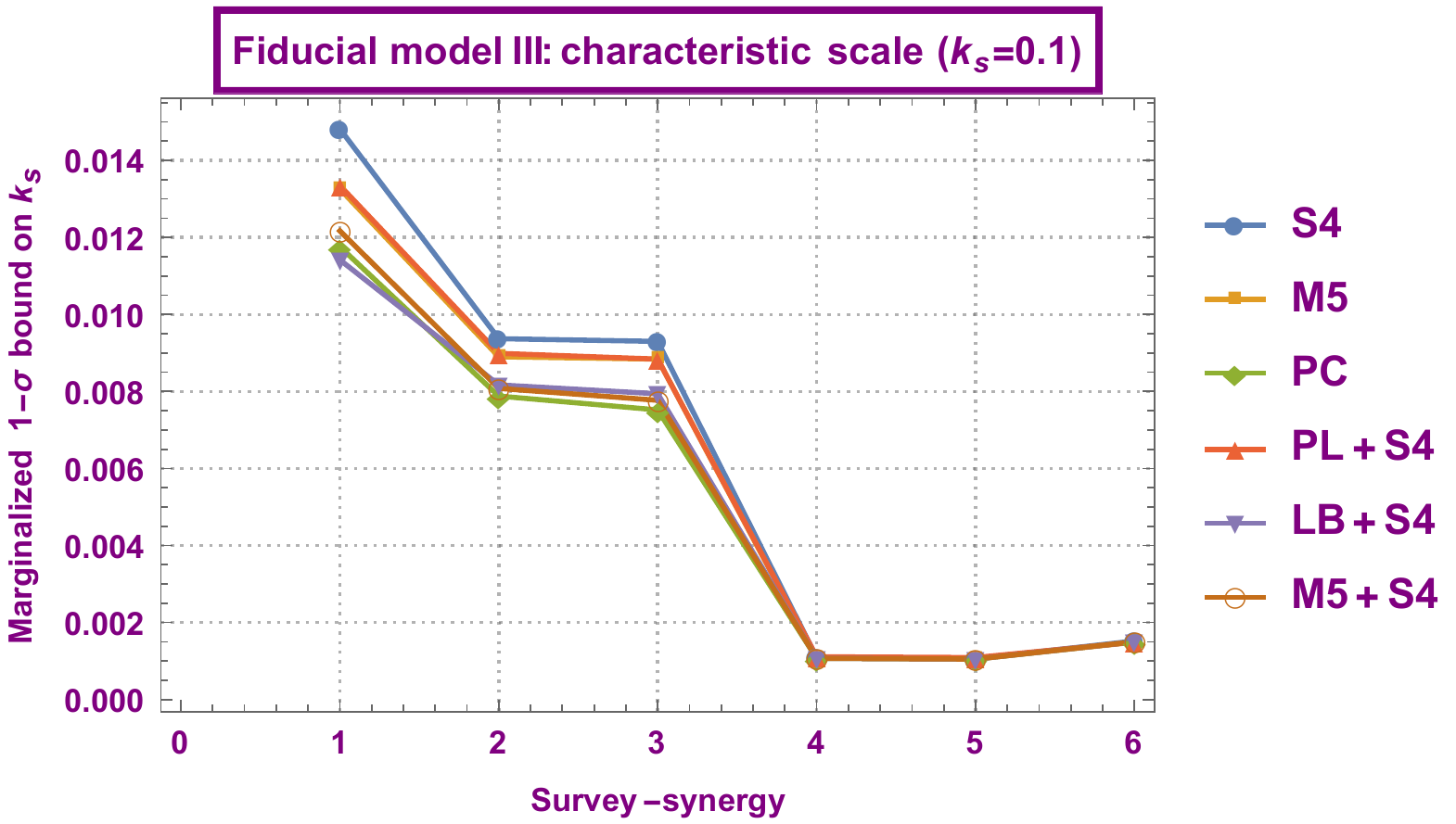}\\
\multicolumn{2}{c}{\includegraphics[width=0.5\textwidth]{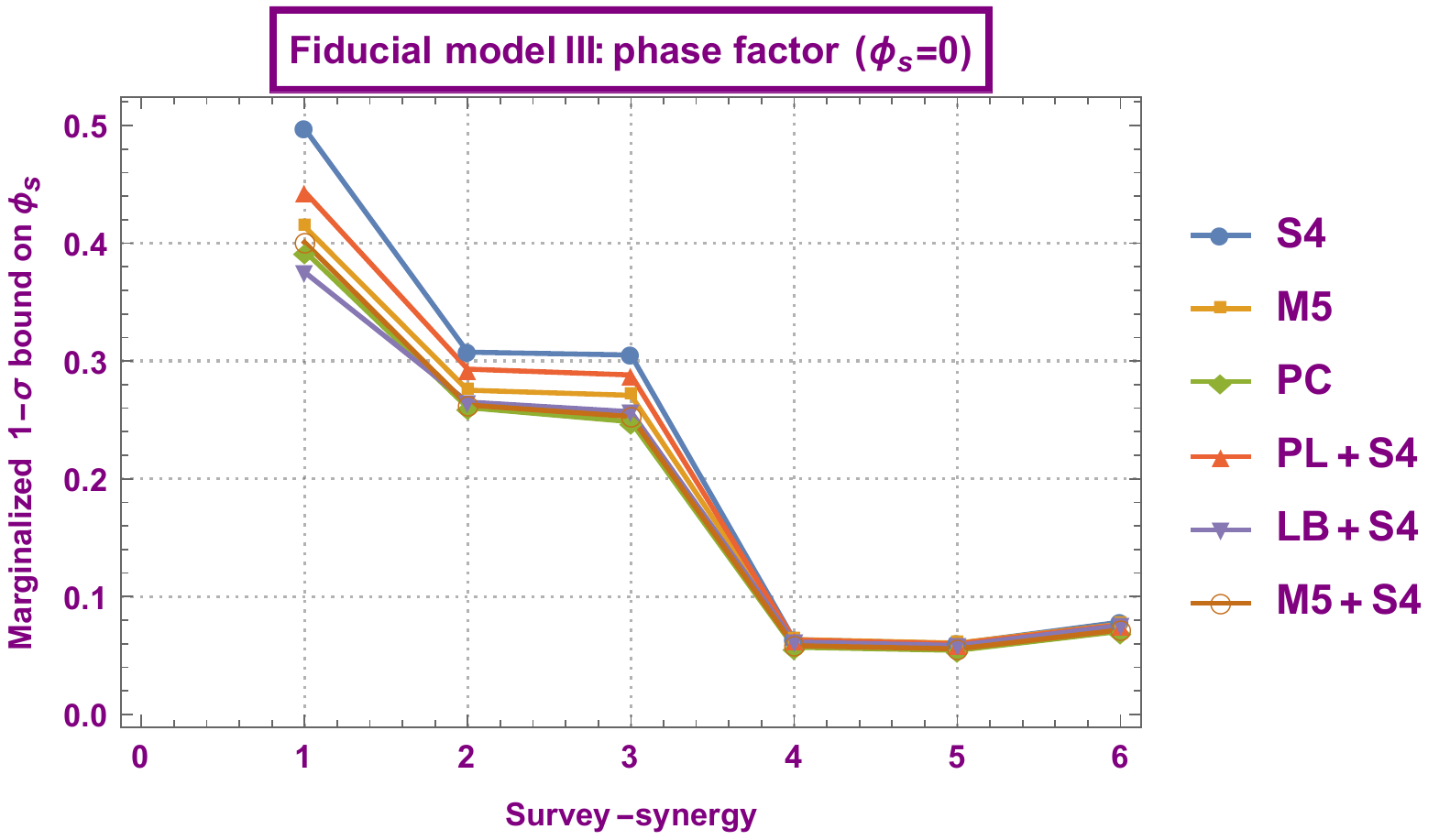}}
\end{array}$
\caption[]{\label{fig:Sharp-3}{\fontfamily{bch}
In the upper \textit{left}, upper \textit{right}, and lower \textit{middle} figures, we illustrate the marginalized 1-$\sigma$ bounds that have been obtained for the \textit{amplitude}~($ {{C_s}} $), \textit{characteristic scale}~(${{k_s}}$), and \textit{phase factor}~(${{\phi_s}}$) of the \textit{sharp feature} signal, respectively. These graphical plots exhibit the found results for the \textit{fiducial model III} of the \textit{sharp feature}, which can be found in table~\ref{table:CMB-S4-Sharp} to~\ref{table:CORE-M5+CMB-S4-Sharp} of appendix~\ref{tables}. On the right side of each plot, the CMB experiments taken into consideration are listed. The numbers \textbf{1}, \textbf{2}, \textbf{3}, \textbf{4}, \textbf{5}, and \textbf{6} along \textbf{X-axes} denote the combinations of cosmological surveys considered in this analysis, namely, \textbf{CMB+SKA1-CS}, \textbf{CMB+SKA2-CS}, \textbf{CMB+SKA1-IM2}, \textbf{CMB+SKA1-IM1}, \textbf{CMB+SKA1-(IM1+IM2)}, \textbf{CMB+EUCLID-(GC+CS)+SKA1-IM2}, respectively. Along \textbf{Y-axes}, we depict the marginalized 1-$\sigma$ errors that are achieved for the feature model parameters.
}}
\end{minipage}
\hfill
\end{mdframed}
\end{figure}

\subsubsection{Fiducial model III: $ {{C_s}}=0.03 $; ${{k_s}}=0.1$; ${{\phi_s}}=0$}
The third scenario of the sharp feature signal, that is, for the fiducial model III, the obtained results for which are exhibited graphically in figure~\ref{fig:Sharp-3}, from which subsequent conclusions can be extracted.
\begin{itemize}[itemsep=-.3em]
\item[\ding{109}] $\textbf{Oscillation amplitude~($\mathbf{{{C_s}}}$):}$ 

Here we have taken into consideration the \textit{left} plot of figure~\ref{fig:Sharp-3}, which portrays the obtained results for the oscillation amplitude~($ {{C_s}} $) of the sharp feature signal. 
\begin{itemize}[itemsep=-.3em]
\item[•] For all survey combinations, the lowest and the highest errors arise from PC~(except for SKA1-(CS), where LB+S4 generates the lowest uncertainty) and S4, respectively. 
\item[•] For fiducial value ${{k_s}}=0.1$ of the characteristic scale, SKA2-(CS) outperforms SKA1-(CS) in constraining the oscillation amplitude~($ {{C_s}} $) for a given CMB experiment.
\item[•] SKA2-(CS) and SKA1-IM2 show almost similar sensitivity for the oscillation amplitude~($ {{C_s}} $).
\item[•] For oscillation amplitude~($ {{C_s}} $), SKA1-(IM1), SKA1-(IM1+IM2) and EUCLID-(CS+GC)+SKA1-(IM2) exhibit nearly identical constraining capacity; however, among them, EUCLID-(CS+GC)+SKA1-(IM2) and SKA1-(IM1) impart the strongest and the weakest constraints, respectively, for a given CMB experiment.   
\item[•] M5, PC, LB+S4 and M5+S4 have comparable constraining capacity for all survey combinations~(except for LB+S4+SKA1-(CS)).
\end{itemize}

\item[\ding{109}] $\textbf{Characteristic scale~($\mathbf{{{k_s}}}$):}$

Here we have summarized the inferences drawn from the \textit{right} plot of figure~\ref{fig:Sharp-3}, which portrays the obtained results for the characteristic scale~(${{k_s}}$) of the sharp feature signal.
\begin{itemize}[itemsep=-.3em]
\item[•] SKA2-(CS) provides better bounds compared to SKA1-(CS) on the characteristic scale~(${{k_s}}$).
\item[•] For a given CMB experiment, SKA2-(CS) and SKA1-(IM2) perform with nearly equal sensitivity in constraining ${{k_s}}$.
\item[•] SKA1-(IM1) significantly improves the bounds on the characteristic scale~(${{k_s}}$) compared with SKA1-(CS), SKA2-(CS), and SKA1-(IM2). When compared with SKA2-(CS) and SKA1-(IM2), SKA1-(IM1) reduces the uncertainties by roughly a factor within 7 to 8.5, which depends on the CMB mission taken into account. In comparison to SKA1-(CS), SKA1-(IM1) shows improvement by nearly a factor of 10.5 to 13.5.
\item[•] When SKA1-(IM2) is added to SKA1-(IM1), a tiny amount of improvement can be observed in the bounds in contrast with the bounds imparted by SKA1-(IM1)-only.
\item[•] The addition of SKA1-(IM2) data with EUCLID-(CS+GC) data produces much better constraints compared with SKA1-(IM2)-alone scenarios. However, EUCLID-(CS+GC)+SKA1-(IM2) can not reach the sensitivity of SKA1-(IM1+IM2) or SKA1-(IM1). The best uncertainties for all CMB experiments arise from SKA1-(IM1+IM2).
\end{itemize}

\item[\ding{109}] $\textbf{Phase factor~($\mathbf{{{\phi_s}}}$):}$

The conclusions derived from the \textit{middle} plot of figure~\ref{fig:Sharp-3}, where the 1-$\sigma$ uncertainties on the phase angle~($ {{\phi_s}} $) of the sharp feature signal are graphically summarized, have been discussed here. 
\begin{itemize}[itemsep=-.3em] 
\item[•] For phase angle~($ {{\phi_s}} $), SKA2-(CS) has a better constraining capacity compared to SKA1-(CS).
\item[•] For all CMB experiments, SKA2-(CS) and SKA1-(IM2) show almost similar sensitivity towards phase angle~($ {{\phi_s}} $).
\item[•] For the phase factor~($ {{\phi_s}} $), both SKA1-(IM1) and SKA1-(IM1+IM2) exhibit almost equal constraining strength for a given CMB experiment.  
\item[•] In the context of phase factor~($ {{\phi_s}} $), the combination of SKA1-(IM2) with EUCLID-(CS+GC) has reduced the uncertainties by a factor of more than 3 compared to the SKA1-(IM2)-alone scenario of all CMB experiments. However, EUCLID-(CS+GC)+SKA1-(IM2) does not meet the sensitivity of SKA1-(IM1+IM2) or SKA1-(IM1); the best uncertainties for all CMB experiments are coming from SKA1-(IM1+IM2).
\end{itemize}
\end{itemize}

\begin{figure}[h!]
\begin{mdframed}
\captionsetup{font=footnotesize}
\center
\begin{minipage}[b1]{1.0\textwidth}
$\begin{array}{rl}
\includegraphics[width=0.5\textwidth]{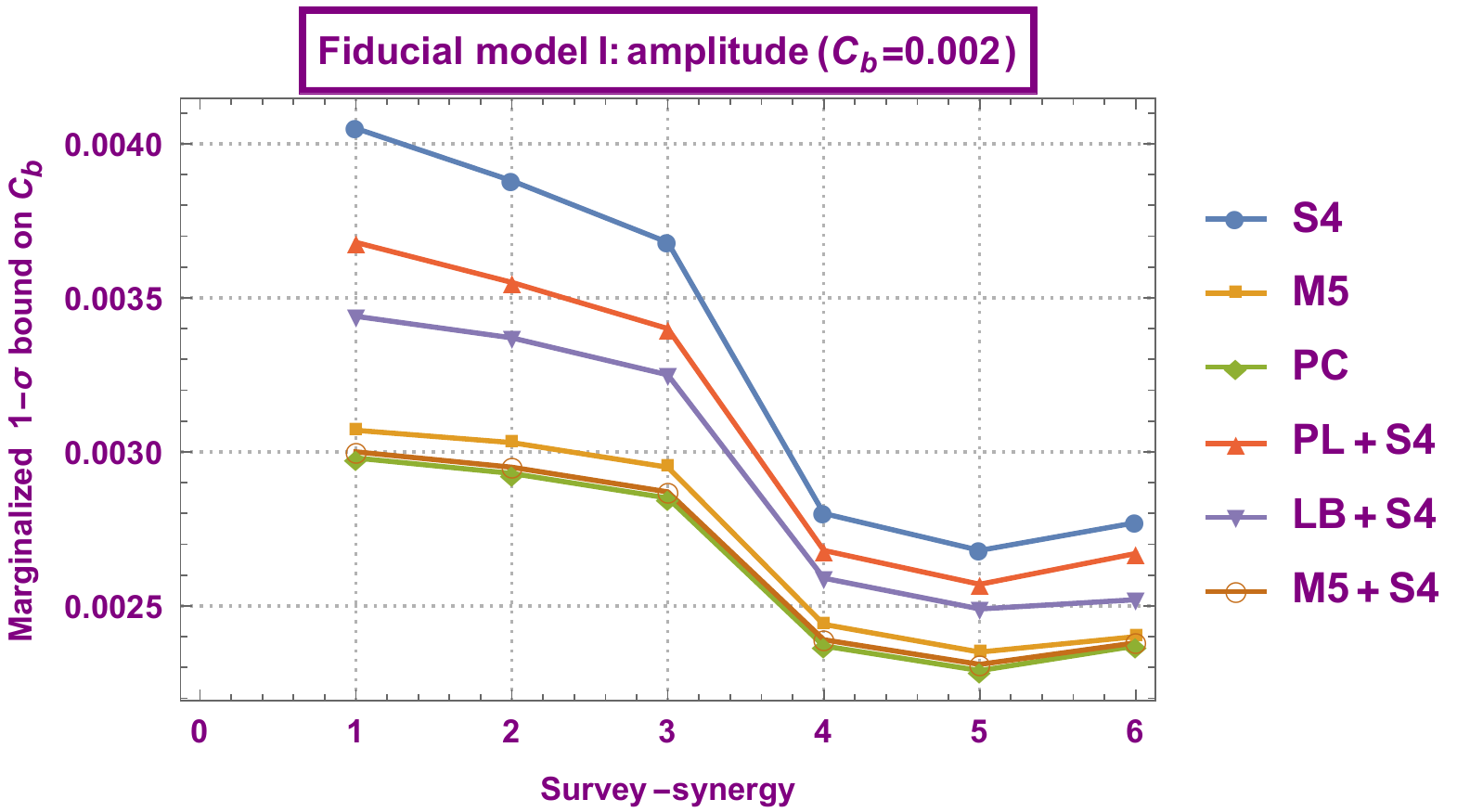} &
\includegraphics[width=0.5\textwidth]{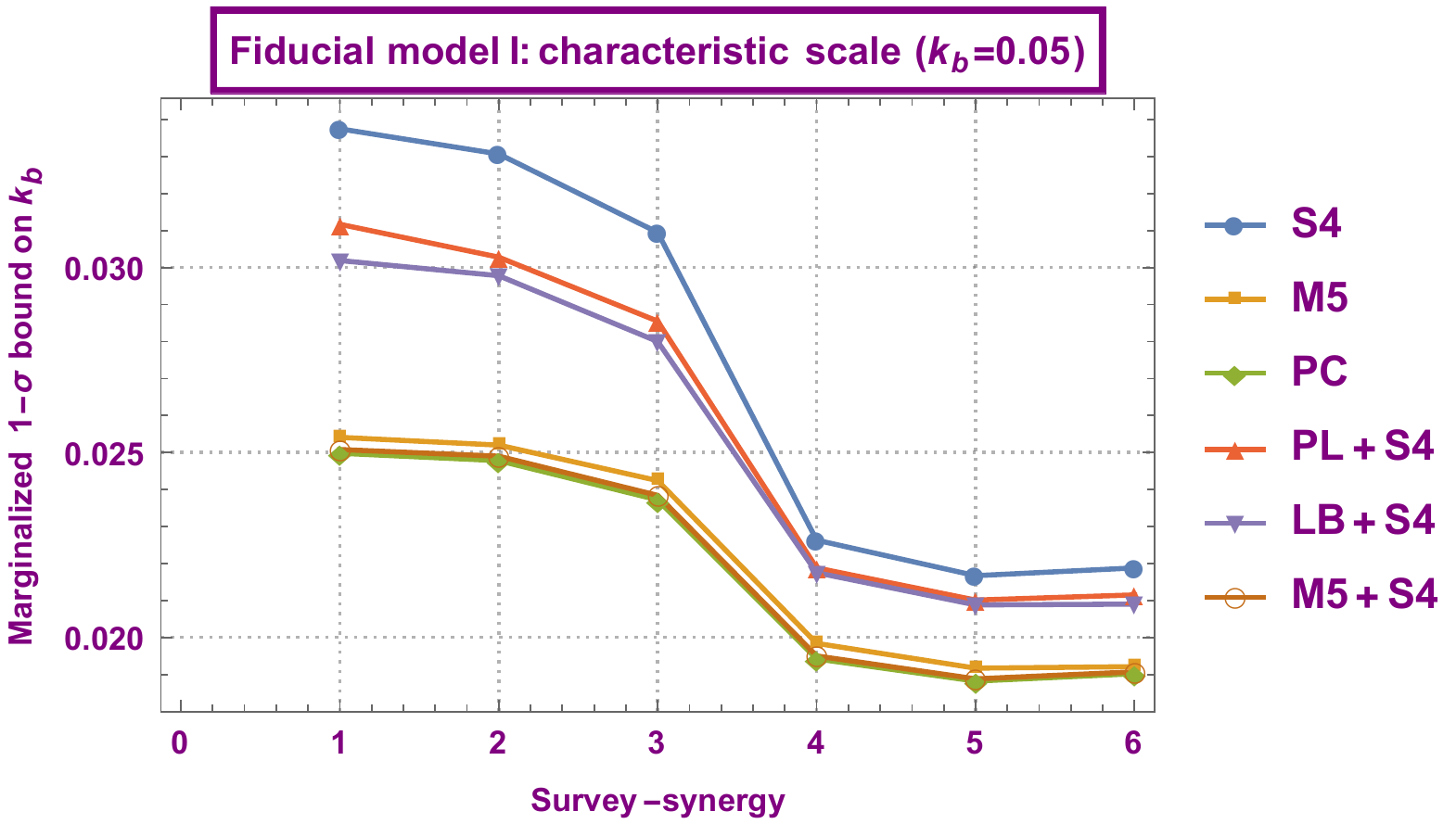}\\
\end{array}$
\caption[]{\label{fig:Bump-1}{\fontfamily{bch}
In the \textit{left}, and \textit{right} figures, we illustrate the marginalized 1-$\sigma$ bounds that have been obtained for the amplitude~($ {{C_b}} $), and characteristic scale~(${{k_b}}$) of the \textit{bump feature} model, respectively. These graphical plots exhibit the found results for the \textit{fiducial model I} of the \textit{bump feature}, which can be found in table~\ref{table:CMB-S4-Bump} to \ref{table:CORE-M5+CMB-S4-Bump} of appendix~\ref{tables}. On the right side of each plot, the CMB experiments taken into consideration are listed. The numbers \textbf{1}, \textbf{2}, \textbf{3}, \textbf{4}, \textbf{5}, and \textbf{6} along \textbf{X-axes} denote the combinations of cosmological surveys considered in this analysis, namely, \textbf{CMB+SKA1-CS}, \textbf{CMB+SKA2-CS}, \textbf{CMB+SKA1-IM2}, \textbf{CMB+SKA1-IM1}, \textbf{CMB+SKA1-(IM1+IM2)}, \textbf{CMB+EUCLID-(GC+CS)+SKA1-IM2}, respectively. Along \textbf{Y-axes}, we depict the marginalized 1-$\sigma$ errors that are achieved for the feature model parameters.
}}
\end{minipage}
\hfill
\end{mdframed}
\end{figure}
\subsection{Feature model II: bump feature}
Now let us continue to interpret the results found for the bump features, which consist of three separate cases for three distinct fiducial models corresponding to three specific sets of fiducial values of model parameters for the bump feature, similar to the previous feature model.

\subsubsection{Fiducial model I: ${{C_b}}=0.002$; ${{k_b}}=0.05$}
The first scenario, which arises for the fiducial model I of bump feature, has been analyzed here; the found results of which have been graphically delineated in figure~\ref{fig:Bump-1}. The inferences derived from the found results of each parameter of the fiducial model are summarized distinctly and step-wise subsequently.
\begin{itemize}[itemsep=-.3em]
\item[\ding{109}] $\textbf{Bump amplitude~($\mathbf{{{C_b}}}$):}$ 

Here, we have examined the \textit{left} plot of figure~\ref{fig:Bump-1}, where the results obtained for the amplitude~(${{C_b}}$) of the bump feature have been delineated. 
\begin{itemize}[itemsep=-.3em]
\item[•] For bump amplitude~(${{C_b}}$), the strongest bounds come from the PC mission, whereas the S4 mission offers the weakest constraints on the amplitude, for all the survey combinations taken into account.
\item[•] For all survey combinations, M5+S4 almost reaches the sensitivity of PC, and even M5 is also giving constraints closer~(although weaker) to the bounds coming from PC or M5+S4.    
\item[•]In regard to the amplitude~(${{C_b}}$) of the bump feature for a given CMB experiment, between the two SKA cosmic shear surveys, the constraining ability of SKA2-(CS) is better compared to SKA1-(CS), and for SKA intensity mapping experiments, the sensitivity of SKA1-(IM1) is stronger than SKA1-(IM2). However, between SKA cosmic shear surveys and SKA intensity mapping surveys, the constraining strength of intensity mapping surveys is more in comparison to cosmic shear surveys of SKA.
\item[•] When we combine two SKA intensity mapping surveys~(SKA1-(IM1+IM2)), it marginally improves the uncertainties on bump amplitude~(${{C_b}}$) in contrast to the bounds that come from SKA1-(IM1) alone, which is true for all CMB experiments.
\item[•] The addition of SKA1-IM1 data with SKA1-IM2 data~(SKA1-(IM1+IM2)) offers the lowest uncertainties on the bump amplitude~(${{C_b}}$) for any given CMB mission under our consideration.
\item[•] For a given CMB experiment, when SKA1-(IM2) data is added to EUCLID-(CS+GC) data, the combination marginally surpasses the sensitivity of SKA1-(IM1)~(except for PC, where the bounds are equal in both cases), and even for LB+S4, the EUCLID-(CS+GC)+SKA1-(IM2) combination reaches the sensitivity of SKA1-(IM1+IM2).
\end{itemize}

\item[\ding{109}] $\textbf{Characteristic scale~($\mathbf{{{k_b}}}$):}$

Here, the \textit{right} plot of figure~\ref{fig:Bump-1} has been examined, which illustrates the obtained results for the characteristic scale~(${{k_b}}$) of the bump feature.

\begin{itemize}[itemsep=-.3em]
\item[•] Regarding the characteristic scale~(${{k_b}}$), the obtained errors are between S4 and PC for all considered survey synergies, where the lowest errors arise from PC. 
\item[•] PC, M5+S4 and M5 exhibit similar performance in constraining the characteristic scale~(${{k_b}}$) for its fiducial value of ${{k_b}}=0.05$, for all survey combinations. Besides, LB+S4 and PL+S4 also show nearly identical sensitivity to the characteristic scale~(${{k_b}}$) for all survey synergies~(except SKA1-(CS), where LB+S4 performs slightly better).
\item[•] For characteristic scale~(${{k_b}}$), SKA1-(CS) and SKA2-(CS) show nearly equal sensitivity for PC, M5+S4, M5 and LB+S4. However, for S4 and PL+S4, SKA2-(CS) marginally outperforms SKA1-(CS).
\item[•] For SKA intensity mapping surveys, the sensitivity of SKA1-(IM1) towards the characteristic scale is stronger than SKA1-(IM2). However, between SKA1-(IM2) and SKA2-(CS), the constraining strength of SKA1-(IM2) is more than SKA2-(CS).
\item[•] For each CMB mission, the combination of SKA1-(IM2) and SKA1-(IM1) yields the best constraints. If we include SKA1-(IM2) data with EUCLID-(CS+GC) data, the combination surpasses the sensitivity of SKA1-(IM1) and almost reaches the sensitivity of SKA1-(IM1+IM2).
\end{itemize}
\end{itemize}

\begin{figure}[h!]
\begin{mdframed}
\captionsetup{font=footnotesize}
\center
\begin{minipage}[b1]{1.0\textwidth}
$\begin{array}{rl}
\includegraphics[width=0.5\textwidth]{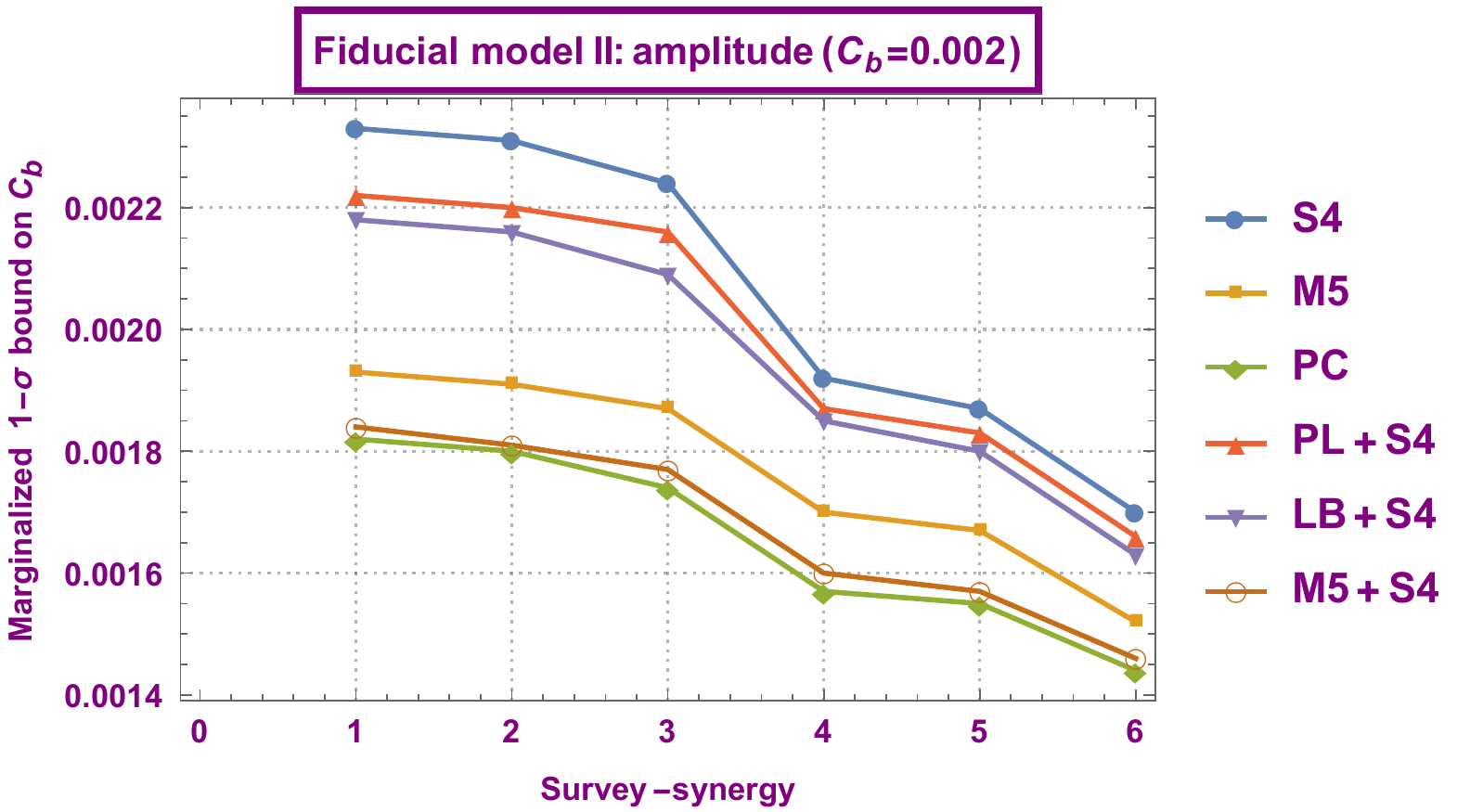} &
\includegraphics[width=0.5\textwidth]{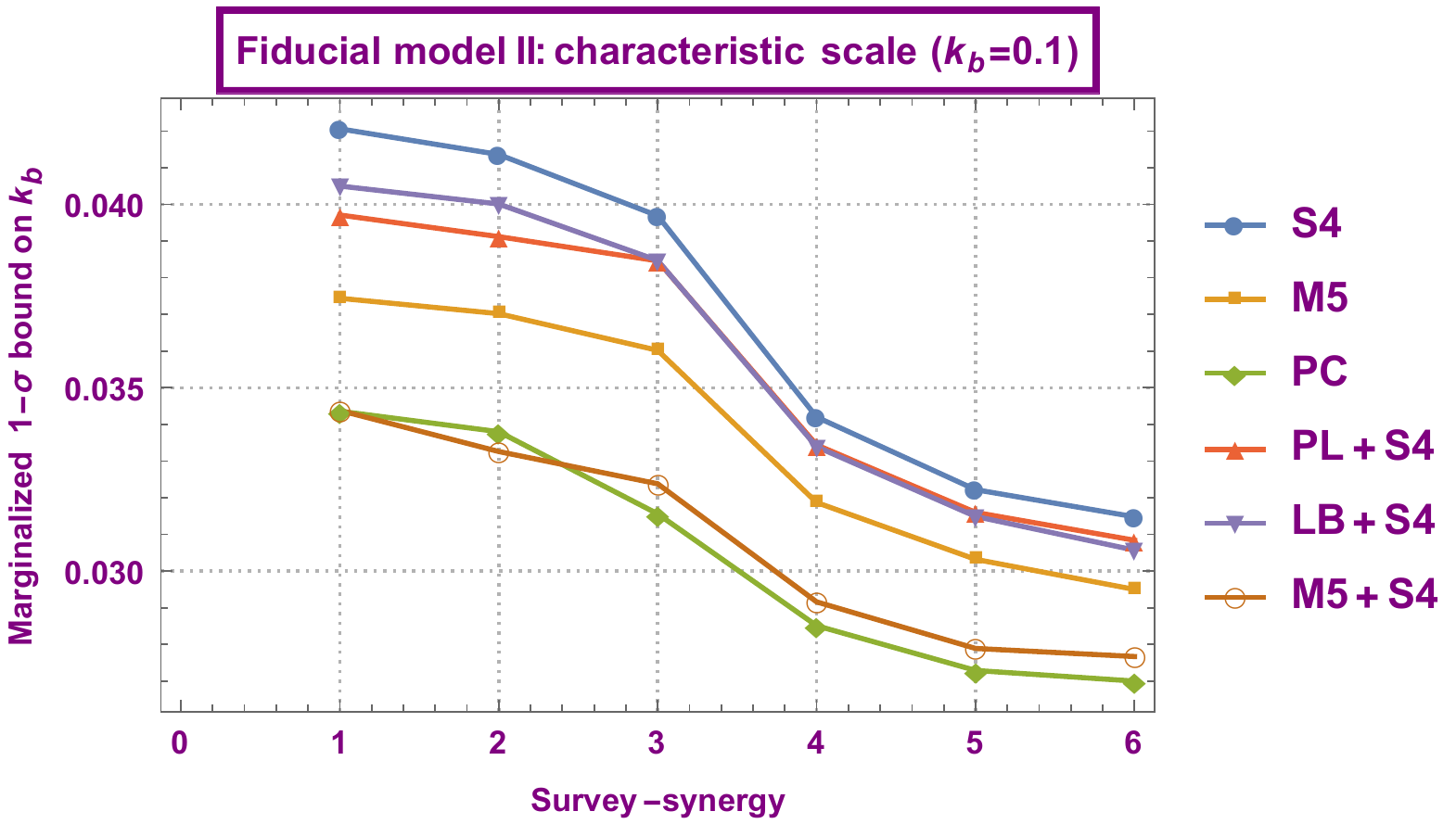}\\
\end{array}$
\caption[]{\label{fig:Bump-2}{\fontfamily{bch}
In the \textit{left}, and \textit{right} figures, we illustrate the marginalized 1-$\sigma$ bounds that have been obtained for the amplitude~($ {{C_b}} $), and characteristic scale~(${{k_b}}$) of the \textit{bump feature} model, respectively. These graphical plots exhibit the found results for the \textit{fiducial model II} of the \textit{bump feature}, which can be found in table~\ref{table:CMB-S4-Bump} to \ref{table:CORE-M5+CMB-S4-Bump} of appendix~\ref{tables}. On the right side of each plot, the CMB experiments taken into consideration are listed. The numbers \textbf{1}, \textbf{2}, \textbf{3}, \textbf{4}, \textbf{5}, and \textbf{6} along \textbf{X-axes} denote the combinations of cosmological surveys considered in this analysis, namely, \textbf{CMB+SKA1-CS}, \textbf{CMB+SKA2-CS}, \textbf{CMB+SKA1-IM2}, \textbf{CMB+SKA1-IM1}, \textbf{CMB+SKA1-(IM1+IM2)}, \textbf{CMB+EUCLID-(GC+CS)+SKA1-IM2}, respectively. Along \textbf{Y-axes}, we depict the marginalized 1-$\sigma$ errors that are achieved for the feature model parameters.
}}
\end{minipage}
\hfill
\end{mdframed}
\end{figure}

\subsubsection{Fiducial model II: ${{C_b}}=0.002$; ${{k_b}}=0.1$}
The next picture of bump feature model arises for the fiducial model II, which has been analyzed here; the found results of which have been graphically delineated in figure~\ref{fig:Bump-2}. The conclusions that can be derived from these results are summarized subsequently for each parameter of the fiducial model individually and step-wise.
\begin{itemize}[itemsep=-.3em]
\item[\ding{109}] $\textbf{Bump amplitude~($\mathbf{{{C_b}}}$):}$ 

Here at first we analyse the \textit{left} plot of figure~\ref{fig:Bump-2}, which graphically describes the obtained results for the amplitude~(${{C_b}}$) of the bump feature model.
\begin{itemize}[itemsep=-.3em]
\item[•] For fiducial model II, the lowest errors on the bump amplitude~(${{C_b}}$) come from PC, and the highest errors from S4, for all survey synergies. 
\item[•] For all survey synergies, the combination of M5 and S4~(M5+S4) nearly matches the sensitivity of PC. Similarly, the PL+S4 and LB+S4 show comparable constraining capacity for the amplitude~(${{C_b}}$) of the bump feature.
\item[•] We find that SKA2-(CS) shows slight improvements in results compared to the bounds that come from SKA1-(CS), which is applicable to all CMB missions.
\item[•] The plot shows that the amplitude~(${{C_b}}$) of the bump feature shows a better sensitivity towards SKA intensity mapping~(IM) experiments compared to SKA cosmic shear~(CS) experiments, which applies to all the CMB missions under our consideration.
\item[•] The constraining capacity of SKA1-(IM1) is stronger than SKA1-(IM2), irrespective of the CMB experiment taken into consideration. Combining SKA1-(IM2) with SKA1-(IM1) further improves the constraints by a tiny amount in comparison to the bounds generated for the SKA1-(IM1)-alone scenario, given a CMB experiment.  
\item[•] The best uncertainty on the bump amplitude~(${{C_b}}$) for a given CMB survey arises from the combination of SKA1-(IM2) and EUCLID-(CS+GC).
\end{itemize}

\item[\ding{109}] $\textbf{Characteristic scale~($\mathbf{{{k_b}}}$):}$

The results achieved for the second parameter of the fiducial model II of bump feature, which is the characteristic scale, have been exhibited in the \textit{right} plot of figure~\ref{fig:Bump-2}, and have been scrutinized subsequently.
\begin{itemize}[itemsep=-.3em]
\item[•] The constraints on the characteristic scale~(${{k_b}}$) are between S4 and PC for all the survey combinations taken into consideration~(except for SKA2-(CS), where the strongest bound is achieved when combined with M5+S4), where the lowest errors are given by PC. 
\item[•] We notice that, for the characteristic scale~(${{k_b}}$) with its fiducial value of ${{k_b}}=0.1$, the sensitivity of PL+S4 follows the sensitivity of LB+S4, except for the two occasions of SKA1-(CS) and SKA2-(CS), where the performance of PL+S4 surpasses LB+S4. For SKA1-(CS), PC and M5+S4 display almost the identical sensitivity towards ${{k_b}}$.
\item[•] Between the two SKA cosmic shear~(CS) surveys, SKA2-(CS) performs slightly better than SKA1-(CS) and, regarding SKA intensity mapping~(IM) surveys, between SKA1-(IM1) and SKA1-(IM2), SKA1-(IM1) has more constraining strength compared to SKA1-(IM2). However, between SKA IM and SKA CS surveys, the characteristic scale shows more sensitivity towards SKA IM surveys compared to SKA CS surveys.
\item[•] The combination of two SKA intensity mapping surveys~(SKA1-(IM1)+SKA1-(IM2)) imparts improved constraints on ${{k_b}}$ compared to the constraints obtainted from SKA1-(IM1) alone for all CMB experiments. However, for a given CMB experiment, the best constraint comes from the survey combination of EUCLID-(CS+GC) and SKA1-(IM2).
\end{itemize}
\end{itemize}

\begin{figure}[h!]
\begin{mdframed}
\captionsetup{font=footnotesize}
\center
\begin{minipage}[b1]{1.0\textwidth}
$\begin{array}{rl}
\includegraphics[width=0.5\textwidth]{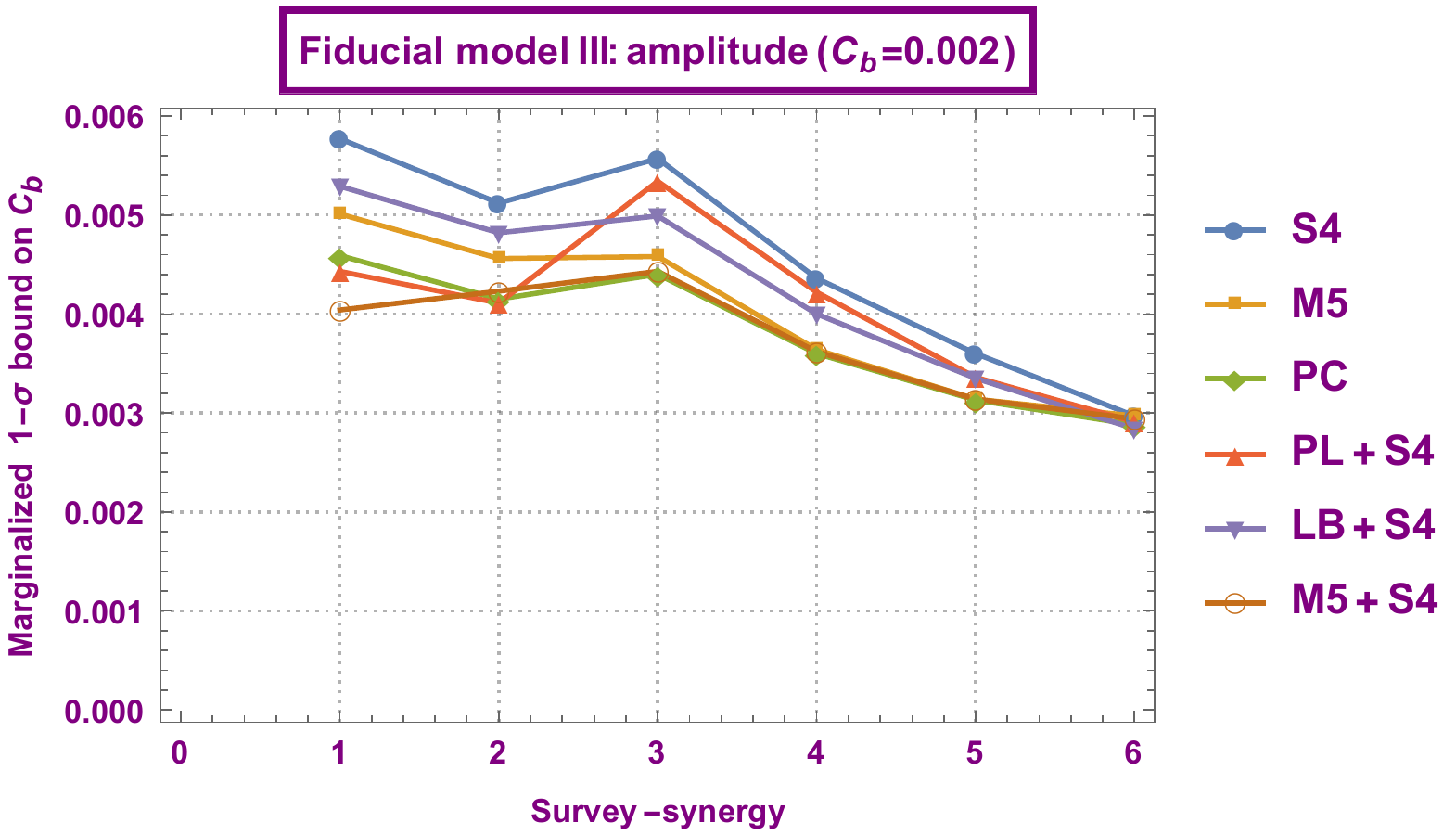} &
\includegraphics[width=0.5\textwidth]{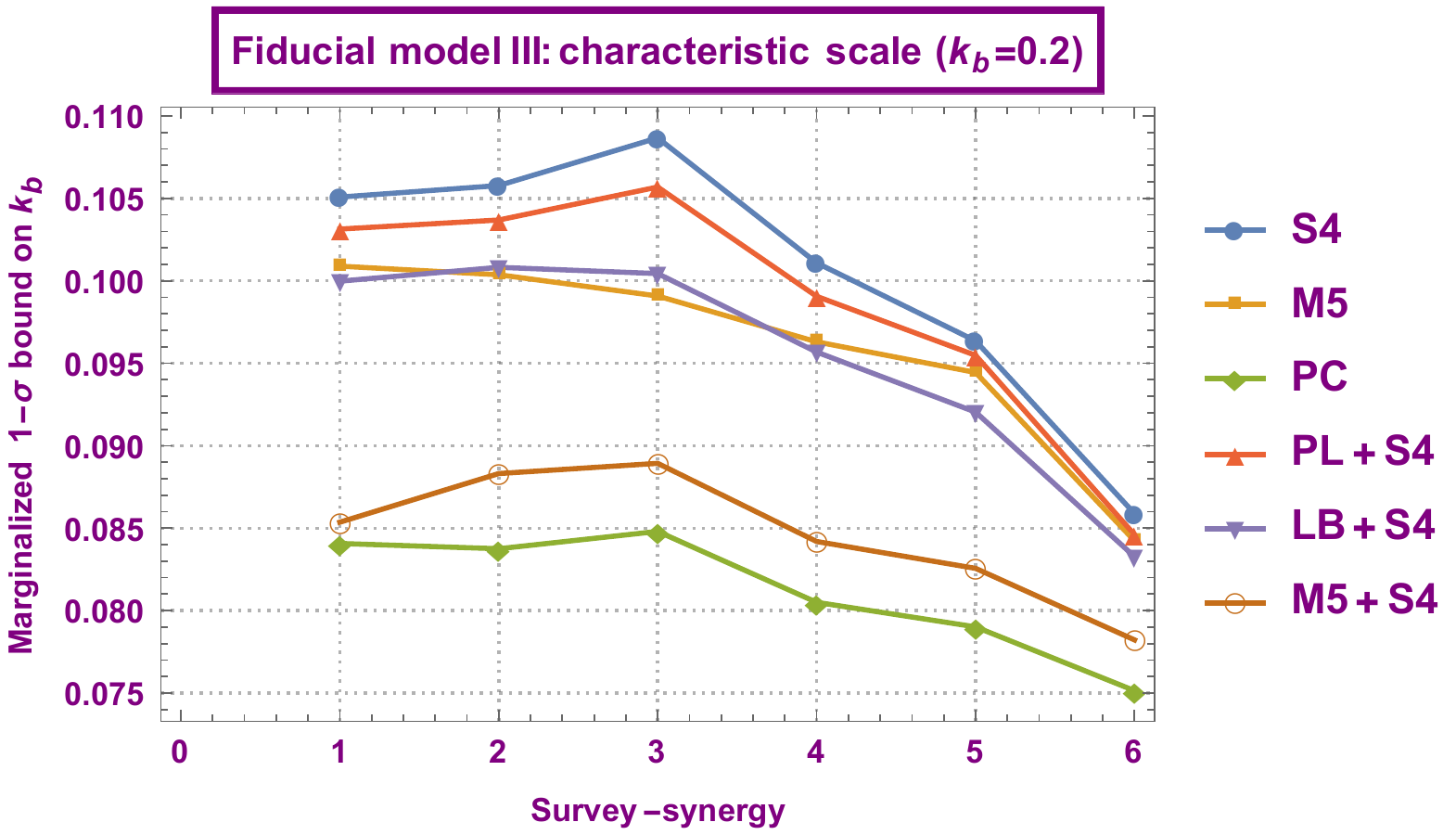}\\
\end{array}$
\caption[]{\label{fig:Bump-3}{\fontfamily{bch}
In the \textit{left}, and \textit{right} figures, we illustrate the marginalized 1-$\sigma$ bounds that have been obtained for the amplitude~($ {{C_b}} $), and characteristic scale~(${{k_b}}$) of the \textit{bump feature} model, respectively. These graphical plots exhibit the found results for the \textit{fiducial model III} of the \textit{bump feature}, which can be found in table~\ref{table:CMB-S4-Bump} to \ref{table:CORE-M5+CMB-S4-Bump} of appendix~\ref{tables}. On the right side of each plot, the CMB experiments taken into consideration are listed. The numbers \textbf{1}, \textbf{2}, \textbf{3}, \textbf{4}, \textbf{5}, and \textbf{6} along \textbf{X-axes} denote the combinations of cosmological surveys considered in this analysis, namely, \textbf{CMB+SKA1-CS}, \textbf{CMB+SKA2-CS}, \textbf{CMB+SKA1-IM2}, \textbf{CMB+SKA1-IM1}, \textbf{CMB+SKA1-(IM1+IM2)}, \textbf{CMB+EUCLID-(GC+CS)+SKA1-IM2}, respectively. Along \textbf{Y-axes}, we depict the marginalized 1-$\sigma$ errors that are achieved for the feature model parameters.
}}
\end{minipage}
\hfill
\end{mdframed}
\end{figure}

\subsubsection{Fiducial model III: ${{C_b}}=0.002$; ${{k_b}}=0.2$}
The final scenario that has been taken under consideration for the bump feature model arises for the fiducial model III, the results obtained for which are displayed in figure~\ref{fig:Bump-3}. Here in the subsequent discussions we have analysed the results thoroughly for the two feature model parameters.
\begin{itemize}[itemsep=-.3em]
\item[\ding{109}] $\textbf{Bump amplitude~($\mathbf{{{C_b}}}$):}$ 

In figure~\ref{fig:Bump-3}, the \textit{left} plot illustrates the obtained results for the amplitude~($ {{C_b}} $) of the bump feature model, which we have examined here in detail.
\begin{itemize}[itemsep=-.3em]
\item[•] The weakest bounds on the bump amplitude~($ {{C_b}} $) have been offered by S4 for any given survey synergy. For EUCLID-(CS+GC)+SKA1-(IM2), the M5 and S4 provide the same bounds on bump amplitude~($ {{C_b}} $). 
\item[•] Regarding the lowest uncertainties on bump amplitude~($ {{C_b}} $) for this particular case of fiducial values, ${{C_b}}=0.002$; ${{k_b}}=0.2$, any particular CMB experiment does not provide the lowest errors for all survey synergies, unlike the other previous cases~(mostly). Instead, for SKA1-(CS), SKA2-(CS) and EUCLID-(CS+GC)+SKA1-(IM2), the lowest bounds come from M5+S4, PL+S4 and LB+S4, respectively; whereas for SKA1-(IM1), SKA1-(IM2) and SKA1-(IM1+IM2), the lowest uncertainties are offered by the PC.
\item[•] SKA2-(CS) shows a better constraining ability compared to SKA1-(CS) for amplitude, which applies to all CMB missions, with the exception of M5+S4.
\item[•] The bump amplitude~($ {{C_b}} $) shows more sensitivity towards SKA2-(CS) compared to SKA1-(IM2) for all CMB experiments. However, for a given CMB experiment, SKA1-(IM2) outperforms SKA1-(CS) except for M5+S4, where M5+S4+SKA1-(CS) provides a stronger bound than M5+S4+SKA1-(IM2).
\item[•] Given a CMB experiment, the bounds that come from SKA1-(IM1) are stronger than the bounds that arise from SKA1-(CS). When comparing SKA1-(IM1) and SKA2-(CS), here also, the performance of SKA1-(IM1) is better than SKA2-(CS)~(except for PL+S4, where PL+S4+SKA2-(CS) provides a smaller uncertainty than PL+S4+SKA1-(IM1)).
\item[•] The combination of SKA1-(IM1) and SKA1-(IM2) provides better constraints on bump amplitude~($ {{C_b}} $) compared to SKA1-(CS), SKA2-(CS), SKA1-(IM1), and SKA1-(IM2). However, the best constraints on bump amplitude~($ {{C_b}} $) come from EUCLID-(CS+GC)+SKA1-(IM2).
\end{itemize}

\item[\ding{109}] $\textbf{Characteristic scale~($\mathbf{{{k_b}}}$):}$

The performance of different survey synergies taken into consideration in constraining the characteristic scale~(${{k_b}}$) of the bump feature has been discussed here, where the corresponding results are depicted in the \textit{right} plot of figure~\ref{fig:Bump-3}.

\begin{itemize}[itemsep=-.3em]
\item[•] The bounds on the characteristic scale~(${{k_b}}$) for this fiducial value, ${{k_b}}=0.2$, are between S4 and PC for all considered survey synergies, where the strongest bounds come from PC. 
\item[•] For all the CMB experiments taken into consideration, SKA1-(CS) shows more sensitivity towards ${{k_b}}$ than SKA2-(CS), except for M5 and PC. For M5 and PC, SKA2-(CS) provides slightly improved bounds on ${{k_b}}$ compared to SKA1-(CS).
\item[•] If we compare SKA1-(CS) with SKA1-(IM2), SKA1-(CS) surpasses the sensitivity of SKA1-(IM2) in constraining the characteristic scale~(${{k_b}}$), except for M5, where SKA1-(IM2) gives a smaller uncertainty.
\item[•] If we compare the performances of SKA2-(CS) and SKA1-(IM2) in constraining $k_B$, depending upon different CMB experiments, they show varied sensitivity. For S4, PC, PL+S4 and M5+S4, the constraining capacity of SKA2-(CS) is higher than SKA1-(IM2), and for M5 and LB+S4, SKA1-(IM2) performed better compared to SKA2-(CS).
\item[•] SKA1-(IM1) provides uncertainties, which are better than the uncertainties that come from SKA1-(CS), SKA2-(CS), and SKA1-(IM2). The combination of SKA1-(IM1) and SKA1-(IM2) provides further improved bounds than SKA1-(IM1); however, the best bounds are given by EUCLID-(CS+GC)+SKA1-(IM2).
\end{itemize}
\end{itemize}

\begin{figure}[h!]
\begin{mdframed}
\captionsetup{font=footnotesize}
\center
\begin{minipage}[b1]{1.0\textwidth}
$\begin{array}{rl}
\includegraphics[width=0.5\textwidth]{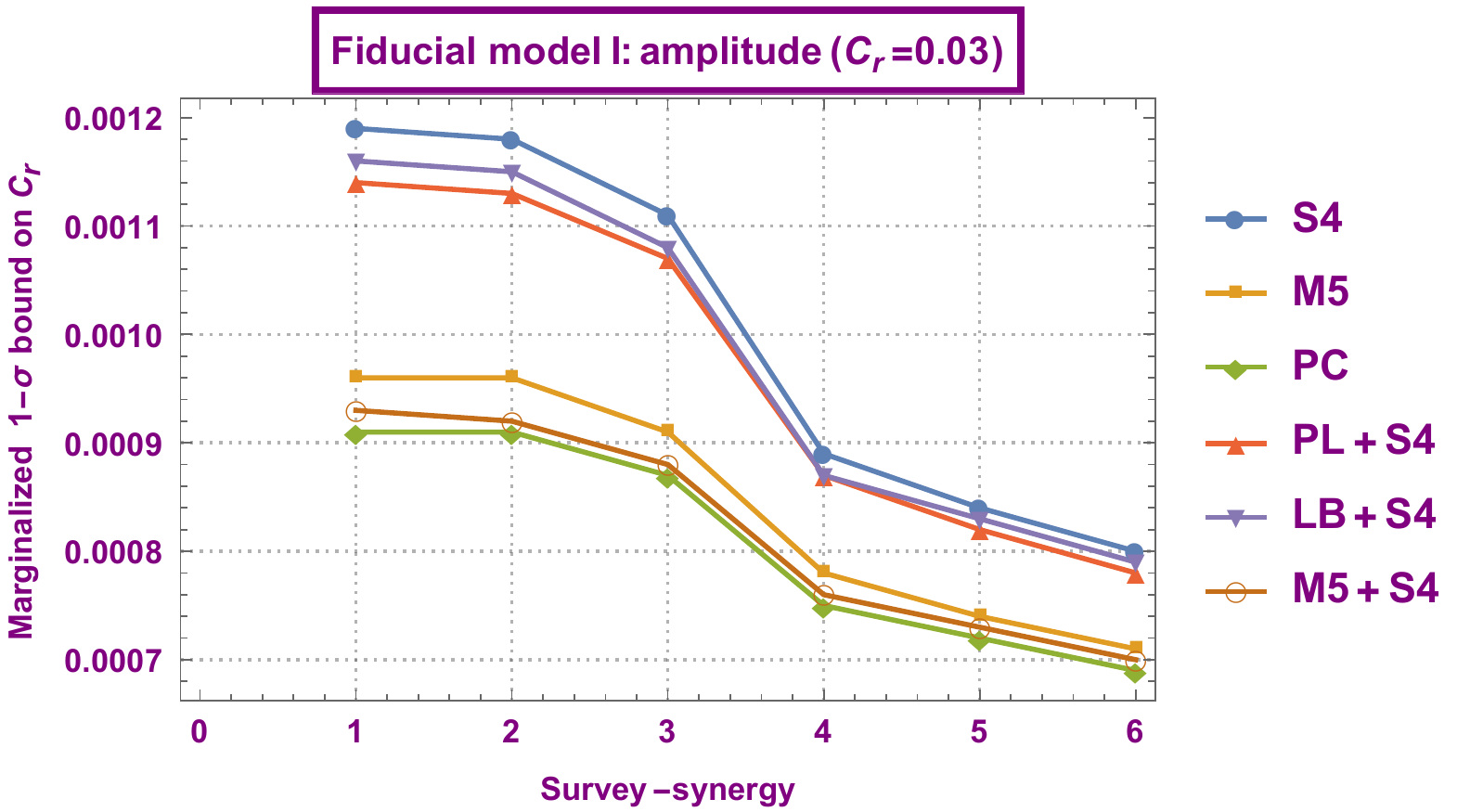} &
\includegraphics[width=0.5\textwidth]{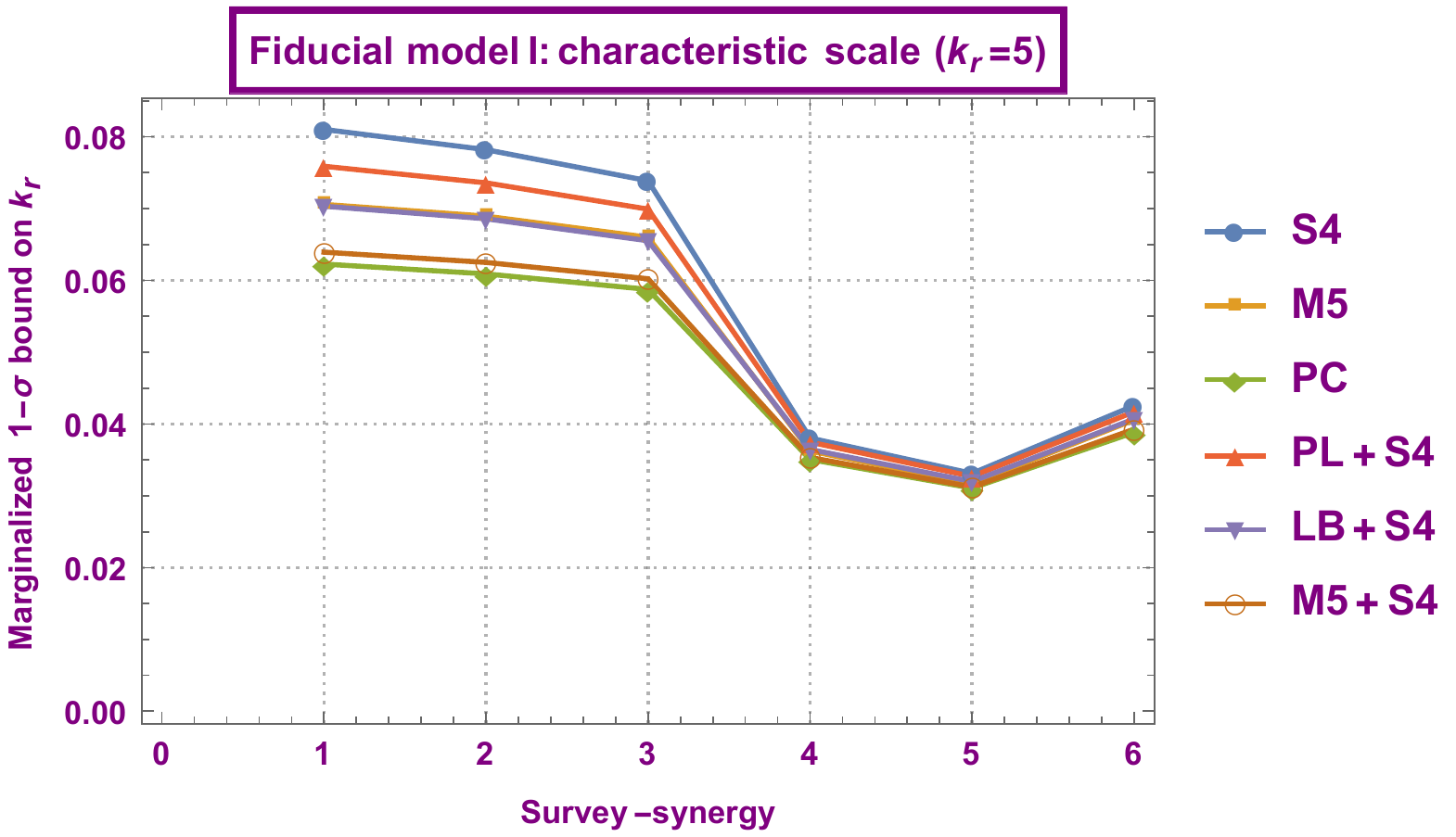}\\
\multicolumn{2}{c}{\includegraphics[width=0.5\textwidth]{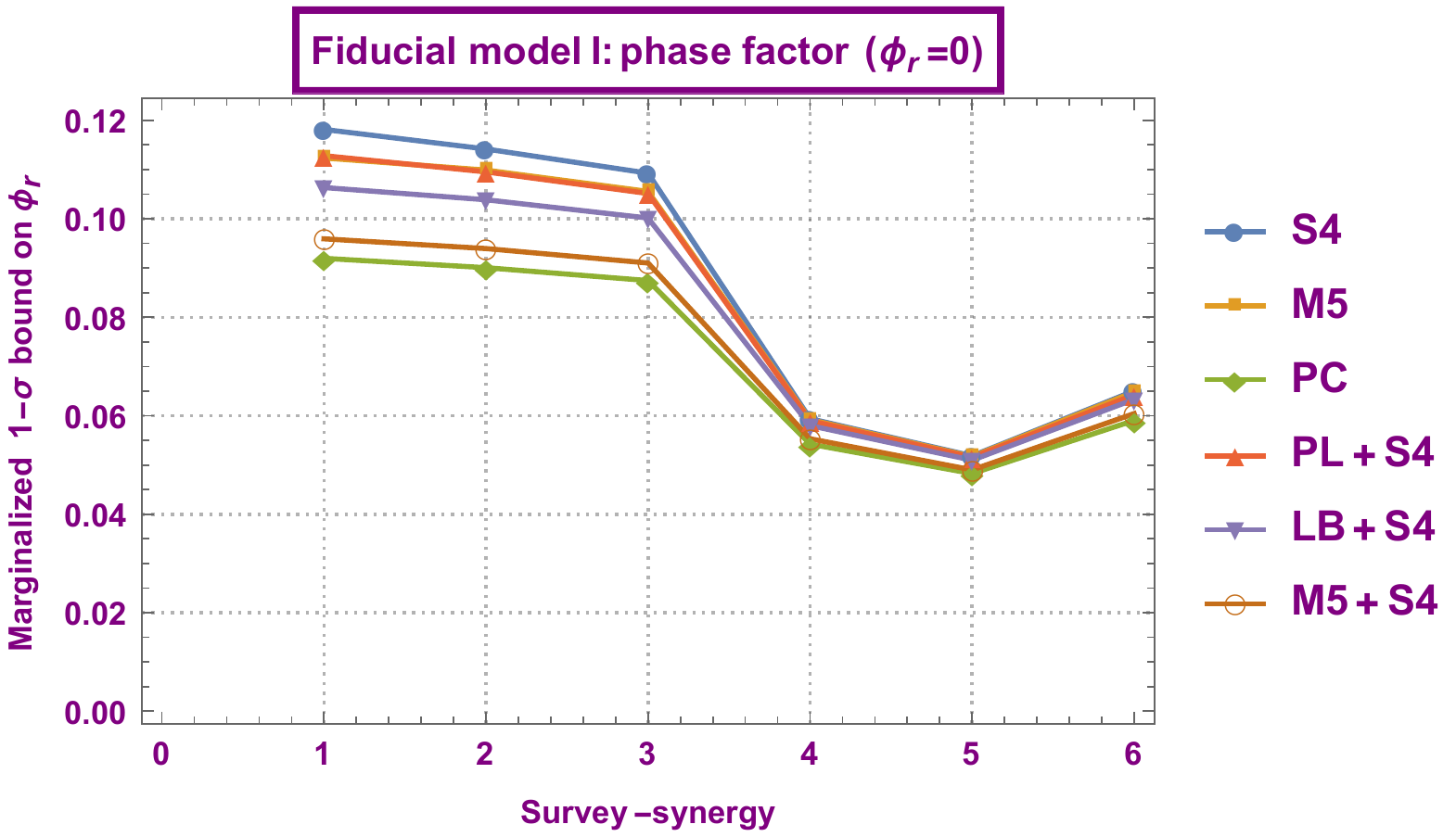}}
\end{array}$
\caption[]{\label{fig:Resonance-1}{\fontfamily{bch}
In the upper \textit{left}, upper \textit{right}, and lower \textit{middle} figures, we illustrate the marginalized 1-$\sigma$ bounds that have been obtained for the \textit{amplitude}~($ {{C_r}} $), \textit{characteristic scale}~(${{k_r}}$), and \textit{phase factor}~(${{\phi_r}}$) of the \textit{resonance feature} signal, respectively. These graphical plots exhibit the found results for the \textit{fiducial model I} of the \textit{resonance feature}, which can be found in table~\ref{table:CMB-S4-Resonance} to~\ref{table:CORE-M5+CMB-S4-Resonance} of appendix~\ref{tables}. On the right side of each plot, the CMB experiments taken into consideration are listed. The numbers \textbf{1}, \textbf{2}, \textbf{3}, \textbf{4}, \textbf{5}, and \textbf{6} along \textbf{X-axes} denote the combinations of cosmological surveys considered in this analysis, namely, \textbf{CMB+SKA1-CS}, \textbf{CMB+SKA2-CS}, \textbf{CMB+SKA1-IM2}, \textbf{CMB+SKA1-IM1}, \textbf{CMB+SKA1-(IM1+IM2)}, \textbf{CMB+EUCLID-(GC+CS)+SKA1-IM2}, respectively. Along \textbf{Y-axes}, we depict the marginalized 1-$\sigma$ errors that are achieved for the feature model parameters.}}
\end{minipage}
\hfill
\end{mdframed}
\end{figure}
\subsection{Feature model III: resonance feature}
Let us now analyze the results of the third case of feature models, which is the resonance feature signal. 
\subsubsection{Fiducial model I: $ {{C_r}}=0.03 $; ${{k_r}}=5$; ${{\phi_r}}=0$}
Let us begin our discussion of the resonance feature with its first case, which is fiducial model I. All the results associated with the fiducial model I of the resonance feature are illustrated in figure~\ref{fig:Resonance-1}, separately for each feature parameter, which allows us to extract the following conclusions from them:
%The first case of the resonance feature is for the fiducial values, $ {{C_r}}=0.03 $; ${{k_r}}=5$; ${{\phi_r}}=0$. All the results for these fiducial values are illustrated in figure~\ref{fig:Resonance-1} individually for each model parameter of the resonance feature, from which the following inferences can be drawn:
\begin{itemize}[itemsep=-.3em]
\item[\ding{109}] $\textbf{Oscillation amplitude~($\mathbf{{{C_r}}}$):}$ 

Here, we have discussed the results of the oscillation amplitude~($ {{C_r}} $) of the resonance feature signal. The found results of the oscillation amplitude~($ {{C_r}} $) are delineated graphically in the \textit{left} plot of figure~\ref{fig:Resonance-1}.
 
\begin{itemize}[itemsep=-.3em]
\item[•] In the context of the oscillation amplitude~($ {{C_r}} $) of the resonance feature, the weakest constraints come from S4 and the strongest constraints come from PC for all the survey synergies.
\item[•] For all surveys combinations, M5+S4 reaches the sensitivity of PC.
\item[•] The best uncertainty on the oscillation amplitude~($ {{C_r}} $) comes from EUCLID-(CS+GC)+SKA1-(IM2) for a given CMB experiment.
\item[•] Towards oscillation amplitude~($ {{C_r}} $), the sensitivity of SKA1-(CS) is almost the same as the sensitivity of SKA2-(CS). Regarding intensity mapping surveys, the constraining ability of SKA1-(IM1) is better than SKA1-(IM2). However, if we compare 
between SKA IM and SKA CS surveys, then both SKA IM surveys provide tighter bounds compared to SKA CS surveys. 
\item[•] When we combine two SKA IM~(SKA1-(IM1+IM2)) surveys, we get further narrower bounds compared to the bounds that come from SKA1-(IM1) alone.
\end{itemize}
\item[\ding{109}] $\textbf{Characteristic scale~($\mathbf{{{k_r}}}$):}$

The \textit{right} plot of figure~\ref{fig:Resonance-1} is devoted to presenting the obtained results for the characteristic scale~(${{k_r}}$) of resonance feature signal, which has been scrutinized here.
\begin{itemize}[itemsep=-.3em]
\item[•] Similar to the oscillation amplitude~(${{C_r}}$), for characteristic scale~(${{k_r}}$) as well, the errors lie between S4 and PC, where the smallest errors come from PC. 
\item[•] For this fiducial value of the characteristic scale~(${{k_r}}=5$), M5, 
and LB+S4 are almost equally sensitive towards ${{k_r}}$. On the other hand, M5 shows a constraining capacity which is comparable to PC. 
\item[•] The best bounds on ${{k_r}}$ come from SKA1-(IM1+IM2) for all CMB experiments.  
\item[•] For this fiducial model of resonance feature, SKA1-(IM1) performs better than EUCLID-(CS+GC)+SKA1-(IM2) in constraining ${{k_r}}$.
\item[•] For the characteristic scale~(${{k_r}}$), SKA1-(CS) provides slightly weaker bounds than SKA2-(CS), whereas SKA2-(CS) provides weaker bounds than SKA1-(IM2). 
\item[•] The constraining strength of SKA1-(IM1) is greater in comparison to SKA1-(IM2) and when they are combined, the combination shows an even better constraining capacity than the individuals.
\end{itemize}
\item[\ding{109}] $\textbf{Phase factor~($\mathbf{{{\phi_r}}}$):}$

The detailed scrutiny of the \textit{middle} plot of figure~\ref{fig:Resonance-1} has been done here, which is summarizing the obtained results of the phase factor~(${{{\phi_r}}}$) in a graphical format. 
\begin{itemize}[itemsep=-.3em] 
\item[•] The uncertainties on phase factor~($ {{\phi_r}} $) are between S4 and PC, where PC provides the lowest uncertainties.
\item[•] Given a CMB experiment, the lowest constraint on the phase angle~($ {{\phi_r}} $) arises for SKA1-(IM1+IM2). 
\item[•] The combination of S4 and Planck almost meets the sensitivity of M5 for all considered LSS surveys.
\item[•] As in the case of characteristic scale~(${{k_r}}$), here also, the comparative behaviour of SKA CS and SKA IM experiments for phase angle~($ {{\phi_r}} $) is similar. The SKA2-(CS) provides better bounds than SKA1-(CS) and SKA1-(IM2) provides further improved bounds than SKA2-(CS).
\item[•] If we consider the comparison between SKA1-(IM1) and SKA1-(IM2), we find that SKA1-(IM1) gives much tighter constraints on $ {{\phi_r}} $ than SKA1-(IM2), and when we add SKA1-(IM2) with SKA1-(IM1), then it further tightens the bounds compared to the bounds coming from SKA1-(IM1).
\item[•] The combination of SKA1-(IM2) and EUCLID-(CS+GC) yields constraints which are better than SKA1-(IM2) but can not surpass the sensitivity of SKA1-(IM1). 
\end{itemize}
\end{itemize}

\begin{figure}[h!]
\begin{mdframed}
\captionsetup{font=footnotesize}
\center
\begin{minipage}[b1]{1.0\textwidth}
$\begin{array}{rl}
\includegraphics[width=0.5\textwidth]{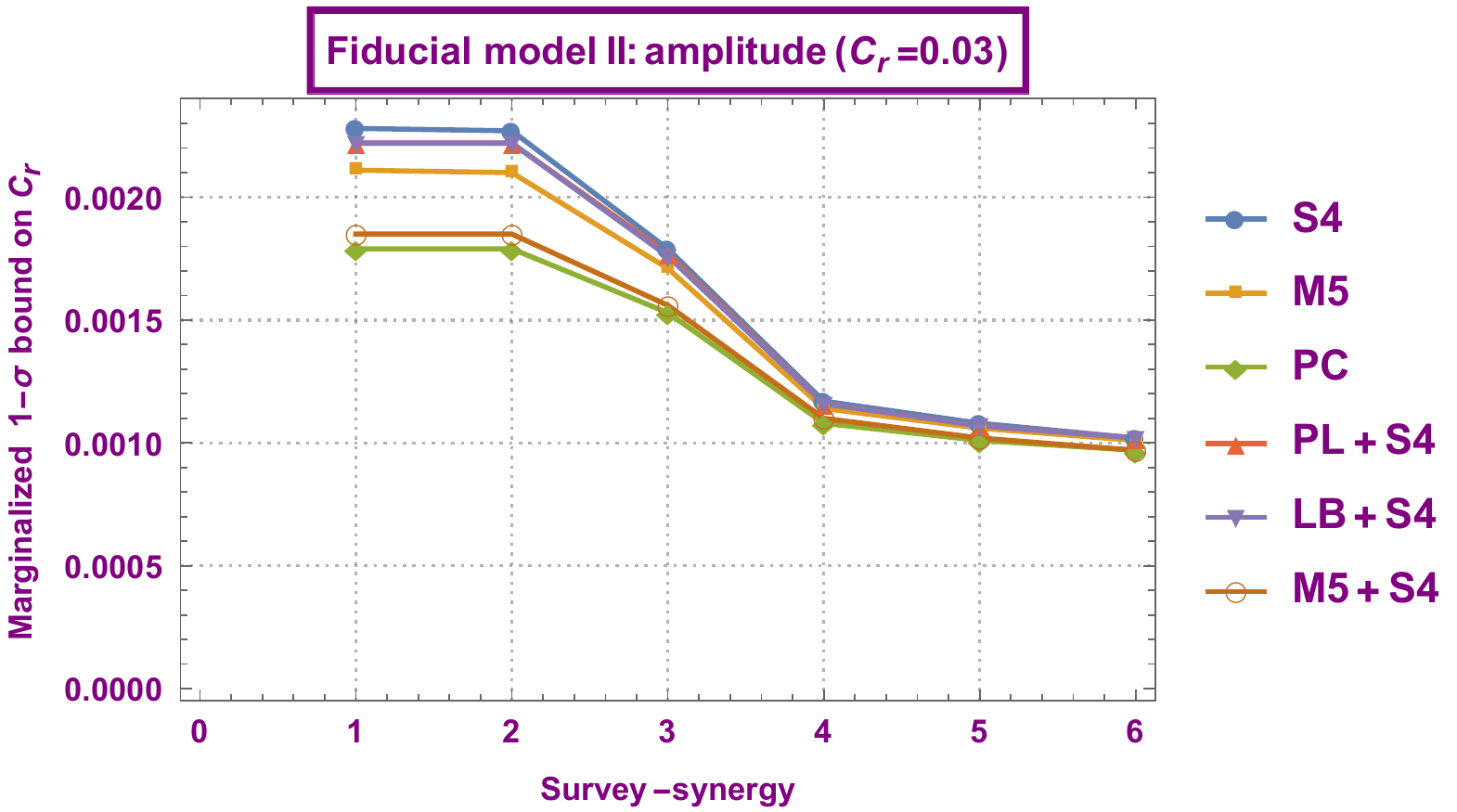} &
\includegraphics[width=0.5\textwidth]{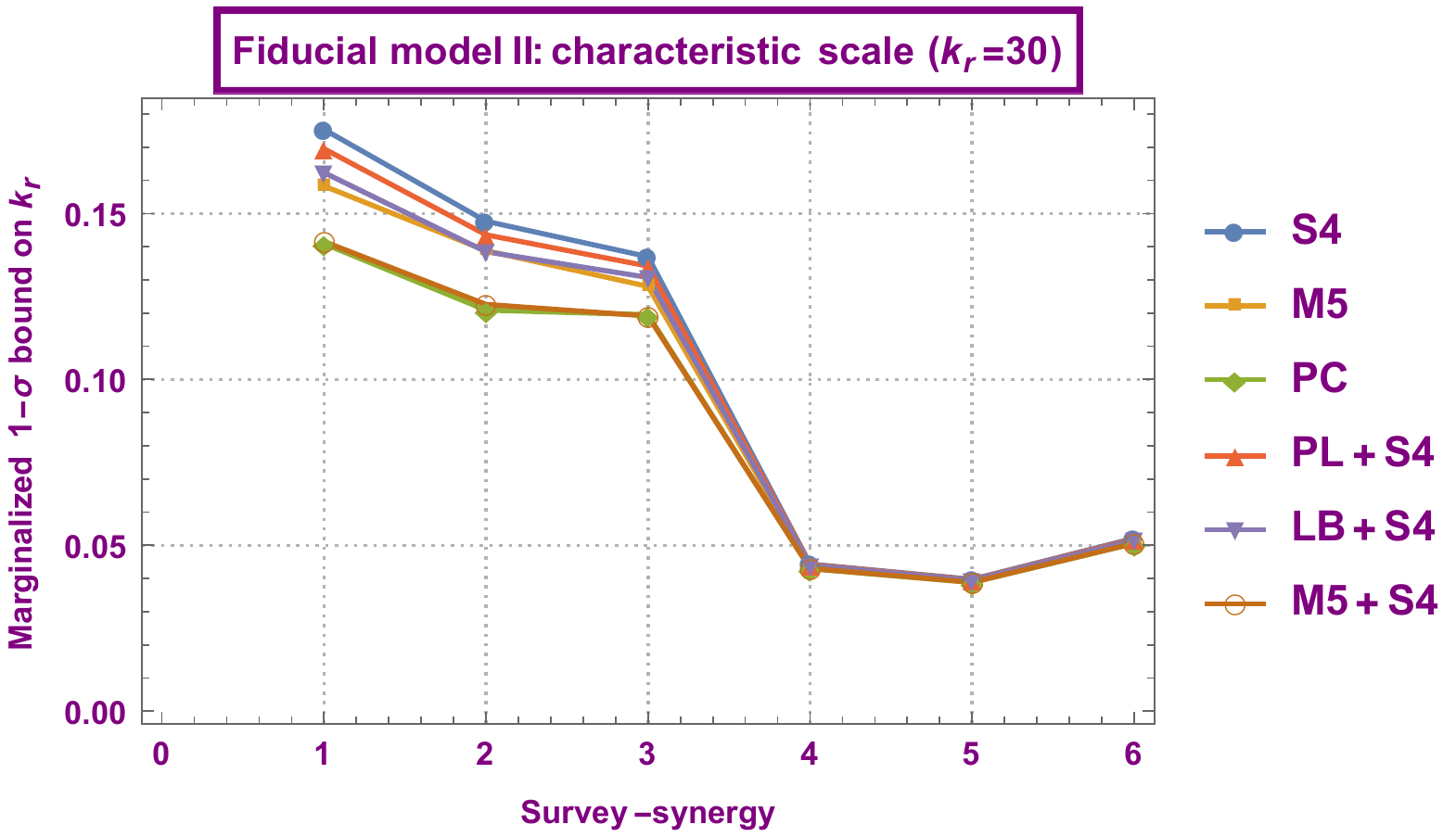}\\
\multicolumn{2}{c}{\includegraphics[width=0.5\textwidth]{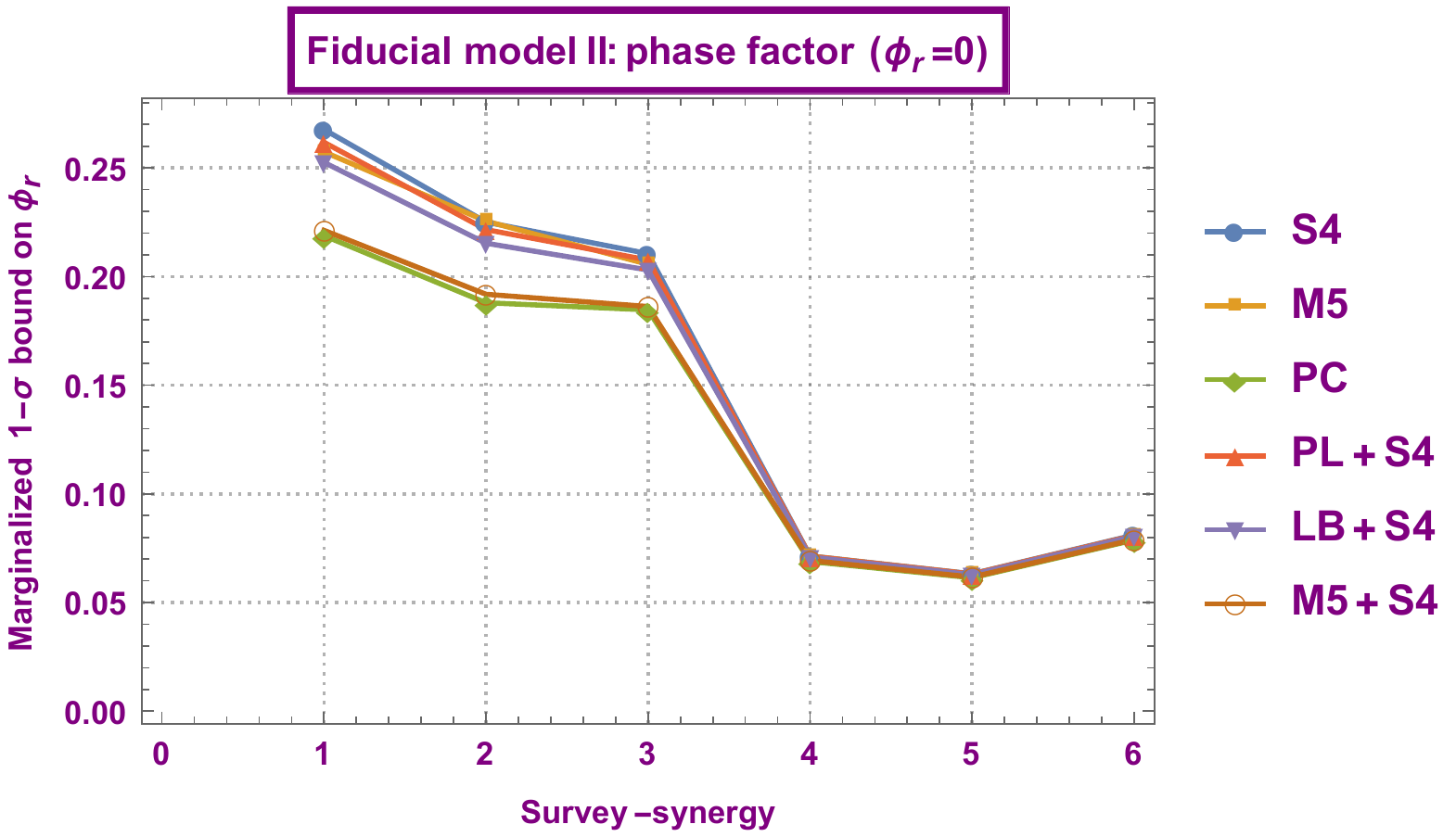}}
\end{array}$
\caption[]{\label{fig:Resonance-2}{\fontfamily{bch}
In the upper \textit{left}, upper \textit{right}, and lower \textit{middle} figures, we illustrate the marginalized 1-$\sigma$ bounds that have been obtained for the \textit{amplitude}~($ {{C_r}} $), \textit{characteristic scale}~(${{k_r}}$), and \textit{phase factor}~(${{\phi_r}}$) of the \textit{resonance feature} signal, respectively. These graphical plots exhibit the found results for the \textit{fiducial model II} of the \textit{resonance feature}, which can be found in table~\ref{table:CMB-S4-Resonance} to~\ref{table:CORE-M5+CMB-S4-Resonance} of appendix~\ref{tables}. On the right side of each plot, the CMB experiments taken into consideration are listed. The numbers \textbf{1}, \textbf{2}, \textbf{3}, \textbf{4}, \textbf{5}, and \textbf{6} along \textbf{X-axes} denote the combinations of cosmological surveys considered in this analysis, namely, \textbf{CMB+SKA1-CS}, \textbf{CMB+SKA2-CS}, \textbf{CMB+SKA1-IM2}, \textbf{CMB+SKA1-IM1}, \textbf{CMB+SKA1-(IM1+IM2)}, \textbf{CMB+EUCLID-(GC+CS)+SKA1-IM2}, respectively. Along \textbf{Y-axes}, we depict the marginalized 1-$\sigma$ errors that are achieved for the feature model parameters.}}
\end{minipage}
\hfill
\end{mdframed}
\end{figure}

\subsubsection{Fiducial model II: ${{C_r}}=0.03$; ${{k_r}}=30$; ${{\phi_r}}=0$}
The fiducial model II, gives rise to the second case of the resonance feature model. The obtained constraints on the feature parameters for this scenario are summarized graphically in figure~\ref{fig:Resonance-2}.  
Below, we consider each plot of figure~\ref{fig:Resonance-2} independently. 
\begin{itemize}[itemsep=-.3em]
\item[\ding{109}] $\textbf{Oscillation amplitude~($\mathbf{{{C_r}}}$):}$ 

The obtained 1-$\sigma$ errors on the oscillation amplitude~(${{C_r}}$) of the logarithmic sinusoidal signal are depicted in the \textit{left} plot of figure~\ref{fig:Resonance-2}, from which we have summarized the following conclusions:
\begin{itemize}[itemsep=-.3em]
\item[•] For this given fiducial model of the resonance feature, the weakest and the strongest constraints on the oscillation amplitude~(${{C_r}}$) come from S4 and PC, respectively.    
\item[•] For oscillation amplitude~(${{C_r}}$), S4, PL+S4 and LB+S4 show comparable constraining strength. Additionally,  PC and M5+S4 also show comparable constraining strength for oscillation amplitude~(${{C_r}}$). 
\item[•] The best error on the oscillation amplitude~(${{C_r}}$) comes 
from EUCLID-(CS+GC)+ SKA1-(IM2) for a given CMB experiment.
\item[•] Regarding the oscillation amplitude~(${{C_r}}$), the sensitivity of SKA1-(CS) is almost the same as that of SKA2-(CS). However, for intensity mapping surveys, the constraining capacity of SKA1-(IM1) is superior to SKA1-(IM2), and between IM and CS surveys, each IM survey offers tighter bounds than CS surveys.
\item[•] The constraining capacity of the combination of SKA1-(IM1) and SKA1-(IM2) is slightly better than SKA1-(IM1). 
\end{itemize}
\item[\ding{109}] $\textbf{Characteristic scale~($\mathbf{{{k_r}}}$):}$

The \textit{right} plot of figure~\ref{fig:Resonance-2} presents the obtained results for the characteristic scale~(${{k_r}}$) of the resonance feature signal. We have examined the plot here subsequently.
\begin{itemize}[itemsep=-.3em]
\item[•] On the characteristic scale~(${{k_r}}$), the found uncertainties lie between S4 and PC, where the smallest errors arise from PC~(except SKA1-(IM2), for which M5+S4 provides the lowest error). 
\item[•] For the fiducial value of ${{k_r}}=30$, PC and M5+S4 show almost the same sensitivity to the characteristic scale~(${{k_r}}$).
\item[•] The characteristic scale~(${{k_r}}$) is more sensitive towards SKA2-(CS) than SKA1-(CS) at this fiducial value~(${{k_r}}=30$), unlike oscillation amplitude~(${{C_r}}$).
\item[•] SKA1-(IM2) slightly improves the constraints compared to SKA2-(CS). However, SKA1-(IM1) significantly improves the constraints compared to the constraints imparted by SKA1-(IM2), and when they are combined~(SKA1-(IM1+IM2)), the errors are even further reduced by a tiny amount.
\item[•] The conjunction of SKA1-(IM2) and EUCLID-(CS+GC) imposes more stringent constraints than SKA1-(IM2), but cannot exceed the sensitivity of SKA1-(IM1) and SKA1-(IM1+IM2).
\end{itemize}
\item[\ding{109}] $\textbf{Phase factor~($\mathbf{{{\phi_r}}}$):}$

The detailed analysis of the \textit{middle} plot of figure~\ref{fig:Resonance-2} has been done here, where we have summarized graphically the obtained results for the phase factor~(${{\phi_r}}$). 
\begin{itemize}[itemsep=-.3em] 
\item[•] For the phase factor~($ {{\phi_r}} $), the uncertainties lie between S4 and PC. S4 imposes the weakest bounds~(except SKA2-(CS), for which M5 imparts the weakest bound).
\item[•] The combination of M5 and S4 nearly meets the sensitivity of PC for all survey combinations.
\item[•] For all CMB missions, the smallest errors on the phase factor~($ {{\phi_r}} $) are obtained when added with SKA1-(IM1+IM2).
\item[•] For this fiducial model of the resonance feature, the phase angle exhibits more sensitivity to SKA2-(CS) than to SKA1-(CS). 
\item[•] For all the adopted CMB missions, the improvement in the bounds is very little for SKA1-(IM2) compared to SKA2-(CS), whereas for SKA1-(IM1) the bounds are significantly improved compared to SKA2-(CS).
\item[•] The sensitivity of the combination of SKA1-(IM2) and EUCLID-(CS+GC) is close enough to SKA1-(IM1) but not exceeding that of SKA1-(IM1).
\end{itemize}
\end{itemize}

\begin{figure}[h!]
\begin{mdframed}
\captionsetup{font=footnotesize}
\center
\begin{minipage}[b1]{1.0\textwidth}
$\begin{array}{rl}
\includegraphics[width=0.5\textwidth]{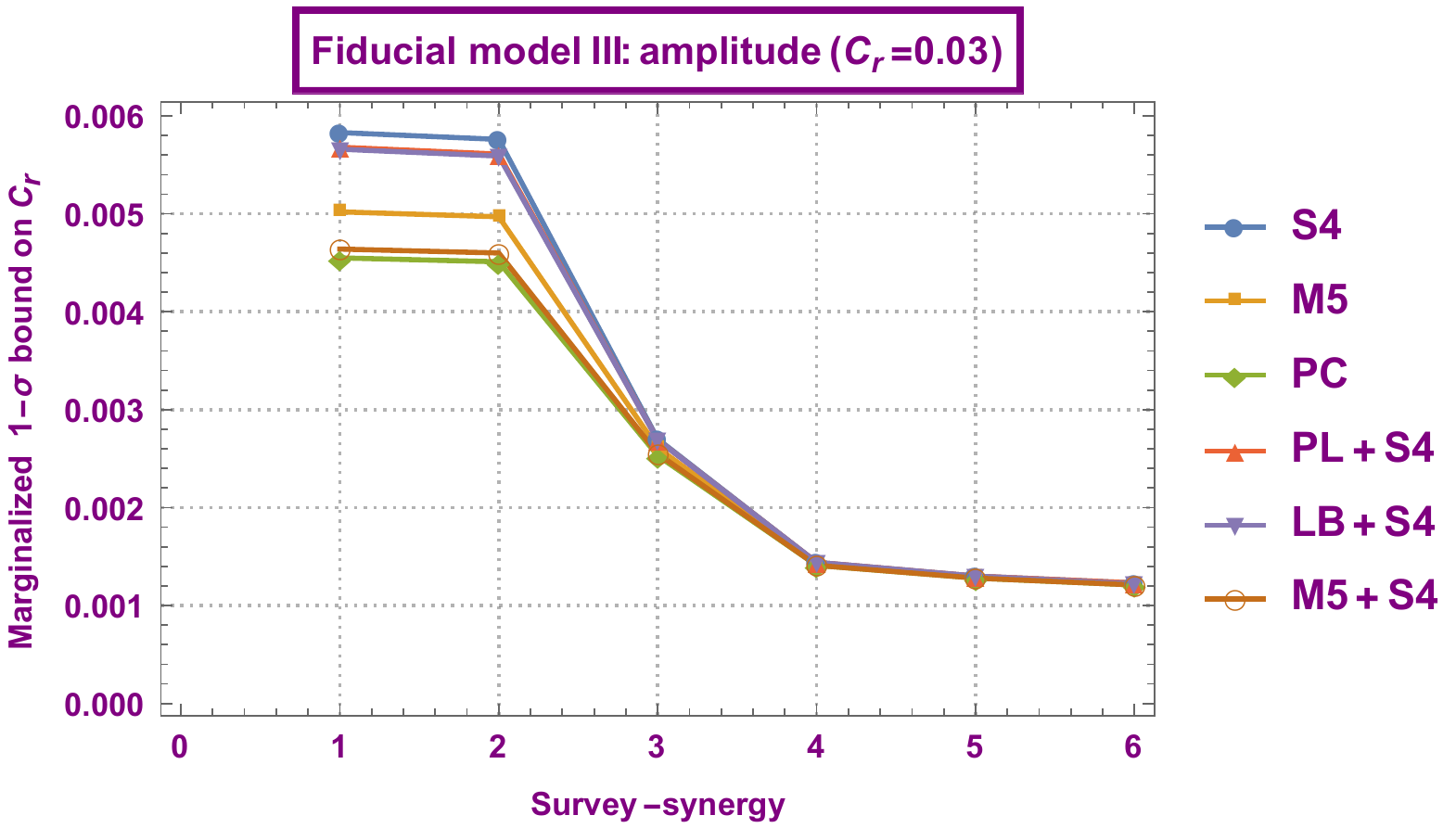} &
\includegraphics[width=0.5\textwidth]{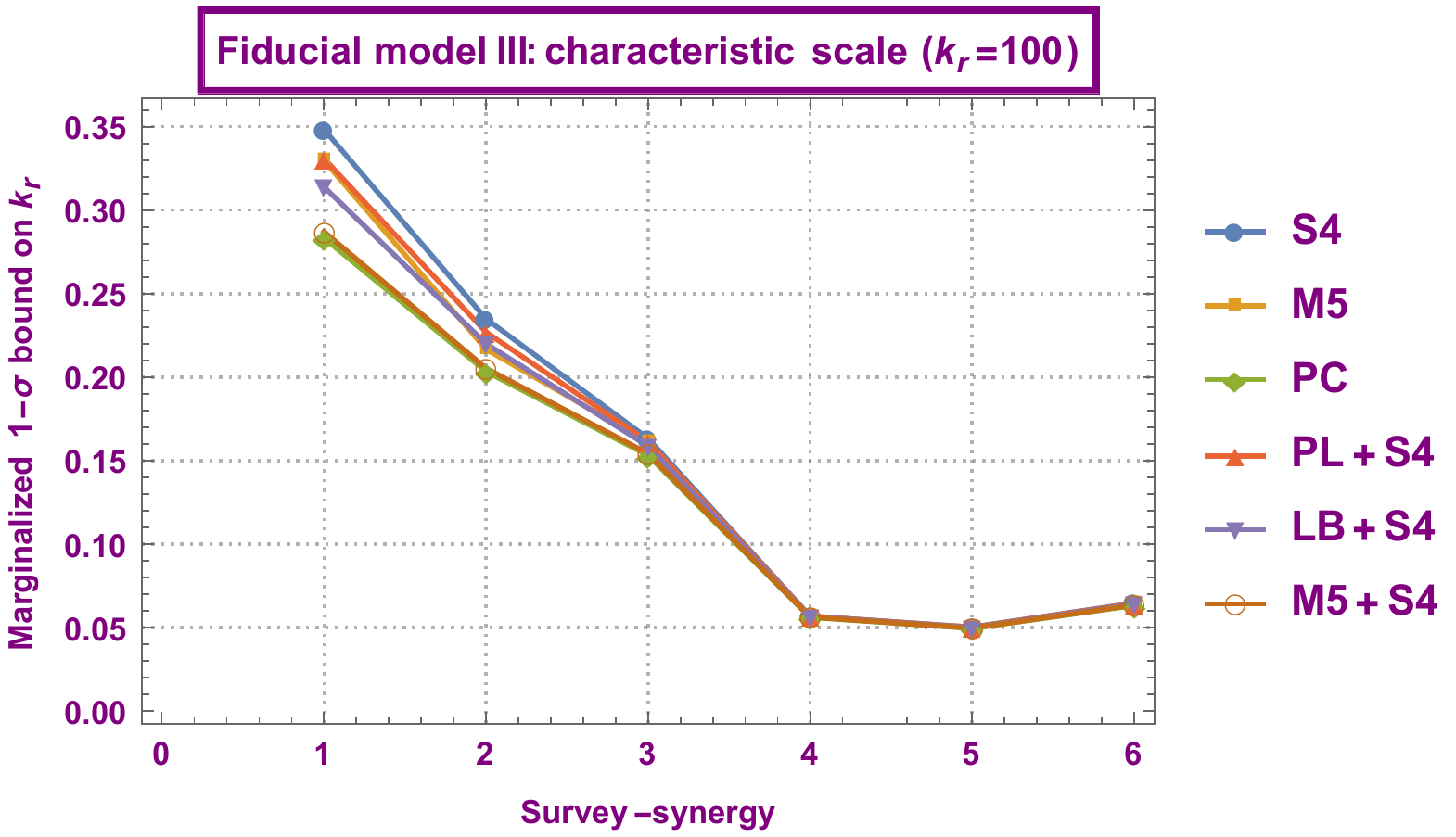}\\
\multicolumn{2}{c}{\includegraphics[width=0.5\textwidth]{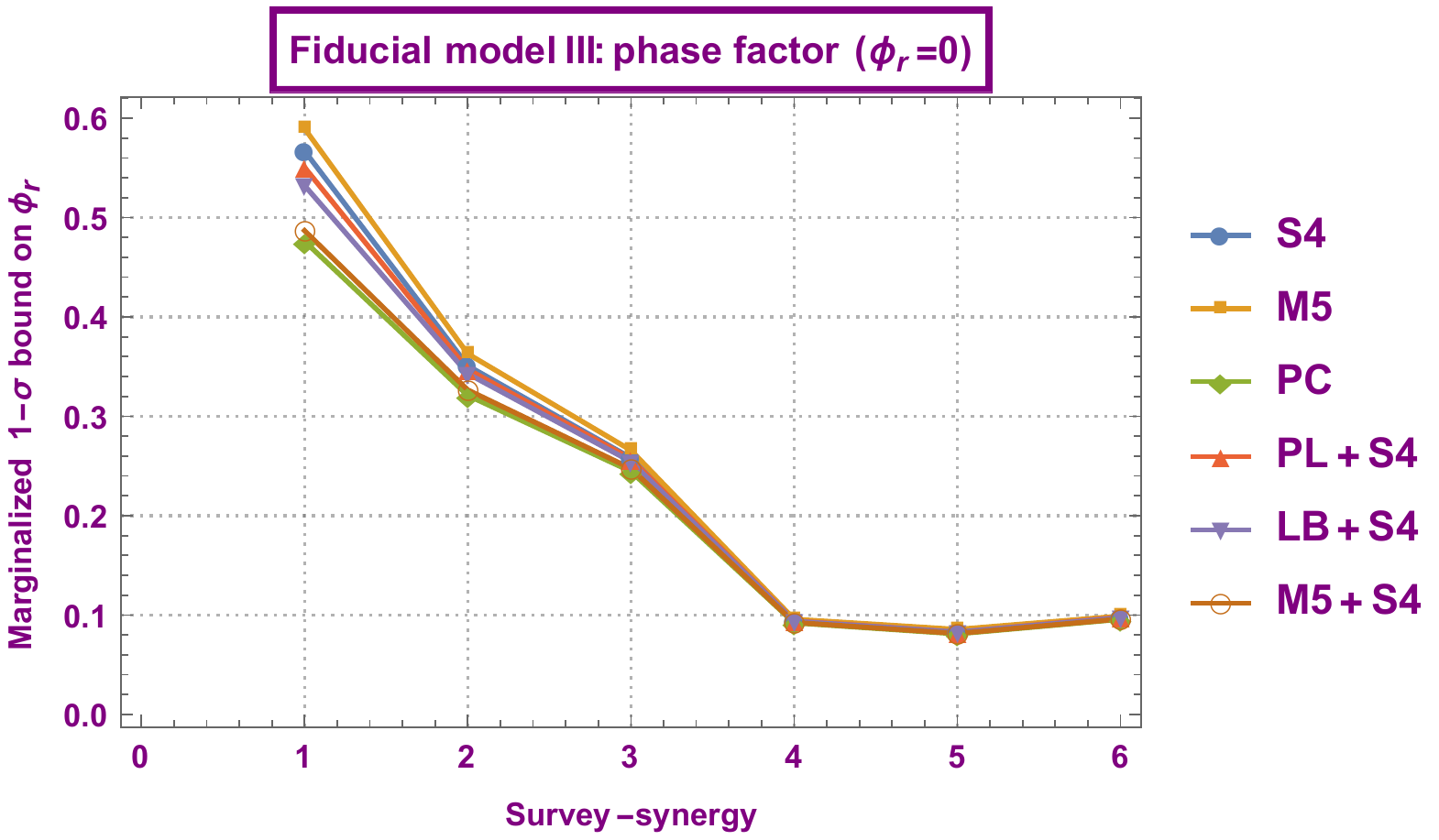}}
\end{array}$
\caption[]{\label{fig:Resonance-3}{\fontfamily{bch}
In the upper \textit{left}, upper \textit{right}, and lower \textit{middle} figures, we illustrate the marginalized 1-$\sigma$ bounds that have been obtained for the \textit{amplitude}~($ {{C_r}} $), \textit{characteristic scale}~(${{k_r}}$), and \textit{phase factor}~(${{\phi_r}}$) of the \textit{resonance feature} signal, respectively. These graphical plots exhibit the found results for the \textit{fiducial model III} of the \textit{resonance feature}, which can be found in table~\ref{table:CMB-S4-Resonance} to~\ref{table:CORE-M5+CMB-S4-Resonance} of appendix~\ref{tables}. On the right side of each plot, the CMB experiments taken into consideration are listed. The numbers \textbf{1}, \textbf{2}, \textbf{3}, \textbf{4}, \textbf{5}, and \textbf{6} along \textbf{X-axes} denote the combinations of cosmological surveys considered in this analysis, namely, \textbf{CMB+SKA1-CS}, \textbf{CMB+SKA2-CS}, \textbf{CMB+SKA1-IM2}, \textbf{CMB+SKA1-IM1}, \textbf{CMB+SKA1-(IM1+IM2)}, \textbf{CMB+EUCLID-(GC+CS)+SKA1-IM2}, respectively. Along \textbf{Y-axes}, we depict the marginalized 1-$\sigma$ errors that are achieved for the feature model parameters.}}
\end{minipage}
\hfill
\end{mdframed}
\end{figure}
\subsubsection{Fiducial model III: ${{C_r}}=0.03$; ${{k_r}}=100$; ${{\phi_r}}=0$}
The final case of the resonance feature signal arises for the fiducial model III. The results for this fiducial model are graphically illustrated in figure~\ref{fig:Resonance-3}. We now discuss the results subsequently.
\begin{itemize}[itemsep=-.3em]
\item[\ding{109}] $\textbf{Oscillation amplitude~($\mathbf{{{C_r}}}$):}$ 

Here we have discussed the \textit{left} plot of figure~\ref{fig:Resonance-3}, where the obtained 1-$\sigma$ uncertainties on the oscillation amplitude~(${{C_r}}$) of the resonance feature signal are shown.
\begin{itemize}[itemsep=-.3em]
\item[•] For a given survey combination, the lowest and the highest bounds on the oscillation amplitude~(${{C_r}}$) are obtained from PC and S4, respectively. 
\item[•] For a given CMB experiment, the best uncertainty is obtained when combined with EUCLID-(CS+GC)+SKA1-(IM2).
\item[•] The SKA2-(CS) imposes almost the same bounds on oscillation amplitude~(${{C_r}}$) as that of SKA1-(CS).
\item[•] When CMB experiments are combined with SKA1-(IM2), a significant improvement in the bounds is observed over the bounds coming from the SKA-CS experiments.
\item[•] Between two intensity mapping surveys of SKA, the constraining strength of SKA1-(IM1) is higher than SKA1-(IM2) and, when combined~(SKA1-(IM1+IM2)), the sensitivity gets enhanced by a tiny amount compared to SKA1-(IM1).
\item[•] The sensitivity of M5 reaches that of PC when combined with S4 for all survey synergies. Besides, S4, LB+S4, and PL+S4 exhibit nearly similar performance in constraining the oscillation amplitude~(${{C_r}}$).
\end{itemize}
\item[\ding{109}] $\textbf{Characteristic scale~($\mathbf{{{k_r}}}$):}$

Here we have analysed the \textit{right} plot of figure~\ref{fig:Resonance-3}, which graphically depicts the obtained results of the characteristic scale~(${{k_r}}$) of the resonance feature signal.
\begin{itemize}[itemsep=-.3em]
\item[•] The S4 and PC impart the smallest and largest errors on the characteristic scale~(${{k_r}}$) for this fiducial model, respectively. 
\item[•] M5+S4 almost reaches the sensitivity of PC in constraining the characteristic scale~(${{k_r}}$). 
\item[•] Between the SKA cosmic shear~(CS) and intensity mapping~(IM) experiments, IM experiments are more sensitive towards ${{k_r}}$ than CS experiments. Between the two SKA-CS surveys taken into consideration, SKA2-(CS) performs better than SKA1-(CS) in constraining ${{k_r}}$; similarly, between the two SKA-IM surveys, SKA1-(IM1) outperforms SKA1-(IM2) in constraining capacity.
\item[•] The addition of SKA1-(IM2) data with SKA1-(IM1) data improves the constraints on ${{k_r}}$ compared to the constraints provided by SKA1-(IM1).
\item[•] The addition of SKA1-(IM2) data with EUCLID-(CS+GC) data imposes constraints on ${{k_r}}$, which is tighter than SKA CS surveys and SKA1-(IM2) survey but cannot outperforms SKA1-(IM1) and SKA1-(IM1+IM2).
\end{itemize}
\item[\ding{109}] $\textbf{Phase factor~($\mathbf{{{\phi_r}}}$):}$

The \textit{middle} plot of figure~\ref{fig:Resonance-3} describing pictorially the 1-$\sigma$ constraints obtained for the phase factor~($ {{\phi_r}} $) has been examined here.
\begin{itemize}[itemsep=-.3em] 
\item[•] Among all the CMB experiments taken into consideration, the best bounds on phase angle~($ {{\phi_r}} $) come from PC and the weakest bounds are given by M5.
\item[•] For all CMB experiments, the lowest uncertainties are obtained from SKA1-(IM1+IM2).
\item[•] The combined performance of M5 and S4 is nearly similar to PC in constraining the phase angle~($ {{\phi_r}} $).
\item[•] SKA2-(CS) imparts more stringent constraints on the phase factor~($ {{\phi_r}} $) compared to SKA1-(CS). 
\item[•] SKA1-(IM1) imposes tighter constraints on the phase factor~($ {{\phi_r}} $) than SKA1-(IM2) does.
\item[•] SKA intensity mapping~(IM) surveys are more sensitive to the phase factor than the SKA cosmic shear~(CS).
\item[•] For the phase factor~($ {{\phi_r}} $), EUCLID-(CS+GC)+SKA1-(IM2) provides slightly weaker bounds compared to SKA1-(IM1). 
\end{itemize}
\end{itemize}

\subsection{General remarks}
It follows from this analysis that among all the CMB experiments (including individual and combination) considered here, in most cases, PICO shows the highest sensitivity to the feature model parameters, and CMB-S4 the lowest (except for a few occasions). This is due to the reason that, among all the CMB experiments taken into consideration, PICO carries a few advantages over other CMB experiments (individual), which includes, availability of a greater number of channels, having higher resolution and sensitivity, having a wide multipole range, and greater sky coverage. Depending upon which CMB experiment (PL, M5, S4 and LB) one is considering, PICO may surpass that with one or more of the aforementioned factors. For instance, in comparison to PICO, CORE-M5 has nearly comparable channel numbers with equal sky fraction and multipole range, and for a few channels, has higher resolution; but PICO has better temperature and polarization sensitivity than CORE-M5 does. Consequently, PICO outperforms CORE-M5 in detecting the features. Similarly, for other CMB experiments, i.e., CMB-S4, LiteBIRD or Planck, PICO significantly surpasses all these experiments in several respects, hence outperforms these experiments individually as well as in combination. Now, for a given LSS survey, if we consider the synergy of the CMB missions rather than the individual CMB missions, the strongest bounds come from the M5+S4 combination (with some exceptions) out of the three CMB combinations (PL+S4, LB+S4, and M5+S4) that are considered in this analysis; it is because CORE-M5 is equipped with higher resolution and sensitivity, as well as has a greater number of frequency channels than Planck and LiteBIRD. Strikingly, the combination of CORE-M5 and CMB-S4 missions reaches the constraining ability of PICO in most cases by compensating for their mutual limitation, as CORE-M5 complements CMB-S4 in number of channels and sky fraction (which are less for CMB-S4), and CMB-S4 complements CORE-M5 in beam resolution and sensitivity (which are significantly stronger for CMB-S4 compared to CORE-M5). Now, among all the LSS surveys taken into consideration in this analysis, be it individual or combined (SKA1-CS, SKA2-CS, SKA1-IM1, SKA1-IM2, SKA1-(IM1+IM2), EUCLID-(CS+GC)+ SKA1-IM2), EUCLID-(CS+GC)+ SKA1-IM2 shows the best sensitivity towards the amplitude of the feature models (with few exceptions). Specifically, the EUCLID-GC survey displays a high sensitivity to power spectrum amplitude. Since EUCLID is equipped with a stronger redshift resolving capacity, making it capable enough to lift the degeneracy between the galaxy bias factor and the amplitude of the perturbations by measuring the redshift space distortions, which in turn allows it to offer a finer measurement of the perturbation amplitude, hence improving its sensitivity towards feature amplitudes. Among SKA surveys, the best bounds are imparted by the synergy of SKA1-IM1 and SKA1-IM2 (SKA1-(IM1+IM2)) and individually by SKA1-IM1 on the feature amplitudes. The sensitivity of different surveys (individual and combined) is found to be dependent on the frequency of the feature models; the best bounds are arising for low frequency models of oscillatory features. To make a comparison of how this analysis has improved the bounds on the feature model parameters in respect of the bounds imparted by the earlier works, these articles~\cite{Chantavat:2010vt,Chen:2016vvw,Ballardini:2016hpi,Palma:2017wxu,Ballardini:2017qwq,Xu:2016kwz,Chen:2016zuu,Chandra:2022utq} can be referred to.}  

{\fontfamily{qpl}
\section{\textbf{Conclusions}}\label{Con}
This article thoroughly explores the competence of upcoming CMB missions in synergy with future LSS surveys in probing features in the primordial power spectrum of inflationary models. Here, we have investigated how different SKA surveys (SKA-CS and SKA-IM) perform in identifying primordial features in combination with forthcoming CMB missions, as well as explored how a completely distinct LSS survey like EUCLID would improve the results when combined with the SKA intensity mapping experiment. %The primordial feature is a scale-dependent rectification in the primordial power spectrum beyond the minimal scenario~\cite{Chen:2010xka,Chluba:2015bqa}. 
The primordial feature is a scale-dependent rectification in the primordial power spectrum beyond the minimal picture~\cite{Chen:2010xka,Chluba:2015bqa}. %The primordial feature is a scale-dependent rectification in the primordial power spectrum beyond the minimal scenario~\cite{Chen:2010xka,Chluba:2015bqa}.
Any decisive identification of primordial features can provide us with a better understanding of primordial physics; i.e., it can reveal the behaviour of inflationary potential, can reveal new energy scales, and it can unravel the intricacies of the inflationary dynamics. Furthermore, detection of primordial features can help us to lift the degeneracy between different inflationary models, along with between inflation and its alternative recipes. The latest search has already advanced our knowledge of the possible cosmological configuration of the Universe, and gives a marginal indication of the presence of features in the primordial power spectrum. The temperature data has shown residuals which do not go in accordance with the standard cosmological scenario, with a statistical significance up to $ 3 \sigma $, whose possible explanation may come from primordial features. Such residuals consist of power suppression around multipoles of $ \ell \sim 20-40 $, and oscillatory structure at higher multipoles around $ \ell \sim 700-800 $ in temperature power spectrum. In the literature, several studies have been conducted to compare many feature models to Planck data~\cite{Chen:2010xka,Slosar:2019gvt,Achucarro:2022qrl}, and despite finding some best-fit candidates which are very intriguing on physical grounds, none of the feature models has been found to be statistically favored over the standard slow-roll inflationary scenario, as feature models introduce extra parameters in the picture.
Cosmic microwave background anisotropies are the basis of observational cosmology and play a pivotal role in most cosmological analyses, which include the search for primordial features as well. CMB observation has several advantages, such as it has access to the largest scales, can be measured very well, and the physics of CMB is completely linear in nature, hence theoretically well tractable. However, there are a few major disadvantages to CMB maps as well, such as CMB measurements lose information in projecting the 3-dimensional information of primordial fluctuations on the 2-dimensional surface, and the precision achievable by a CMB survey is cosmic variance-limited~\cite{Beutler:2019ojk}. Future CMB missions aim to advance our present knowledge of the Universe by several degrees, by delivering more promising data in contrast to those obtained from the Planck mission, since they will achieve an unprecedented precision in measuring E modes, that will eventually reduce the error bars on EE and TE power spectra to the level of cosmic variance. %Upcoming CMB experiments promise to improve our current understanding of the Universe to varied orders of magnitude, by providing more sensitive data compared to what we have from Planck18, since they will achieve an unprecedented precision in measuring E modes that will consequently reduce the error bars on EE and TE power spectra to the level of cosmic variance. 
Due to the narrower transfer functions in multipole space, E-mode polarization power spectra of these future CMB missions will be much more sensitive than temperature spectra towards features. Forthcoming CMB missions will thus provide us with new possibilities for estimating the statistical significance of features by constraining their effects on T-mode and E-mode together. It is not very clear yet whether these signals in the CMB power spectra are caused by statistical factors or whether they have some fundamental physical origin; however, it is expected that future CMB experiments will improve our understanding on this.

On considering the indications of the data and the interesting physical information that their detection would reveal, we have done a thorough study on the status of the features as targets for the major next-generation CMB missions in synergy with upcoming SKA surveys. In this work, we have considered five different CMB experiments, such as Planck, CMB-S4, LiteBIRD, CORE-M5 and PICO, where except Planck, the rest of the missions are next-generation missions. Among these CMB missions, only PICO has been considered individually. For CMB-S4 and CORE-M5, we have assumed them individually as well as in combination (CMB-S4+CORE-M5). Planck and LiteBIRD have been considered only in conjunction with the CMB-S4 experiment, not individually. All these mentioned CMB missions and their assumed combinations are studied in synergy with the most ambitious next generation LSS surveys, i.e., the SKA surveys. Here, we have considered both the 21cm intensity mapping surveys (SKA-IM) and the weak lensing surveys (SKA-CS) of SKA. For the SKA-IM survey, we have considered two distinct configurations, namely SKA1-IM1 and SKA1-IM2, individually as well as in combination (SKA1-IM1, SKA1-IM2, and SKA1-IM1+SKA1-IM2). Similarly, for the SKA-CS surveys, we have considered two distinct configurations, namely SKA1-CS and SKA2-CS. Here, we have not considered the combination of SKA1-CS, and SKA2-CS, because in weak lensing measurements, we use information all the way down to redshift zero. %For SKA surveys, we have considered two distinct cosmic shear configurations for the SKA-CS experiment, namely SKA1-CS and SKA2-CS, and similarly, for the SKA-IM experiment, we have considered two distinct configurations, such as SKA1-IM1 and SKA1-IM2, individually as well as in combination (SKA1-IM1+SKA1-IM2).
In addition to SKA surveys, we have also studied how the EUCLID surveys alter results when combined with the SKA intensity mapping survey (EUCLID-(CS+GC)+SKA1-IM2). Here, we have only considered the SKA1-IM2 configuration with EUCLID to avoid double-counting of information due to the overlap of redshift ranges. To realize this in a systematic fashion, three distinctive features that characterize three separate classes of models, specifically, bump feature, sharp feature signal, and resonance feature signal, have been adopted here, and on them the Fisher matrix forecast method has been applied for the above-mentioned synergies of the next-generation CMB and LSS surveys. There are previous studies in the literature in this direction~\cite{Chantavat:2010vt,Chen:2016vvw,Ballardini:2016hpi,Palma:2017wxu,Ballardini:2017qwq,Xu:2016kwz,Chen:2016zuu,Chandra:2022utq}. However, our analysis differs in many respects when compared to earlier studies. To begin with, they in general differ in the prescriptions of employing the Fisher forecast method to obtain the bounds on the cosmological parameters; here we have utilized the MontyPython package to evaluate the Fisher matrix straight from the log-likelihood function by producing mock data to mimic real data, whereas the prescription of previous works is to express the Fisher matrix as a function of derivatives of observable quantities with respect the model parameters. In addition to that, for a given experiment, our analysis differs in the chosen feature models for investigation, list of model parameters included, adopted instrumental sensitivities and experimental specifications, list of systemic errors taken into consideration and marginalized over. Furthermore, our analysis differs in the assumption of fiducial cosmology, where we have assumed the six parameters $\Lambda \text{CDM}$ model as our fiducial baseline cosmology with the latest bestfit values for the cosmological parameters coming from the Planck 2018 final data release. Thus, collectively, the differences range from adopted next-generation surveys and their combinations, computational methodology, adopted experimental specifications and instrumental sensitivity, number of systematic errors taken into account, to latest fiducial values.

We have expounded all the results obtained in detail, which can be summarized by stating that PICO and CMB-S4 show the highest and lowest sensitivity to feature model parameters, except for a few instances; the combination of CORE-M5 and CMB-S4 achieves the sensitivity of PICO in most cases by compensating for their mutual limitation and EUCLID-(CS+GC)+ SKA1-IM2 shows the best sensitivity towards the amplitude of the feature models, with few exceptions. Besides, among SKA surveys, the best bounds come from the SKA1-IM1+SKA1-IM2 combination, and individually from SKA1-IM1 on the feature amplitudes. It is further observed that the sensitivity of different surveys (individual and combined) relies on the frequency of the feature models and that the low frequency models of oscillatory features show the best sensitivity.

In summary, this work presents a comparative study of different CMB experiments in combination with SKA surveys for different strengths and weaknesses of both. This makes our analysis robust, encompassing almost all the possibilities to explore.}

\appendix

{\fontfamily{qpl}
\section{\textbf{Tables}}\label{tables}
\setlength{\tabcolsep}{4.5pt} % Default value: 6pt
\renewcommand{\arraystretch}{1.5} % Default value: 1
\newcolumntype{C}[1]{>{\Centering}m{#1}}
\renewcommand\tabularxcolumn[1]{C{#1}}
\begin{minipage}{\linewidth}
\centering
\captionsetup{font=footnotesize}
\begin{tabular}{|c|c|c|c|c|c|c|c|}
\hline
\textbf{Models}            & \textbf{Parameters}                                                       & \textbf{\begin{tabular}[c]{@{}c@{}}SKA1\\ (CS)\end{tabular}} & \textbf{\begin{tabular}[c]{@{}c@{}}SKA2\\ (CS)\end{tabular}} & \textbf{\begin{tabular}[c]{@{}c@{}}SKA1\\ (IM1)\end{tabular}}            & \textbf{\begin{tabular}[c]{@{}c@{}}SKA1\\ (IM2)\end{tabular}} & \textbf{\begin{tabular}[c]{@{}c@{}}SKA1\\ (IM1+\\ IM2)\end{tabular}}           & \textbf{\begin{tabular}[c]{@{}c@{}}Euclid\\ (CS+GC)+\\ SKA1 (IM2)\end{tabular}}       \\ \hline
\textbf{\begin{tabular}[c]{@{}c@{}}Bump Feature\\ ${k_b}=0.05$\end{tabular}}           & \begin{tabular}[c]{@{}c@{}}$10^{5}\times\sigma \left(\omega_{\mathrm{b}}\right)$\\ $\sigma \left(\omega_{\mathrm{cdm}}\right)$\\ $\sigma \left( H_0 \right)$\\ $10^{11}\times\sigma \left( A_\mathrm{s} \right)$\\ $\sigma \left( n_\mathrm{s} \right)$\\ $\sigma \left( \tau_{\mathrm{reio}} \right)$\\ $\sigma \left({C_b}\right)$\\ $\sigma \left({k_b}\right)$\end{tabular}     & \begin{tabular}[c]{@{}c@{}}3.70804
\\ 0.00024
\\ 0.10114
\\ 1.06302
\\ 0.00216
\\ 0.00234
\\ 0.00405
\\ 0.03375
\end{tabular}     & \begin{tabular}[c]{@{}c@{}}3.59120
\\ 0.00016
\\ 0.06615
\\ 0.83861
\\ 0.00206
\\ 0.00182
\\ 0.00388
\\ 0.03307
\end{tabular}     & \begin{tabular}[c]{@{}c@{}}3.23790
\\ 0.00010
\\ 0.04263
\\ 0.64741
\\ 0.00183
\\ 0.00153
\\ 0.00280
\\ 0.02263
\end{tabular}     & \begin{tabular}[c]{@{}c@{}}3.33047
\\ 0.00014
\\ 0.05713
\\ 0.76186
\\ 0.00203
\\ 0.00176
\\ 0.00368
\\ 0.03095
\end{tabular}     & \begin{tabular}[c]{@{}c@{}}3.19784
\\ 0.00009
\\ 0.03670
\\ 0.61816
\\ 0.00181
\\ 0.00146
\\ 0.00268
\\ 0.02167
\end{tabular}     & \begin{tabular}[c]{@{}c@{}}3.24983
\\ 0.00010
\\ 0.04292
\\ 0.72094
\\ 0.00180
\\ 0.00151
\\ 0.00277
\\ 0.02188
\end{tabular}     \\ \hline
\textbf{\begin{tabular}[c]{@{}c@{}}Bump Feature\\ ${k_b}=0.1$\end{tabular}}           & \begin{tabular}[c]{@{}c@{}}$10^{5}\times\sigma \left(\omega_{\mathrm{b}}\right)$\\ $\sigma \left(\omega_{\mathrm{cdm}}\right)$\\ $\sigma \left( H_0 \right)$\\ $10^{11}\times\sigma \left( A_\mathrm{s} \right)$\\ $\sigma \left( n_\mathrm{s} \right)$\\ $\sigma \left( \tau_{\mathrm{reio}} \right)$\\ $\sigma \left({C_b}\right)$\\ $\sigma \left({k_b}\right)$\end{tabular}     & \begin{tabular}[c]{@{}c@{}}3.86135
\\ 0.00022
\\ 0.09303
\\ 0.87517
\\ 0.00186
\\ 0.00218
\\ 0.00233
\\ 0.04206
\end{tabular}     & \begin{tabular}[c]{@{}c@{}}3.81542
\\ 0.00016
\\ 0.06346
\\ 0.73402
\\ 0.00179
\\ 0.00181
\\ 0.00231
\\ 0.04136
\end{tabular}     & \begin{tabular}[c]{@{}c@{}}3.52789
\\ 0.00010
\\ 0.04369
\\ 0.61461
\\ 0.00163
\\ 0.00155
\\ 0.00192
\\ 0.03420
\end{tabular}     & \begin{tabular}[c]{@{}c@{}}3.63564
\\ 0.00014
\\ 0.05722
\\ 0.70054
\\ 0.00180
\\ 0.00179
\\ 0.00224
\\ 0.03969
\end{tabular}     & \begin{tabular}[c]{@{}c@{}}3.48648
\\ 0.00009
\\ 0.03809
\\ 0.58469
\\ 0.00160
\\ 0.00148
\\ 0.00187
\\ 0.03222
\end{tabular}     & \begin{tabular}[c]{@{}c@{}}3.35745
\\ 0.00010
\\ 0.04264
\\ 0.62869
\\ 0.00154
\\ 0.00151
\\ 0.00170
\\ 0.03148
\end{tabular}     \\ \hline
\textbf{\begin{tabular}[c]{@{}c@{}}Bump Feature\\ ${k_b}=0.2$\end{tabular}}           & \begin{tabular}[c]{@{}c@{}}$10^{5}\times\sigma \left(\omega_{\mathrm{b}}\right)$\\ $\sigma \left(\omega_{\mathrm{cdm}}\right)$\\ $\sigma \left( H_0 \right)$\\ $10^{11}\times\sigma \left( A_\mathrm{s} \right)$\\ $\sigma \left( n_\mathrm{s} \right)$\\ $\sigma \left( \tau_{\mathrm{reio}} \right)$\\ $\sigma \left({C_b}\right)$\\ $\sigma \left({k_b}\right)$\end{tabular}     & \begin{tabular}[c]{@{}c@{}}4.04185
\\ 0.00024
\\ 0.09969
\\ 0.93517
\\ 0.00319
\\ 0.00233
\\ 0.00577
\\ 0.10508
\end{tabular}     & \begin{tabular}[c]{@{}c@{}}4.08722
\\ 0.00017
\\ 0.06594
\\ 0.76095
\\ 0.00273
\\ 0.00184
\\ 0.00512
\\ 0.10578
\end{tabular}     & \begin{tabular}[c]{@{}c@{}}3.68875
\\ 0.00012
\\ 0.04530
\\ 0.65366
\\ 0.00267
\\ 0.00162
\\ 0.00436
\\ 0.10109
\end{tabular}     & \begin{tabular}[c]{@{}c@{}}3.97089
\\ 0.00016
\\ 0.05962
\\ 0.80941
\\ 0.00316
\\ 0.00194
\\ 0.00557
\\ 0.10867
\end{tabular}     & \begin{tabular}[c]{@{}c@{}}3.52145
\\ 0.00010
\\ 0.04015
\\ 0.60796
\\ 0.00241
\\ 0.00155
\\ 0.00360
\\ 0.09636
\end{tabular}     & \begin{tabular}[c]{@{}c@{}}3.37082
\\ 0.00011
\\ 0.04709
\\ 0.60970
\\ 0.00203
\\ 0.00155
\\ 0.00297
\\ 0.08588
\end{tabular}     \\ \hline
\end{tabular}\par
\bigskip
\parbox{16.4cm}{\captionof{table}{{\fontfamily{bch}
The possible 1-$\sigma$ sensitivity of \textbf{CMB-S4} for bump feature in synergy with SKA1(CS), SKA2(CS), SKA1(IM1), SKA1(IM2), SKA1(IM1+IM2), and Euclid(CS+GC)+SKA1(IM2).}}} \label{table:CMB-S4-Bump}
\end{minipage}

\setlength{\tabcolsep}{4.8pt} % Default value: 6pt
\renewcommand{\arraystretch}{1.5} % Default value: 1
\newcolumntype{C}[1]{>{\Centering}m{#1}}
\renewcommand\tabularxcolumn[1]{C{#1}}
\begin{minipage}{\linewidth}
\centering
\captionsetup{font=footnotesize}
\begin{tabular}{|c|c|c|c|c|c|c|c|}
\hline
\textbf{Models}            & \textbf{Parameters}                                                       & \textbf{\begin{tabular}[c]{@{}c@{}}SKA1\\ (CS)\end{tabular}} & \textbf{\begin{tabular}[c]{@{}c@{}}SKA2\\ (CS)\end{tabular}} & \textbf{\begin{tabular}[c]{@{}c@{}}SKA1\\ (IM1)\end{tabular}}            & \textbf{\begin{tabular}[c]{@{}c@{}}SKA1\\ (IM2)\end{tabular}} & \textbf{\begin{tabular}[c]{@{}c@{}}SKA1\\ (IM1+\\ IM2)\end{tabular}} & \textbf{\begin{tabular}[c]{@{}c@{}}Euclid\\ (CS+GC)+\\ SKA1 (IM2)\end{tabular}}       \\ \hline
\textbf{\begin{tabular}[c]{@{}c@{}}Bump Feature\\ ${k_b}=0.05$\end{tabular}}           & \begin{tabular}[c]{@{}c@{}}$10^{5}\times\sigma \left(\omega_{\mathrm{b}}\right)$\\ $\sigma \left(\omega_{\mathrm{cdm}}\right)$\\ $\sigma \left( H_0 \right)$\\ $10^{11}\times\sigma \left( A_\mathrm{s} \right)$\\ $\sigma \left( n_\mathrm{s} \right)$\\ $\sigma \left( \tau_{\mathrm{reio}} \right)$\\ $\sigma \left({C_b}\right)$\\ $\sigma \left({k_b}\right)$\end{tabular}     & \begin{tabular}[c]{@{}c@{}}4.04137
\\ 0.00017
\\ 0.07301
\\ 0.61364
\\ 0.00172
\\ 0.00131
\\ 0.00307
\\ 0.02541
\end{tabular}     & \begin{tabular}[c]{@{}c@{}}3.94162
\\ 0.00012
\\ 0.05354
\\ 0.59062
\\ 0.00166
\\ 0.00128
\\ 0.00303
\\ 0.02520
\end{tabular}     & \begin{tabular}[c]{@{}c@{}}3.79359
\\ 0.00009
\\ 0.04127
\\ 0.52691
\\ 0.00158
\\ 0.00119
\\ 0.00244
\\ 0.01984
\end{tabular}     & \begin{tabular}[c]{@{}c@{}}3.93386
\\ 0.00012
\\ 0.05475
\\ 0.58291
\\ 0.00168
\\ 0.00130
\\ 0.00295
\\ 0.02424
\end{tabular}     & \begin{tabular}[c]{@{}c@{}}3.72677
\\ 0.00008
\\ 0.03573
\\ 0.51300
\\ 0.00157
\\ 0.00116
\\ 0.00235
\\ 0.01917
\end{tabular}     & \begin{tabular}[c]{@{}c@{}}3.67971
\\ 0.00009
\\ 0.04041
\\ 0.57288
\\ 0.00157
\\ 0.00120
\\ 0.00240
\\ 0.01921
\end{tabular}     \\ \hline
\textbf{\begin{tabular}[c]{@{}c@{}}Bump Feature\\ ${k_b}=0.1$\end{tabular}}           & \begin{tabular}[c]{@{}c@{}}$10^{5}\times\sigma \left(\omega_{\mathrm{b}}\right)$\\ $\sigma \left(\omega_{\mathrm{cdm}}\right)$\\ $\sigma \left( H_0 \right)$\\ $10^{11}\times\sigma \left( A_\mathrm{s} \right)$\\ $\sigma \left( n_\mathrm{s} \right)$\\ $\sigma \left( \tau_{\mathrm{reio}} \right)$\\ $\sigma \left({C_b}\right)$\\ $\sigma \left({k_b}\right)$\end{tabular}     & \begin{tabular}[c]{@{}c@{}}4.33190
\\ 0.00017
\\ 0.07334
\\ 0.56758
\\ 0.00160
\\ 0.00134
\\ 0.00193
\\ 0.03744
\end{tabular}     & \begin{tabular}[c]{@{}c@{}}4.22221
\\ 0.00012
\\ 0.05404
\\ 0.54584
\\ 0.00154
\\ 0.00132
\\ 0.00191
\\ 0.03702
\end{tabular}     & \begin{tabular}[c]{@{}c@{}}4.05329
\\ 0.00009
\\ 0.04242
\\ 0.50055
\\ 0.00147
\\ 0.00120
\\ 0.00170
\\ 0.03187
\end{tabular}     & \begin{tabular}[c]{@{}c@{}}4.18499
\\ 0.00012
\\ 0.05525
\\ 0.54378
\\ 0.00156
\\ 0.00132
\\ 0.00187
\\ 0.03602
\end{tabular}     & \begin{tabular}[c]{@{}c@{}}3.98895
\\ 0.00008
\\ 0.03728
\\ 0.48834
\\ 0.00145
\\ 0.00117
\\ 0.00167
\\ 0.03032
\end{tabular}     & \begin{tabular}[c]{@{}c@{}}3.79742
\\ 0.00009
\\ 0.04060
\\ 0.51383
\\ 0.00139
\\ 0.00121
\\ 0.00152
\\ 0.02950
\end{tabular}     \\ \hline
\textbf{\begin{tabular}[c]{@{}c@{}}Bump Feature\\ ${k_b}=0.2$\end{tabular}}           & \begin{tabular}[c]{@{}c@{}}$10^{5}\times\sigma \left(\omega_{\mathrm{b}}\right)$\\ $\sigma \left(\omega_{\mathrm{cdm}}\right)$\\ $\sigma \left( H_0 \right)$\\ $10^{11}\times\sigma \left( A_\mathrm{s} \right)$\\ $\sigma \left( n_\mathrm{s} \right)$\\ $\sigma \left( \tau_{\mathrm{reio}} \right)$\\ $\sigma \left({C_b}\right)$\\ $\sigma \left({k_b}\right)$\end{tabular}     & \begin{tabular}[c]{@{}c@{}}4.48930
\\ 0.00018
\\ 0.07258
\\ 0.62625
\\ 0.00264
\\ 0.00135
\\ 0.00501
\\ 0.10089
\end{tabular}     & \begin{tabular}[c]{@{}c@{}}4.37099
\\ 0.00013
\\ 0.05328
\\ 0.59552
\\ 0.00236
\\ 0.00132
\\ 0.00456
\\ 0.10037
\end{tabular}     & \begin{tabular}[c]{@{}c@{}}3.96582
\\ 0.00010
\\ 0.04196
\\ 0.53385
\\ 0.00220
\\ 0.00123
\\ 0.00364
\\ 0.09630
\end{tabular}     & \begin{tabular}[c]{@{}c@{}}4.24163
\\ 0.00014
\\ 0.05485
\\ 0.62097
\\ 0.00253
\\ 0.00137
\\ 0.00458
\\ 0.09908
\end{tabular}     & \begin{tabular}[c]{@{}c@{}}3.84368
\\ 0.00009
\\ 0.03743
\\ 0.51023
\\ 0.00206
\\ 0.00120
\\ 0.00314
\\ 0.09444
\end{tabular}     & \begin{tabular}[c]{@{}c@{}}3.67867
\\ 0.00010
\\ 0.04259
\\ 0.50090
\\ 0.00195
\\ 0.00120
\\ 0.00297
\\ 0.08423
\end{tabular}     \\ \hline
\end{tabular}\par
\bigskip
\parbox{16.55cm}{\captionof{table}{{\fontfamily{bch}The possible 1-$\sigma$ sensitivity of \textbf{CORE-M5} for bump feature in synergy with SKA1(CS), SKA2(CS), SKA1(IM1), SKA1(IM2), SKA1(IM1+IM2), and Euclid(CS+GC)+SKA1(IM2).}}} \label{table:CORE-M5-Bump}
\end{minipage}

\setlength{\tabcolsep}{4.8pt} % Default value: 6pt
\renewcommand{\arraystretch}{1.5} % Default value: 1
\newcolumntype{C}[1]{>{\Centering}m{#1}}
\renewcommand\tabularxcolumn[1]{C{#1}}
\begin{minipage}{\linewidth}
\centering
\captionsetup{font=footnotesize}
\begin{tabular}{|c|c|c|c|c|c|c|c|}
\hline
\textbf{Models}            & \textbf{Parameters}                                                       & \textbf{\begin{tabular}[c]{@{}c@{}}SKA1\\ (CS)\end{tabular}} & \textbf{\begin{tabular}[c]{@{}c@{}}SKA2\\ (CS)\end{tabular}} & \textbf{\begin{tabular}[c]{@{}c@{}}SKA1\\ (IM1)\end{tabular}}            & \textbf{\begin{tabular}[c]{@{}c@{}}SKA1\\ (IM2)\end{tabular}} & \textbf{\begin{tabular}[c]{@{}c@{}}SKA1\\ (IM1+\\ IM2)\end{tabular}}           & \textbf{\begin{tabular}[c]{@{}c@{}}Euclid\\ (CS+GC)+\\ SKA1 (IM2)\end{tabular}}        \\ \hline
\textbf{\begin{tabular}[c]{@{}c@{}}Bump Feature\\ ${k_b}=0.05$\end{tabular}}           & \begin{tabular}[c]{@{}c@{}}$10^{5}\times\sigma \left(\omega_{\mathrm{b}}\right)$\\ $\sigma \left(\omega_{\mathrm{cdm}}\right)$\\ $\sigma \left( H_0 \right)$\\ $10^{11}\times\sigma \left( A_\mathrm{s} \right)$\\ $\sigma \left( n_\mathrm{s} \right)$\\ $\sigma \left( \tau_{\mathrm{reio}} \right)$\\ $\sigma \left({C_b}\right)$\\ $\sigma \left({k_b}\right)$\end{tabular}     & \begin{tabular}[c]{@{}c@{}}3.13800
\\ 0.00015
\\ 0.05879
\\ 0.54724
\\ 0.00172
\\ 0.00119
\\ 0.00298
\\ 0.02498
\end{tabular}     & \begin{tabular}[c]{@{}c@{}}3.08561
\\ 0.00011
\\ 0.04718
\\ 0.54161
\\ 0.00167
\\ 0.00118
\\ 0.00293
\\ 0.02478
\end{tabular}     & \begin{tabular}[c]{@{}c@{}}2.76889
\\ 0.00008
\\ 0.03559
\\ 0.47898
\\ 0.00146
\\ 0.00107
\\ 0.00237
\\ 0.01942
\end{tabular}     & \begin{tabular}[c]{@{}c@{}}2.82763
\\ 0.00011
\\ 0.04631
\\ 0.52427
\\ 0.00156
\\ 0.00115
\\ 0.00285
\\ 0.02372
\end{tabular}     & \begin{tabular}[c]{@{}c@{}}2.74731
\\ 0.00007
\\ 0.03097
\\ 0.47009
\\ 0.00144
\\ 0.00105
\\ 0.00229
\\ 0.01882
\end{tabular}     & \begin{tabular}[c]{@{}c@{}}2.8915
\\ 0.00008
\\ 0.03509
\\ 0.49598
\\ 0.00159
\\ 0.00106
\\ 0.00237
\\ 0.01902
\end{tabular}     \\ \hline
\textbf{\begin{tabular}[c]{@{}c@{}}Bump Feature\\ ${k_b}=0.1$\end{tabular}}           & \begin{tabular}[c]{@{}c@{}}$10^{5}\times\sigma \left(\omega_{\mathrm{b}}\right)$\\ $\sigma \left(\omega_{\mathrm{cdm}}\right)$\\ $\sigma \left( H_0 \right)$\\ $10^{11}\times\sigma \left( A_\mathrm{s} \right)$\\ $\sigma \left( n_\mathrm{s} \right)$\\ $\sigma \left( \tau_{\mathrm{reio}} \right)$\\ $\sigma \left({C_b}\right)$\\ $\sigma \left({k_b}\right)$\end{tabular}     & \begin{tabular}[c]{@{}c@{}}3.86537
\\ 0.00014
\\ 0.05855
\\ 0.63295
\\ 0.00202
\\ 0.00121
\\ 0.00182
\\ 0.03436
\end{tabular}     & \begin{tabular}[c]{@{}c@{}}3.71980
\\ 0.00011
\\ 0.04850
\\ 0.60179
\\ 0.00190
\\ 0.00118
\\ 0.00180
\\ 0.03380
\end{tabular}     & \begin{tabular}[c]{@{}c@{}}2.99705
\\ 0.00008
\\ 0.03624
\\ 0.45150
\\ 0.00130
\\ 0.00108
\\ 0.00157
\\ 0.02851
\end{tabular}     & \begin{tabular}[c]{@{}c@{}}3.06133
\\ 0.00011
\\ 0.04626
\\ 0.48372
\\ 0.00139
\\ 0.00117
\\ 0.00174
\\ 0.03156
\end{tabular}     & \begin{tabular}[c]{@{}c@{}}2.98083
\\ 0.00007
\\ 0.03202
\\ 0.44127
\\ 0.00128
\\ 0.00106
\\ 0.00155
\\ 0.02729
\end{tabular}     & \begin{tabular}[c]{@{}c@{}}3.00300
\\ 0.00008
\\ 0.03503
\\ 0.44524
\\ 0.00138
\\ 0.00107
\\ 0.00144
\\ 0.02700
\end{tabular}     \\ \hline
\textbf{\begin{tabular}[c]{@{}c@{}}Bump Feature\\ ${k_b}=0.2$\end{tabular}}           & \begin{tabular}[c]{@{}c@{}}$10^{5}\times\sigma \left(\omega_{\mathrm{b}}\right)$\\ $\sigma \left(\omega_{\mathrm{cdm}}\right)$\\ $\sigma \left( H_0 \right)$\\ $10^{11}\times\sigma \left( A_\mathrm{s} \right)$\\ $\sigma \left( n_\mathrm{s} \right)$\\ $\sigma \left( \tau_{\mathrm{reio}} \right)$\\ $\sigma \left({C_b}\right)$\\ $\sigma \left({k_b}\right)$\end{tabular}     & \begin{tabular}[c]{@{}c@{}}3.38861
\\ 0.00015
\\ 0.05898
\\ 0.55688
\\ 0.00261
\\ 0.00124
\\ 0.00459
\\ 0.08407
\end{tabular}     & \begin{tabular}[c]{@{}c@{}}3.35148
\\ 0.00012
\\ 0.04714
\\ 0.53616
\\ 0.00235
\\ 0.00122
\\ 0.00415
\\ 0.08375
\end{tabular}     & \begin{tabular}[c]{@{}c@{}}3.10513
\\ 0.00010
\\ 0.03746
\\ 0.48524
\\ 0.00210
\\ 0.00112
\\ 0.00360
\\ 0.08050
\end{tabular}     & \begin{tabular}[c]{@{}c@{}}3.28614
\\ 0.00013
\\ 0.04804
\\ 0.55134
\\ 0.00242
\\ 0.00123
\\ 0.00440
\\ 0.08480
\end{tabular}     & \begin{tabular}[c]{@{}c@{}}2.98583
\\ 0.00009
\\ 0.03304
\\ 0.45509
\\ 0.00195
\\ 0.00108
\\ 0.00313
\\ 0.07901
\end{tabular}     & \begin{tabular}[c]{@{}c@{}}3.0284
\\ 0.00009
\\ 0.03697
\\ 0.44960
\\ 0.00206
\\ 0.00111
\\ 0.00288
\\ 0.07514
\end{tabular}     \\ \hline
\end{tabular}\par
\bigskip
\parbox{16.55cm}{\captionof{table}{{\fontfamily{bch}The possible 1-$\sigma$ sensitivity of \textbf{PICO} for bump feature in synergy with SKA1(CS), SKA2(CS), SKA1(IM1), SKA1(IM2), SKA1(IM1+IM2), and Euclid(CS+GC)+SKA1(IM2).}}} \label{table:PICO-Bump}
\end{minipage}

\setlength{\tabcolsep}{4.8pt} % Default value: 6pt
\renewcommand{\arraystretch}{1.5} % Default value: 1
\newcolumntype{C}[1]{>{\Centering}m{#1}}
\renewcommand\tabularxcolumn[1]{C{#1}}
\begin{minipage}{\linewidth}
\centering
\captionsetup{font=footnotesize}
\begin{tabular}{|c|c|c|c|c|c|c|c|}
\hline
\textbf{Models}            & \textbf{Parameters}                                                       & \textbf{\begin{tabular}[c]{@{}c@{}}SKA1\\ (CS)\end{tabular}} & \textbf{\begin{tabular}[c]{@{}c@{}}SKA2\\ (CS)\end{tabular}} & \textbf{\begin{tabular}[c]{@{}c@{}}SKA1\\ (IM1)\end{tabular}}            & \textbf{\begin{tabular}[c]{@{}c@{}}SKA1\\ (IM2)\end{tabular}} & \textbf{\begin{tabular}[c]{@{}c@{}}SKA1\\ (IM1+\\ IM2)\end{tabular}}           & \textbf{\begin{tabular}[c]{@{}c@{}}Euclid\\ (CS+GC)+\\ SKA1 (IM2)\end{tabular}}       \\ \hline
\textbf{\begin{tabular}[c]{@{}c@{}}Bump Feature\\ ${k_b}=0.05$\end{tabular}}           & \begin{tabular}[c]{@{}c@{}}$10^{5}\times\sigma \left(\omega_{\mathrm{b}}\right)$\\ $\sigma \left(\omega_{\mathrm{cdm}}\right)$\\ $\sigma \left( H_0 \right)$\\ $10^{11}\times\sigma \left( A_\mathrm{s} \right)$\\ $\sigma \left( n_\mathrm{s} \right)$\\ $\sigma \left( \tau_{\mathrm{reio}} \right)$\\ $\sigma \left({C_b}\right)$\\ $\sigma \left({k_b}\right)$\end{tabular}     & \begin{tabular}[c]{@{}c@{}}3.55116
\\ 0.00022
\\ 0.09236
\\ 0.91945
\\ 0.00200
\\ 0.00203
\\ 0.00368
\\ 0.03117
\end{tabular}     & \begin{tabular}[c]{@{}c@{}}3.50269
\\ 0.00015
\\ 0.06147
\\ 0.72692
\\ 0.00198
\\ 0.00166
\\ 0.00355
\\ 0.03028
\end{tabular}     & \begin{tabular}[c]{@{}c@{}}3.24564
\\ 0.00010
\\ 0.04359
\\ 0.64736
\\ 0.00174
\\ 0.00153
\\ 0.00268
\\ 0.02188
\end{tabular}     & \begin{tabular}[c]{@{}c@{}}3.34366
\\ 0.00014
\\ 0.05934
\\ 0.76225
\\ 0.00189
\\ 0.00179
\\ 0.00340
\\ 0.02855
\end{tabular}     & \begin{tabular}[c]{@{}c@{}}3.20200
\\ 0.00009
\\ 0.03732
\\ 0.62098
\\ 0.00172
\\ 0.00147
\\ 0.00257
\\ 0.02101
\end{tabular}     & \begin{tabular}[c]{@{}c@{}}3.22950
\\ 0.00010
\\ 0.04236
\\ 0.65974
\\ 0.00180
\\ 0.00148
\\ 0.00267
\\ 0.02115
\end{tabular}     \\ \hline
\textbf{\begin{tabular}[c]{@{}c@{}}Bump Feature\\ ${k_b}=0.1$\end{tabular}}           & \begin{tabular}[c]{@{}c@{}}$10^{5}\times\sigma \left(\omega_{\mathrm{b}}\right)$\\ $\sigma \left(\omega_{\mathrm{cdm}}\right)$\\ $\sigma \left( H_0 \right)$\\ $10^{11}\times\sigma \left( A_\mathrm{s} \right)$\\ $\sigma \left( n_\mathrm{s} \right)$\\ $\sigma \left( \tau_{\mathrm{reio}} \right)$\\ $\sigma \left({C_b}\right)$\\ $\sigma \left({k_b}\right)$\end{tabular}     & \begin{tabular}[c]{@{}c@{}}3.96163
\\ 0.00021
\\ 0.09142
\\ 0.79421
\\ 0.00168
\\ 0.00194
\\ 0.00222
\\ 0.03971
\end{tabular}     & \begin{tabular}[c]{@{}c@{}}3.85794
\\ 0.00014
\\ 0.05992
\\ 0.66153
\\ 0.00162
\\ 0.00158
\\ 0.00220
\\ 0.03912
\end{tabular}     & \begin{tabular}[c]{@{}c@{}}3.55113
\\ 0.00010
\\ 0.04392
\\ 0.58458
\\ 0.00155
\\ 0.00146
\\ 0.00187
\\ 0.03344
\end{tabular}     & \begin{tabular}[c]{@{}c@{}}3.67431
\\ 0.00013
\\ 0.05748
\\ 0.65917
\\ 0.00168
\\ 0.00168
\\ 0.00216
\\ 0.03846
\end{tabular}     & \begin{tabular}[c]{@{}c@{}}3.50757
\\ 0.00009
\\ 0.03849
\\ 0.56085
\\ 0.00152
\\ 0.00141
\\ 0.00183
\\ 0.03160
\end{tabular}     & \begin{tabular}[c]{@{}c@{}}3.39021
\\ 0.00010
\\ 0.04287
\\ 0.60177
\\ 0.00147
\\ 0.00145
\\ 0.00166
\\ 0.03084\end{tabular}     \\ \hline
\textbf{\begin{tabular}[c]{@{}c@{}}Bump Feature\\ ${k_b}=0.2$\end{tabular}}           & \begin{tabular}[c]{@{}c@{}}$10^{5}\times\sigma \left(\omega_{\mathrm{b}}\right)$\\ $\sigma \left(\omega_{\mathrm{cdm}}\right)$\\ $\sigma \left( H_0 \right)$\\ $10^{11}\times\sigma \left( A_\mathrm{s} \right)$\\ $\sigma \left( n_\mathrm{s} \right)$\\ $\sigma \left( \tau_{\mathrm{reio}} \right)$\\ $\sigma \left({C_b}\right)$\\ $\sigma \left({k_b}\right)$\end{tabular}     & \begin{tabular}[c]{@{}c@{}}4.05012
\\ 0.00022
\\ 0.09426
\\ 0.81148
\\ 0.00237
\\ 0.00198
\\ 0.00443
\\ 0.10314
\end{tabular}     & \begin{tabular}[c]{@{}c@{}}4.01606
\\ 0.00015
\\ 0.05983
\\ 0.67139
\\ 0.00214
\\ 0.00158
\\ 0.00411
\\ 0.10369
\end{tabular}     & \begin{tabular}[c]{@{}c@{}}3.65382
\\ 0.00011
\\ 0.04380
\\ 0.62287
\\ 0.00251
\\ 0.00151
\\ 0.00422
\\ 0.09906
\end{tabular}     & \begin{tabular}[c]{@{}c@{}}3.93495
\\ 0.00015
\\ 0.05724
\\ 0.75706
\\ 0.00293
\\ 0.00177
\\ 0.00534
\\ 0.10569
\end{tabular}     & \begin{tabular}[c]{@{}c@{}}3.45170
\\ 0.00010
\\ 0.03889
\\ 0.57781
\\ 0.00221
\\ 0.00144
\\ 0.00336
\\ 0.09548
\end{tabular}     & \begin{tabular}[c]{@{}c@{}}3.30689
\\ 0.00011
\\ 0.04522
\\ 0.57641
\\ 0.00194
\\ 0.00145
\\ 0.00291
\\ 0.08460
\end{tabular}     \\ \hline
\end{tabular}\par
\bigskip
\parbox{16.55cm}{\captionof{table}{{\fontfamily{bch}The possible 1-$\sigma$ sensitivity of \textbf{Planck+CMB-S4} for bump feature in synergy with SKA1(CS), SKA2(CS), SKA1(IM1), SKA1(IM2), SKA1(IM1+IM2), and Euclid(CS+GC)+SKA1(IM2).}}} \label{table:Planck+CMB-S4-Bump}
\end{minipage}

\setlength{\tabcolsep}{4.8pt} % Default value: 6pt
\renewcommand{\arraystretch}{1.5} % Default value: 1
\newcolumntype{C}[1]{>{\Centering}m{#1}}
\renewcommand\tabularxcolumn[1]{C{#1}}
\begin{minipage}{\linewidth}
\centering
\captionsetup{font=footnotesize}
\begin{tabular}{|c|c|c|c|c|c|c|c|}
\hline
\textbf{Models}            & \textbf{Parameters}                                                       & \textbf{\begin{tabular}[c]{@{}c@{}}SKA1\\ (CS)\end{tabular}} & \textbf{\begin{tabular}[c]{@{}c@{}}SKA2\\ (CS)\end{tabular}} & \textbf{\begin{tabular}[c]{@{}c@{}}SKA1\\ (IM1)\end{tabular}}            & \textbf{\begin{tabular}[c]{@{}c@{}}SKA1\\ (IM2)\end{tabular}} & \textbf{\begin{tabular}[c]{@{}c@{}}SKA1\\ (IM1+\\ IM2)\end{tabular}}           & \textbf{\begin{tabular}[c]{@{}c@{}}Euclid\\ (CS+GC)+\\ SKA1 (IM2)\end{tabular}}        \\ \hline
\textbf{\begin{tabular}[c]{@{}c@{}}Bump Feature\\ ${k_b}=0.05$\end{tabular}}           & \begin{tabular}[c]{@{}c@{}}$10^{5}\times\sigma \left(\omega_{\mathrm{b}}\right)$\\ $\sigma \left(\omega_{\mathrm{cdm}}\right)$\\ $\sigma \left( H_0 \right)$\\ $10^{11}\times\sigma \left( A_\mathrm{s} \right)$\\ $\sigma \left( n_\mathrm{s} \right)$\\ $\sigma \left( \tau_{\mathrm{reio}} \right)$\\ $\sigma \left({C_b}\right)$\\ $\sigma \left({k_b}\right)$\end{tabular}     & \begin{tabular}[c]{@{}c@{}}3.37032
\\ 0.00019
\\ 0.08018
\\ 0.69741
\\ 0.00180
\\ 0.00150
\\ 0.00344
\\ 0.03019
\end{tabular}     & \begin{tabular}[c]{@{}c@{}}3.37023
\\ 0.00013
\\ 0.05376
\\ 0.62484
\\ 0.00174
\\ 0.00133
\\ 0.00337
\\ 0.02978
\end{tabular}     & \begin{tabular}[c]{@{}c@{}}3.10107
\\ 0.00009
\\ 0.03906
\\ 0.53124
\\ 0.00159
\\ 0.00120
\\ 0.00259
\\ 0.02175
\end{tabular}     & \begin{tabular}[c]{@{}c@{}}3.18408
\\ 0.00013
\\ 0.05167
\\ 0.59245
\\ 0.00172
\\ 0.00131
\\ 0.00325
\\ 0.02800
\end{tabular}     & \begin{tabular}[c]{@{}c@{}}3.06305
\\ 0.00008
\\ 0.03385
\\ 0.51499
\\ 0.00158
\\ 0.00117
\\ 0.00249
\\ 0.02088
\end{tabular}     & \begin{tabular}[c]{@{}c@{}}3.06767
\\ 0.00009
\\ 0.03865
\\ 0.57400
\\ 0.00157
\\ 0.00119
\\ 0.00252
\\ 0.02090
\end{tabular}     \\ \hline
\textbf{\begin{tabular}[c]{@{}c@{}}Bump Feature\\ ${k_b}=0.1$\end{tabular}}           & \begin{tabular}[c]{@{}c@{}}$10^{5}\times\sigma \left(\omega_{\mathrm{b}}\right)$\\ $\sigma \left(\omega_{\mathrm{cdm}}\right)$\\ $\sigma \left( H_0 \right)$\\ $10^{11}\times\sigma \left( A_\mathrm{s} \right)$\\ $\sigma \left( n_\mathrm{s} \right)$\\ $\sigma \left( \tau_{\mathrm{reio}} \right)$\\ $\sigma \left({C_b}\right)$\\ $\sigma \left({k_b}\right)$\end{tabular}     & \begin{tabular}[c]{@{}c@{}}3.58611
\\ 0.00019
\\ 0.07750
\\ 0.61201
\\ 0.00162
\\ 0.00147
\\ 0.00218
\\ 0.04050
\end{tabular}     & \begin{tabular}[c]{@{}c@{}}3.61360
\\ 0.00013
\\ 0.05239
\\ 0.55967
\\ 0.00157
\\ 0.00133
\\ 0.00216
\\ 0.04001
\end{tabular}     & \begin{tabular}[c]{@{}c@{}}3.38830
\\ 0.00009
\\ 0.04009
\\ 0.50491
\\ 0.00144
\\ 0.00121
\\ 0.00185
\\ 0.03339
\end{tabular}     & \begin{tabular}[c]{@{}c@{}}3.47468
\\ 0.00013
\\ 0.05178
\\ 0.54847
\\ 0.00156
\\ 0.00132
\\ 0.00209
\\ 0.03847
\end{tabular}     & \begin{tabular}[c]{@{}c@{}}3.35017
\\ 0.00008
\\ 0.03523
\\ 0.48867
\\ 0.00143
\\ 0.00118
\\ 0.00180
\\ 0.03149
\end{tabular}     & \begin{tabular}[c]{@{}c@{}}3.20818
\\ 0.00009
\\ 0.03884
\\ 0.51179
\\ 0.00138
\\ 0.00120
\\ 0.00163
\\ 0.03057
\end{tabular}     \\ \hline
\textbf{\begin{tabular}[c]{@{}c@{}}Bump Feature\\ ${k_b}=0.2$\end{tabular}}           & \begin{tabular}[c]{@{}c@{}}$10^{5}\times\sigma \left(\omega_{\mathrm{b}}\right)$\\ $\sigma \left(\omega_{\mathrm{cdm}}\right)$\\ $\sigma \left( H_0 \right)$\\ $10^{11}\times\sigma \left( A_\mathrm{s} \right)$\\ $\sigma \left( n_\mathrm{s} \right)$\\ $\sigma \left( \tau_{\mathrm{reio}} \right)$\\ $\sigma \left({C_b}\right)$\\ $\sigma \left({k_b}\right)$\end{tabular}     & \begin{tabular}[c]{@{}c@{}}3.79989
\\ 0.00020
\\ 0.07950
\\ 0.66037
\\ 0.00271
\\ 0.00151
\\ 0.00529
\\ 0.09998
\end{tabular}     & \begin{tabular}[c]{@{}c@{}}3.89553
\\ 0.00014
\\ 0.05408
\\ 0.60101
\\ 0.00239
\\ 0.00135
\\ 0.00482
\\ 0.10082
\end{tabular}     & \begin{tabular}[c]{@{}c@{}}3.53483
\\ 0.00011
\\ 0.04127
\\ 0.53940
\\ 0.00226
\\ 0.00125
\\ 0.00400
\\ 0.09568
\end{tabular}     & \begin{tabular}[c]{@{}c@{}}3.75939
\\ 0.00014
\\ 0.05367
\\ 0.62934
\\ 0.00260
\\ 0.00139
\\ 0.00499
\\ 0.10044
\end{tabular}     & \begin{tabular}[c]{@{}c@{}}3.38441
\\ 0.00010
\\ 0.03684
\\ 0.50720
\\ 0.00206
\\ 0.00121
\\ 0.00335
\\ 0.09204
\end{tabular}     & \begin{tabular}[c]{@{}c@{}}3.22296
\\ 0.00010
\\ 0.04204
\\ 0.49755
\\ 0.00182
\\ 0.00121
\\ 0.00284
\\ 0.08327
\end{tabular}     \\ \hline
\end{tabular}\par
\bigskip
\parbox{16.53cm}{\captionof{table}{{\fontfamily{bch}The possible 1-$\sigma$ sensitivity of \textbf{LiteBIRD+CMB-S4} for bump feature in synergy with SKA1(CS), SKA2(CS), SKA1(IM1), SKA1(IM2), SKA1(IM1+IM2), and Euclid(CS+GC)+SKA1(IM2).}}} \label{table:LiteBIRD+CMB-S4-Bump}
\end{minipage}

\setlength{\tabcolsep}{4.8pt} % Default value: 6pt
\renewcommand{\arraystretch}{1.5} % Default value: 1
\newcolumntype{C}[1]{>{\Centering}m{#1}}
\renewcommand\tabularxcolumn[1]{C{#1}}
\begin{minipage}{\linewidth}
\centering
\captionsetup{font=footnotesize}
\begin{tabular}{|c|c|c|c|c|c|c|c|}
\hline
\textbf{Models}            & \textbf{Parameters}                                                       & \textbf{\begin{tabular}[c]{@{}c@{}}SKA1\\ (CS)\end{tabular}} & \textbf{\begin{tabular}[c]{@{}c@{}}SKA2\\ (CS)\end{tabular}} & \textbf{\begin{tabular}[c]{@{}c@{}}SKA1\\ (IM1)\end{tabular}}            & \textbf{\begin{tabular}[c]{@{}c@{}}SKA1\\ (IM2)\end{tabular}} & \textbf{\begin{tabular}[c]{@{}c@{}}SKA1\\ (IM1+\\ IM2)\end{tabular}}           & \textbf{\begin{tabular}[c]{@{}c@{}}Euclid\\ (CS+GC)+\\ SKA1 (IM2)\end{tabular}}       \\ \hline
\textbf{\begin{tabular}[c]{@{}c@{}}Bump Feature\\ ${k_b}=0.05$\end{tabular}}           & \begin{tabular}[c]{@{}c@{}}$10^{5}\times\sigma \left(\omega_{\mathrm{b}}\right)$\\ $\sigma \left(\omega_{\mathrm{cdm}}\right)$\\ $\sigma \left( H_0 \right)$\\ $10^{11}\times\sigma \left( A_\mathrm{s} \right)$\\ $\sigma \left( n_\mathrm{s} \right)$\\ $\sigma \left( \tau_{\mathrm{reio}} \right)$\\ $\sigma \left({C_b}\right)$\\ $\sigma \left({k_b}\right)$\end{tabular}     & \begin{tabular}[c]{@{}c@{}}3.07306
\\ 0.00015
\\ 0.06062
\\ 0.55638
\\ 0.00171
\\ 0.00121
\\ 0.00300
\\ 0.02508
\end{tabular}     & \begin{tabular}[c]{@{}c@{}}3.05924
\\ 0.00011
\\ 0.04784
\\ 0.54524
\\ 0.00168
\\ 0.00119
\\ 0.00295
\\ 0.02490
\end{tabular}     & \begin{tabular}[c]{@{}c@{}}2.77239
\\ 0.00009
\\ 0.03608
\\ 0.48923
\\ 0.00148
\\ 0.00109
\\ 0.00239
\\ 0.01950
\end{tabular}     & \begin{tabular}[c]{@{}c@{}}2.83103
\\ 0.00012
\\ 0.04726
\\ 0.53424
\\ 0.00159
\\ 0.00118
\\ 0.00287
\\ 0.02384
\end{tabular}     & \begin{tabular}[c]{@{}c@{}}2.74504
\\ 0.00007
\\ 0.03118
\\ 0.47666
\\ 0.00147
\\ 0.00107
\\ 0.00231
\\ 0.01888
\end{tabular}     & \begin{tabular}[c]{@{}c@{}}2.86030
\\ 0.00008
\\ 0.03555
\\ 0.50667
\\ 0.00159
\\ 0.00109
\\ 0.00238
\\ 0.01907
\end{tabular}     \\ \hline
\textbf{\begin{tabular}[c]{@{}c@{}}Bump Feature\\ ${k_b}=0.1$\end{tabular}}           & \begin{tabular}[c]{@{}c@{}}$10^{5}\times\sigma \left(\omega_{\mathrm{b}}\right)$\\ $\sigma \left(\omega_{\mathrm{cdm}}\right)$\\ $\sigma \left( H_0 \right)$\\ $10^{11}\times\sigma \left( A_\mathrm{s} \right)$\\ $\sigma \left( n_\mathrm{s} \right)$\\ $\sigma \left( \tau_{\mathrm{reio}} \right)$\\ $\sigma \left({C_b}\right)$\\ $\sigma \left({k_b}\right)$\end{tabular}     & \begin{tabular}[c]{@{}c@{}}3.41347
\\ 0.00015
\\ 0.05913
\\ 0.55862
\\ 0.00170
\\ 0.00122
\\ 0.00184
\\ 0.03438
\end{tabular}     & \begin{tabular}[c]{@{}c@{}}3.29353
\\ 0.00012
\\ 0.04689
\\ 0.49377
\\ 0.00151
\\ 0.00120
\\ 0.00181
\\ 0.03326
\end{tabular}     & \begin{tabular}[c]{@{}c@{}}3.02381
\\ 0.00009
\\ 0.03690
\\ 0.46207
\\ 0.00133
\\ 0.00111
\\ 0.00160
\\ 0.02916
\end{tabular}     & \begin{tabular}[c]{@{}c@{}}3.08755
\\ 0.00011
\\ 0.04724
\\ 0.49250
\\ 0.00143
\\ 0.00120
\\ 0.00177
\\ 0.03238
\end{tabular}     & \begin{tabular}[c]{@{}c@{}}2.99682
\\ 0.00007
\\ 0.03232
\\ 0.44945
\\ 0.00132
\\ 0.00108
\\ 0.00157
\\ 0.02789
\end{tabular}     & \begin{tabular}[c]{@{}c@{}}2.99929
\\ 0.00008
\\ 0.03565
\\ 0.45625
\\ 0.00140
\\ 0.00110
\\ 0.00146
\\ 0.02767
\end{tabular}     \\ \hline
\textbf{\begin{tabular}[c]{@{}c@{}}Bump Feature\\ ${k_b}=0.2$\end{tabular}}           & \begin{tabular}[c]{@{}c@{}}$10^{5}\times\sigma \left(\omega_{\mathrm{b}}\right)$\\ $\sigma \left(\omega_{\mathrm{cdm}}\right)$\\ $\sigma \left( H_0 \right)$\\ $10^{11}\times\sigma \left( A_\mathrm{s} \right)$\\ $\sigma \left( n_\mathrm{s} \right)$\\ $\sigma \left( \tau_{\mathrm{reio}} \right)$\\ $\sigma \left({C_b}\right)$\\ $\sigma \left({k_b}\right)$\end{tabular}     & \begin{tabular}[c]{@{}c@{}}3.55693
\\ 0.00016
\\ 0.06075
\\ 0.54720
\\ 0.00249
\\ 0.00122
\\ 0.00404
\\ 0.08536
\end{tabular}     & \begin{tabular}[c]{@{}c@{}}3.39791
\\ 0.00012
\\ 0.04778
\\ 0.54474
\\ 0.00237
\\ 0.00124
\\ 0.00423
\\ 0.08831
\end{tabular}     & \begin{tabular}[c]{@{}c@{}}3.14578
\\ 0.00010
\\ 0.03792
\\ 0.49833
\\ 0.00213
\\ 0.00115
\\ 0.00362
\\ 0.08419
\end{tabular}     & \begin{tabular}[c]{@{}c@{}}3.32161
\\ 0.00013
\\ 0.04889
\\ 0.56265
\\ 0.00244
\\ 0.00127
\\ 0.00443
\\ 0.08893
\end{tabular}     & \begin{tabular}[c]{@{}c@{}}3.03895
\\ 0.00009
\\ 0.03385
\\ 0.47290
\\ 0.00197
\\ 0.00112
\\ 0.00314
\\ 0.08256
\end{tabular}     & \begin{tabular}[c]{@{}c@{}}3.02031
\\ 0.00009
\\ 0.03758
\\ 0.46321
\\ 0.00207
\\ 0.00114
\\ 0.00294
\\ 0.07822
\end{tabular}     \\ \hline
\end{tabular}\par
\bigskip
\parbox{16.5cm}{\captionof{table}{{\fontfamily{bch}The possible 1-$\sigma$ sensitivity of \textbf{CORE-M5+CMB-S4} for bump feature in synergy with SKA1(CS), SKA2(CS), SKA1(IM1), SKA1(IM2), SKA1(IM1+IM2), and Euclid(CS+GC)+SKA1(IM2).}}} \label{table:CORE-M5+CMB-S4-Bump}
\end{minipage}

\setlength{\tabcolsep}{4.85pt} % Default value: 6pt
\renewcommand{\arraystretch}{1.5} % Default value: 1
\newcolumntype{C}[1]{>{\Centering}m{#1}}
\renewcommand\tabularxcolumn[1]{C{#1}}
\begin{minipage}{\linewidth}
\centering
\captionsetup{font=footnotesize}
\begin{tabular}{|c|c|c|c|c|c|c|c|}
\hline
\textbf{Models}            & \textbf{Parameters}                                                       & \textbf{\begin{tabular}[c]{@{}c@{}}SKA1\\ (CS)\end{tabular}} & \textbf{\begin{tabular}[c]{@{}c@{}}SKA2\\ (CS)\end{tabular}} & \textbf{\begin{tabular}[c]{@{}c@{}}SKA1\\ (IM1)\end{tabular}}            & \textbf{\begin{tabular}[c]{@{}c@{}}SKA1\\ (IM2)\end{tabular}} & \textbf{\begin{tabular}[c]{@{}c@{}}SKA1\\ (IM1+\\ IM2)\end{tabular}}           & \textbf{\begin{tabular}[c]{@{}c@{}}Euclid\\ (CS+GC)+\\ SKA1 (IM2)\end{tabular}}       \\ \hline
\textbf{\begin{tabular}[c]{@{}c@{}}Sharp Feature\\ ${k_s}=0.004$\end{tabular}}      & \begin{tabular}[c]{@{}c@{}}$10^{5}\times\sigma \left(\omega_{\mathrm{b}}\right)$\\ $\sigma \left(\omega_{\mathrm{cdm}}\right)$\\ $\sigma \left( H_0 \right)$\\ $10^{11}\times\sigma \left( A_\mathrm{s} \right)$\\ $\sigma \left( n_\mathrm{s} \right)$\\ $\sigma \left( \tau_{\mathrm{reio}} \right)$\\ $\sigma \left({C_s}\right)$\\ $10^{5}\times\sigma \left({k_s}\right)$\\ $\sigma \left( {\phi_s} \right)$\end{tabular} & \begin{tabular}[c]{@{}c@{}}3.33359
\\ 0.00026
\\ 0.10705
\\ 0.87660
\\ 0.00164
\\ 0.00227
\\ 0.00338
\\ 0.98039
\\ 0.15982
\end{tabular} & \begin{tabular}[c]{@{}c@{}}3.26685
\\ 0.00020
\\ 0.07748
\\ 0.72092
\\ 0.00161
\\ 0.00183
\\ 0.00333
\\ 0.57719
\\ 0.13651
\end{tabular} & \begin{tabular}[c]{@{}c@{}}3.05623
\\ 0.00009
\\ 0.03222
\\ 0.56324
\\ 0.00143
\\ 0.00141
\\ 0.00122
\\ 0.20296
\\ 0.05714
\end{tabular} & \begin{tabular}[c]{@{}c@{}}3.23277
\\ 0.00012
\\ 0.04393
\\ 0.66264
\\ 0.00157
\\ 0.00165
\\ 0.00223
\\ 0.48894
\\ 0.11828
\end{tabular} & \begin{tabular}[c]{@{}c@{}}2.98983
\\ 0.00008
\\ 0.02807
\\ 0.53496
\\ 0.00138
\\ 0.00136
\\ 0.00111
\\ 0.18655
\\ 0.05432
\end{tabular} & \begin{tabular}[c]{@{}c@{}}2.95296
\\ 0.00009
\\ 0.03580
\\ 0.54232
\\ 0.00136
\\ 0.00140
\\ 0.00105
\\ 0.25265
\\ 0.06069
\end{tabular} \\ \hline
\textbf{\begin{tabular}[c]{@{}c@{}}Sharp Feature\\ ${k_s}=0.03$\end{tabular}}          & \begin{tabular}[c]{@{}c@{}}$10^{5}\times\sigma \left(\omega_{\mathrm{b}}\right)$\\ $\sigma \left(\omega_{\mathrm{cdm}}\right)$\\ $\sigma \left( H_0 \right)$\\ $10^{11}\times\sigma \left( A_\mathrm{s} \right)$\\ $\sigma \left( n_\mathrm{s} \right)$\\ $\sigma \left( \tau_{\mathrm{reio}} \right)$\\ $\sigma \left({C_s}\right)$\\ $\sigma \left({k_s}\right)$\\ $\sigma \left( {\phi_s} \right)$\end{tabular} & \begin{tabular}[c]{@{}c@{}}3.20773
\\ 0.00026
\\ 0.10303
\\ 0.85103
\\ 0.00164
\\ 0.00230
\\ 0.00122
\\ 0.00035
\\ 0.10756
\end{tabular} & \begin{tabular}[c]{@{}c@{}}3.20048
\\ 0.00018
\\ 0.07066
\\ 0.69130
\\ 0.00157
\\ 0.00184
\\ 0.00121
\\ 0.00030
\\ 0.09396
\end{tabular} & \begin{tabular}[c]{@{}c@{}}3.18374
\\ 0.00010
\\ 0.04306
\\ 0.61032
\\ 0.00145
\\ 0.00154
\\ 0.00092
\\ 0.00009
\\ 0.04200
\end{tabular} & \begin{tabular}[c]{@{}c@{}}3.31133
\\ 0.00014
\\ 0.05765
\\ 0.72368
\\ 0.00159
\\ 0.00186
\\ 0.00112
\\ 0.00032
\\ 0.09473
\end{tabular} & \begin{tabular}[c]{@{}c@{}}3.14667
\\ 0.00009
\\ 0.03621
\\ 0.58268
\\ 0.00141
\\ 0.00148
\\ 0.00088
\\ 0.00008
\\ 0.03907
\end{tabular} & \begin{tabular}[c]{@{}c@{}}2.98117

\\ 0.00010
\\ 0.03865
\\ 0.57749
\\ 0.00133
\\ 0.00156
\\ 0.00083
\\ 0.00009
\\ 0.03861\end{tabular} \\ \hline
\textbf{\begin{tabular}[c]{@{}c@{}}Sharp Feature\\ ${k_s}=0.1$\end{tabular}}      & \begin{tabular}[c]{@{}c@{}}$10^{5}\times\sigma \left(\omega_{\mathrm{b}}\right)$\\ $\sigma \left(\omega_{\mathrm{cdm}}\right)$\\ $\sigma \left( H_0 \right)$\\ $10^{11}\times\sigma \left( A_\mathrm{s} \right)$\\ $\sigma \left( n_\mathrm{s} \right)$\\ $\sigma \left( \tau_{\mathrm{reio}} \right)$\\ $\sigma \left({C_s}\right)$\\ $\sigma \left({k_s}\right)$\\ $\sigma \left( {\phi_s} \right)$\end{tabular} & \begin{tabular}[c]{@{}c@{}}4.05591
\\ 0.00034
\\ 0.12905
\\ 1.29376
\\ 0.00831
\\ 0.00305
\\ 0.00588
\\ 0.01482
\\ 0.49743
\end{tabular} & \begin{tabular}[c]{@{}c@{}}4.09036
\\ 0.00019
\\ 0.07215
\\ 0.87828
\\ 0.00528
\\ 0.00235
\\ 0.00390
\\ 0.00937
\\ 0.30755
\end{tabular} & \begin{tabular}[c]{@{}c@{}}3.67834
\\ 0.00010
\\ 0.04500
\\ 0.63253
\\ 0.00255
\\ 0.00166
\\ 0.00159
\\ 0.00109
\\ 0.06350
\end{tabular} & \begin{tabular}[c]{@{}c@{}}3.95155
\\ 0.00018
\\ 0.06692
\\ 0.86385
\\ 0.00551
\\ 0.00240
\\ 0.00386
\\ 0.00930
\\ 0.30515
\end{tabular} & \begin{tabular}[c]{@{}c@{}}3.62702
\\ 0.00009
\\ 0.03900
\\ 0.60409
\\ 0.00244
\\ 0.00158
\\ 0.00150
\\ 0.00106
\\ 0.06018
\end{tabular} & \begin{tabular}[c]{@{}c@{}}3.34753
\\ 0.00010
\\ 0.04272
\\ 0.65192
\\ 0.00275
\\ 0.00152
\\ 0.00144
\\ 0.00152
\\ 0.07835
\end{tabular} \\ \hline
\end{tabular}\par
\bigskip
\parbox{16.55cm}{\captionof{table}{{\fontfamily{bch}The possible 1-$\sigma$ sensitivity of \textbf{CMB-S4} for sharp feature signal in synergy with SKA1(CS), SKA2(CS), SKA1(IM1), SKA1(IM2), SKA1(IM1+IM2), and Euclid(CS+GC)+SKA1(IM2).}}} \label{table:CMB-S4-Sharp} 
\end{minipage}

\setlength{\tabcolsep}{4.85pt} % Default value: 6pt
\renewcommand{\arraystretch}{1.5} % Default value: 1
\newcolumntype{C}[1]{>{\Centering}m{#1}}
\renewcommand\tabularxcolumn[1]{C{#1}}
\begin{minipage}{\linewidth}
\centering
\captionsetup{font=footnotesize}
\begin{tabular}{|c|c|c|c|c|c|c|c|}
\hline
\textbf{Models}            & \textbf{Parameters}                                                       & \textbf{\begin{tabular}[c]{@{}c@{}}SKA1\\ (CS)\end{tabular}} & \textbf{\begin{tabular}[c]{@{}c@{}}SKA2\\ (CS)\end{tabular}} & \textbf{\begin{tabular}[c]{@{}c@{}}SKA1\\ (IM1)\end{tabular}}            & \textbf{\begin{tabular}[c]{@{}c@{}}SKA1\\ (IM2)\end{tabular}} & \textbf{\begin{tabular}[c]{@{}c@{}}SKA1\\ (IM1+\\ IM2)\end{tabular}}           & \textbf{\begin{tabular}[c]{@{}c@{}}Euclid\\ (CS+GC)+\\ SKA1 (IM2)\end{tabular}}       \\ \hline
\textbf{\begin{tabular}[c]{@{}c@{}}Sharp Feature\\ ${k_s}=0.004$\end{tabular}}      & \begin{tabular}[c]{@{}c@{}}$10^{5}\times\sigma \left(\omega_{\mathrm{b}}\right)$\\ $\sigma \left(\omega_{\mathrm{cdm}}\right)$\\ $\sigma \left( H_0 \right)$\\ $10^{11}\times\sigma \left( A_\mathrm{s} \right)$\\ $\sigma \left( n_\mathrm{s} \right)$\\ $\sigma \left( \tau_{\mathrm{reio}} \right)$\\ $\sigma \left({C_s}\right)$\\ $10^{5}\times\sigma \left({k_s}\right)$\\ $\sigma \left( {\phi_s} \right)$\end{tabular} & \begin{tabular}[c]{@{}c@{}}3.77132
\\ 0.00021
\\ 0.08604
\\ 0.57130
\\ 0.00141
\\ 0.00134
\\ 0.00273
\\ 1.12566
\\ 0.14173
\end{tabular} & \begin{tabular}[c]{@{}c@{}}3.64088
\\ 0.00018
\\ 0.07220
\\ 0.54750
\\ 0.00136
\\ 0.00130
\\ 0.00271
\\ 0.62665
\\ 0.11373
\end{tabular} & \begin{tabular}[c]{@{}c@{}}3.47258
\\ 0.00009
\\ 0.03236
\\ 0.48172
\\ 0.00126
\\ 0.00115
\\ 0.00118
\\ 0.19961
\\ 0.05508
\end{tabular} & \begin{tabular}[c]{@{}c@{}}3.70363
\\ 0.00011
\\ 0.04372
\\ 0.54580
\\ 0.00135
\\ 0.00128
\\ 0.00201
\\ 0.46771
\\ 0.10340
\end{tabular} & \begin{tabular}[c]{@{}c@{}}3.40187
\\ 0.00008
\\ 0.02845
\\ 0.46363
\\ 0.00122
\\ 0.00112
\\ 0.00108
\\ 0.18319
\\ 0.05266
\end{tabular} & \begin{tabular}[c]{@{}c@{}}3.41682
\\ 0.00009
\\ 0.03581
\\ 0.46738
\\ 0.00121
\\ 0.00115
\\ 0.00102
\\ 0.24619
\\ 0.05876\end{tabular} \\ \hline
\textbf{\begin{tabular}[c]{@{}c@{}}Sharp Feature\\ ${k_s}=0.03$\end{tabular}}          & \begin{tabular}[c]{@{}c@{}}$10^{5}\times\sigma \left(\omega_{\mathrm{b}}\right)$\\ $\sigma \left(\omega_{\mathrm{cdm}}\right)$\\ $\sigma \left( H_0 \right)$\\ $10^{11}\times\sigma \left( A_\mathrm{s} \right)$\\ $\sigma \left( n_\mathrm{s} \right)$\\ $\sigma \left( \tau_{\mathrm{reio}} \right)$\\ $\sigma \left({C_s}\right)$\\ $\sigma \left({k_s}\right)$\\ $\sigma \left( {\phi_s} \right)$\end{tabular} & \begin{tabular}[c]{@{}c@{}}3.67480
\\ 0.00019
\\ 0.07826
\\ 0.53783
\\ 0.00136
\\ 0.00132
\\ 0.00107
\\ 0.00035
\\ 0.09421
\end{tabular} & \begin{tabular}[c]{@{}c@{}}3.52518
\\ 0.00014
\\ 0.06048
\\ 0.52227
\\ 0.00132
\\ 0.00129
\\ 0.00106
\\ 0.00030
\\ 0.08371
\end{tabular} & \begin{tabular}[c]{@{}c@{}}3.62575
\\ 0.00009
\\ 0.04110
\\ 0.48965
\\ 0.00129
\\ 0.00119
\\ 0.00085
\\ 0.00009
\\ 0.03807
\end{tabular} & \begin{tabular}[c]{@{}c@{}}3.75336
\\ 0.00012
\\ 0.05405
\\ 0.54207
\\ 0.00136
\\ 0.00131
\\ 0.00100
\\ 0.00032
\\ 0.08542
\end{tabular} & \begin{tabular}[c]{@{}c@{}}3.58470
\\ 0.00008
\\ 0.03465
\\ 0.47526
\\ 0.00126
\\ 0.00116
\\ 0.00082
\\ 0.00008
\\ 0.03567
\end{tabular} & \begin{tabular}[c]{@{}c@{}}3.30583
\\ 0.00009
\\ 0.03707
\\ 0.47094
\\ 0.00118
\\ 0.00119
\\ 0.00078
\\ 0.00009
\\ 0.03649\end{tabular} \\ \hline
\textbf{\begin{tabular}[c]{@{}c@{}}Sharp Feature\\ ${k_s}=0.1$\end{tabular}}      & \begin{tabular}[c]{@{}c@{}}$10^{5}\times\sigma \left(\omega_{\mathrm{b}}\right)$\\ $\sigma \left(\omega_{\mathrm{cdm}}\right)$\\ $\sigma \left( H_0 \right)$\\ $10^{11}\times\sigma \left( A_\mathrm{s} \right)$\\ $\sigma \left( n_\mathrm{s} \right)$\\ $\sigma \left( \tau_{\mathrm{reio}} \right)$\\ $\sigma \left({C_s}\right)$\\ $\sigma \left({k_s}\right)$\\ $\sigma \left( {\phi_s} \right)$\end{tabular} & \begin{tabular}[c]{@{}c@{}}4.57094
\\ 0.00028
\\ 0.10213
\\ 1.04240
\\ 0.00619
\\ 0.00176
\\ 0.00484
\\ 0.01326
\\ 0.41418
\end{tabular} & \begin{tabular}[c]{@{}c@{}}4.37443
\\ 0.00018
\\ 0.06916
\\ 0.79919
\\ 0.00431
\\ 0.00152
\\ 0.00333
\\ 0.00890
\\ 0.27527
\end{tabular} & \begin{tabular}[c]{@{}c@{}}4.08577
\\ 0.00009
\\ 0.04204
\\ 0.51486
\\ 0.00229
\\ 0.00125
\\ 0.00138
\\ 0.00108
\\ 0.06357
\end{tabular} & \begin{tabular}[c]{@{}c@{}}4.32006
\\ 0.00017
\\ 0.06605
\\ 0.80478
\\ 0.00437
\\ 0.00153
\\ 0.00325
\\ 0.00884
\\ 0.27092
\end{tabular} & \begin{tabular}[c]{@{}c@{}}4.02111
\\ 0.00008
\\ 0.03688
\\ 0.49995
\\ 0.00222
\\ 0.00121
\\ 0.00131
\\ 0.00105
\\ 0.06058
\end{tabular} & \begin{tabular}[c]{@{}c@{}}3.66293
\\ 0.00009
\\ 0.03980
\\ 0.54211
\\ 0.00245
\\ 0.00121
\\ 0.00128
\\ 0.00149
\\ 0.07660
\end{tabular} \\ \hline
\end{tabular}\par
\bigskip
\parbox{16.55cm}{\captionof{table}{{\fontfamily{bch}The possible 1-$\sigma$ sensitivity of \textbf{CORE-M5} for sharp feature signal in synergy with SKA1(CS), SKA2(CS), SKA1(IM1), SKA1(IM2), SKA1(IM1+IM2), and Euclid(CS+GC)+SKA1(IM2).}}} \label{table:CORE-M5-Sharp} 
\end{minipage}

\setlength{\tabcolsep}{4.85pt} % Default value: 6pt
\renewcommand{\arraystretch}{1.5} % Default value: 1
\newcolumntype{C}[1]{>{\Centering}m{#1}}
\renewcommand\tabularxcolumn[1]{C{#1}}
\begin{minipage}{\linewidth}
\centering
\captionsetup{font=footnotesize}
\begin{tabular}{|c|c|c|c|c|c|c|c|}
\hline
\textbf{Models}            & \textbf{Parameters}                                                       & \textbf{\begin{tabular}[c]{@{}c@{}}SKA1\\ (CS)\end{tabular}} & \textbf{\begin{tabular}[c]{@{}c@{}}SKA2\\ (CS)\end{tabular}} & \textbf{\begin{tabular}[c]{@{}c@{}}SKA1\\ (IM1)\end{tabular}}            & \textbf{\begin{tabular}[c]{@{}c@{}}SKA1\\ (IM2)\end{tabular}} & \textbf{\begin{tabular}[c]{@{}c@{}}SKA1\\ (IM1+\\ IM2)\end{tabular}}           & \textbf{\begin{tabular}[c]{@{}c@{}}Euclid\\ (CS+GC)+\\ SKA1 (IM2)\end{tabular}}       \\ \hline
\textbf{\begin{tabular}[c]{@{}c@{}}Sharp Feature\\ ${k_s}=0.004$\end{tabular}}      & \begin{tabular}[c]{@{}c@{}}$10^{5}\times\sigma \left(\omega_{\mathrm{b}}\right)$\\ $\sigma \left(\omega_{\mathrm{cdm}}\right)$\\ $\sigma \left( H_0 \right)$\\ $10^{11}\times\sigma \left( A_\mathrm{s} \right)$\\ $\sigma \left( n_\mathrm{s} \right)$\\ $\sigma \left( \tau_{\mathrm{reio}} \right)$\\ $\sigma \left({C_s}\right)$\\ $10^{5}\times\sigma \left({k_s}\right)$\\ $\sigma \left( {\phi_s} \right)$\end{tabular} & \begin{tabular}[c]{@{}c@{}}2.79515
\\ 0.00016
\\ 0.06425
\\ 0.49253
\\ 0.00129
\\ 0.00118
\\ 0.00258
\\ 0.84427
\\ 0.12635
\end{tabular} & \begin{tabular}[c]{@{}c@{}}2.75049
\\ 0.00016
\\ 0.06044
\\ 0.48627
\\ 0.00126
\\ 0.00117
\\ 0.00257
\\ 0.53424
\\ 0.10778
\end{tabular} & \begin{tabular}[c]{@{}c@{}}2.63415
\\ 0.00008
\\ 0.02942
\\ 0.42190
\\ 0.00116
\\ 0.00102
\\ 0.00116
\\ 0.19314
\\ 0.05375
\end{tabular} & \begin{tabular}[c]{@{}c@{}}2.73235
\\ 0.00010
\\ 0.03829
\\ 0.46210
\\ 0.00123
\\ 0.00110
\\ 0.00195
\\ 0.45137
\\ 0.09964
\end{tabular} & \begin{tabular}[c]{@{}c@{}}2.59294
\\ 0.00007
\\ 0.02557
\\ 0.40817
\\ 0.00113
\\ 0.00099
\\ 0.00107
\\ 0.17680
\\ 0.05137
\end{tabular} & \begin{tabular}[c]{@{}c@{}}2.55283
\\ 0.00008
\\ 0.03208
\\ 0.40717
\\ 0.00110
\\ 0.00101
\\ 0.00101
\\ 0.24216
\\ 0.05693\end{tabular} \\ \hline
\textbf{\begin{tabular}[c]{@{}c@{}}Sharp Feature\\ ${k_s}=0.03$\end{tabular}}          & \begin{tabular}[c]{@{}c@{}}$10^{5}\times\sigma \left(\omega_{\mathrm{b}}\right)$\\ $\sigma \left(\omega_{\mathrm{cdm}}\right)$\\ $\sigma \left( H_0 \right)$\\ $10^{11}\times\sigma \left( A_\mathrm{s} \right)$\\ $\sigma \left( n_\mathrm{s} \right)$\\ $\sigma \left( \tau_{\mathrm{reio}} \right)$\\ $\sigma \left({C_s}\right)$\\ $\sigma \left({k_s}\right)$\\ $\sigma \left( {\phi_s} \right)$\end{tabular} & \begin{tabular}[c]{@{}c@{}}2.71848
\\ 0.00017
\\ 0.06476
\\ 0.46881
\\ 0.00125
\\ 0.00118
\\ 0.00095
\\ 0.00027
\\ 0.08042
\end{tabular} & \begin{tabular}[c]{@{}c@{}}2.71824
\\ 0.00013
\\ 0.05255
\\ 0.47821
\\ 0.00119
\\ 0.00113
\\ 0.00095
\\ 0.00025
\\ 0.07552
\end{tabular} & \begin{tabular}[c]{@{}c@{}}2.70806
\\ 0.00009
\\ 0.03665
\\ 0.44147
\\ 0.00116
\\ 0.00108
\\ 0.00078
\\ 0.00009
\\ 0.03725
\end{tabular} & \begin{tabular}[c]{@{}c@{}}2.77653
\\ 0.00012
\\ 0.04760
\\ 0.48398
\\ 0.00123
\\ 0.00119
\\ 0.00090
\\ 0.00026
\\ 0.07574
\end{tabular} & \begin{tabular}[c]{@{}c@{}}2.6756
\\ 0.00007
\\ 0.03003
\\ 0.42441
\\ 0.00114
\\ 0.00104
\\ 0.00076
\\ 0.00008
\\ 0.03488
\end{tabular} & \begin{tabular}[c]{@{}c@{}}2.53623
\\ 0.00008
\\ 0.03270
\\ 0.41275
\\ 0.00106
\\ 0.00107
\\ 0.00073
\\ 0.00009
\\ 0.03428
\end{tabular} \\ \hline
\textbf{\begin{tabular}[c]{@{}c@{}}Sharp Feature\\ ${k_s}=0.1$\end{tabular}}      & \begin{tabular}[c]{@{}c@{}}$10^{5}\times\sigma \left(\omega_{\mathrm{b}}\right)$\\ $\sigma \left(\omega_{\mathrm{cdm}}\right)$\\ $\sigma \left( H_0 \right)$\\ $10^{11}\times\sigma \left( A_\mathrm{s} \right)$\\ $\sigma \left( n_\mathrm{s} \right)$\\ $\sigma \left( \tau_{\mathrm{reio}} \right)$\\ $\sigma \left({C_s}\right)$\\ $\sigma \left({k_s}\right)$\\ $\sigma \left( {\phi_s} \right)$\end{tabular} & \begin{tabular}[c]{@{}c@{}}3.66396
\\ 0.00027
\\ 0.09126
\\ 0.95842
\\ 0.00631
\\ 0.00174
\\ 0.00472
\\ 0.01176
\\ 0.39335
\end{tabular} & \begin{tabular}[c]{@{}c@{}}3.42067
\\ 0.00016
\\ 0.05948
\\ 0.73013
\\ 0.00432
\\ 0.00145
\\ 0.00320
\\ 0.00788
\\ 0.26048
\end{tabular} & \begin{tabular}[c]{@{}c@{}}3.10908
\\ 0.00008
\\ 0.03677
\\ 0.46925
\\ 0.00208
\\ 0.00114
\\ 0.00133
\\ 0.00107
\\ 0.05708
\end{tabular} & \begin{tabular}[c]{@{}c@{}}3.31696
\\ 0.00016
\\ 0.05854
\\ 0.70549
\\ 0.00428
\\ 0.00145
\\ 0.00305
\\ 0.00752
\\ 0.24853
\end{tabular} & \begin{tabular}[c]{@{}c@{}}3.04827
\\ 0.00007
\\ 0.03141
\\ 0.45085
\\ 0.00201
\\ 0.00109
\\ 0.00126
\\ 0.00105
\\ 0.05441
\end{tabular} & \begin{tabular}[c]{@{}c@{}}2.80293
\\ 0.00008
\\ 0.03539
\\ 0.48601
\\ 0.00222
\\ 0.00108
\\ 0.00127
\\ 0.00150
\\ 0.07062
\end{tabular} \\ \hline
\end{tabular}\par
\bigskip
\parbox{16.5cm}{\captionof{table}{{\fontfamily{bch}The possible 1-$\sigma$ sensitivity of \textbf{PICO} for sharp feature signal in synergy with SKA1(CS), SKA2(CS), SKA1(IM1), SKA1(IM2), SKA1(IM1+IM2), and Euclid(CS+GC)+SKA1(IM2).}}} \label{table:PICO-Sharp}
\end{minipage}

\setlength{\tabcolsep}{4.85pt} % Default value: 6pt
\renewcommand{\arraystretch}{1.5} % Default value: 1
\newcolumntype{C}[1]{>{\Centering}m{#1}}
\renewcommand\tabularxcolumn[1]{C{#1}}
\begin{minipage}{\linewidth}
\centering
\captionsetup{font=footnotesize}
\begin{tabular}{|c|c|c|c|c|c|c|c|}
\hline
\textbf{Models}            & \textbf{Parameters}                                                       & \textbf{\begin{tabular}[c]{@{}c@{}}SKA1\\ (CS)\end{tabular}} & \textbf{\begin{tabular}[c]{@{}c@{}}SKA2\\ (CS)\end{tabular}} & \textbf{\begin{tabular}[c]{@{}c@{}}SKA1\\ (IM1)\end{tabular}}            & \textbf{\begin{tabular}[c]{@{}c@{}}SKA1\\ (IM2)\end{tabular}} & \textbf{\begin{tabular}[c]{@{}c@{}}SKA1\\ (IM1+\\ IM2)\end{tabular}}           & \textbf{\begin{tabular}[c]{@{}c@{}}Euclid\\ (CS+GC)+\\ SKA1 (IM2)\end{tabular}}       \\ \hline
\textbf{\begin{tabular}[c]{@{}c@{}}Sharp Feature\\ ${k_s}=0.004$\end{tabular}}      & \begin{tabular}[c]{@{}c@{}}$10^{5}\times\sigma \left(\omega_{\mathrm{b}}\right)$\\ $\sigma \left(\omega_{\mathrm{cdm}}\right)$\\ $\sigma \left( H_0 \right)$\\ $10^{11}\times\sigma \left( A_\mathrm{s} \right)$\\ $\sigma \left( n_\mathrm{s} \right)$\\ $\sigma \left( \tau_{\mathrm{reio}} \right)$\\ $\sigma \left({C_s}\right)$\\ $10^{5}\times\sigma \left({k_s}\right)$\\ $\sigma \left( {\phi_s} \right)$\end{tabular} & \begin{tabular}[c]{@{}c@{}}3.22755
\\ 0.00024
\\ 0.09965
\\ 0.77378
\\ 0.00154
\\ 0.00198
\\ 0.00324
\\ 0.96716
\\ 0.15218
\end{tabular} & \begin{tabular}[c]{@{}c@{}}3.17715
\\ 0.00019
\\ 0.07399
\\ 0.66166
\\ 0.00152
\\ 0.00165
\\ 0.00321
\\ 0.57044
\\ 0.13023
\end{tabular} & \begin{tabular}[c]{@{}c@{}}3.04679
\\ 0.00009
\\ 0.03254
\\ 0.56132
\\ 0.00136
\\ 0.00140
\\ 0.00121
\\ 0.19744
\\ 0.05621
\end{tabular} & \begin{tabular}[c]{@{}c@{}}3.21640
\\ 0.00011
\\ 0.04408
\\ 0.66422
\\ 0.00148
\\ 0.00163
\\ 0.00219
\\ 0.42454
\\ 0.10766
\end{tabular} & \begin{tabular}[c]{@{}c@{}}2.98176
\\ 0.00008
\\ 0.02834
\\ 0.53568
\\ 0.00132
\\ 0.00135
\\ 0.00110
\\ 0.18220
\\ 0.05352
\end{tabular} & \begin{tabular}[c]{@{}c@{}}2.95341
\\ 0.00009
\\ 0.03605
\\ 0.53418
\\ 0.00129
\\ 0.00136
\\ 0.00105
\\ 0.24349
\\ 0.05930\end{tabular} \\ \hline
\textbf{\begin{tabular}[c]{@{}c@{}}Sharp Feature\\ ${k_s}=0.03$\end{tabular}}          & \begin{tabular}[c]{@{}c@{}}$10^{5}\times\sigma \left(\omega_{\mathrm{b}}\right)$\\ $\sigma \left(\omega_{\mathrm{cdm}}\right)$\\ $\sigma \left( H_0 \right)$\\ $10^{11}\times\sigma \left( A_\mathrm{s} \right)$\\ $\sigma \left( n_\mathrm{s} \right)$\\ $\sigma \left( \tau_{\mathrm{reio}} \right)$\\ $\sigma \left({C_s}\right)$\\ $\sigma \left({k_s}\right)$\\ $\sigma \left( {\phi_s} \right)$\end{tabular} & \begin{tabular}[c]{@{}c@{}}3.12403
\\ 0.00024
\\ 0.09598
\\ 0.75157
\\ 0.00153
\\ 0.00200
\\ 0.00118
\\ 0.00034
\\ 0.10155
\end{tabular} & \begin{tabular}[c]{@{}c@{}}3.11476
\\ 0.00017
\\ 0.06682
\\ 0.63397
\\ 0.00147
\\ 0.00166
\\ 0.00117
\\ 0.00029
\\ 0.08963
\end{tabular} & \begin{tabular}[c]{@{}c@{}}3.12999
\\ 0.00010
\\ 0.04288
\\ 0.60491
\\ 0.00138
\\ 0.00152
\\ 0.00090
\\ 0.00009
\\ 0.04143
\end{tabular} & \begin{tabular}[c]{@{}c@{}}3.24746
\\ 0.00014
\\ 0.05778
\\ 0.73427
\\ 0.00151
\\ 0.00187
\\ 0.00110
\\ 0.00036
\\ 0.10312
\end{tabular} & \begin{tabular}[c]{@{}c@{}}3.09587
\\ 0.00009
\\ 0.03597
\\ 0.57801
\\ 0.00135
\\ 0.00146
\\ 0.00086
\\ 0.00008
\\ 0.03857
\end{tabular} & \begin{tabular}[c]{@{}c@{}}2.98386
\\ 0.00010
\\ 0.03864
\\ 0.55819
\\ 0.00131
\\ 0.00145
\\ 0.00082
\\ 0.00010
\\ 0.03951
\end{tabular} \\ \hline
\textbf{\begin{tabular}[c]{@{}c@{}}Sharp Feature\\ ${k_s}=0.1$\end{tabular}}      & \begin{tabular}[c]{@{}c@{}}$10^{5}\times\sigma \left(\omega_{\mathrm{b}}\right)$\\ $\sigma \left(\omega_{\mathrm{cdm}}\right)$\\ $\sigma \left( H_0 \right)$\\ $10^{11}\times\sigma \left( A_\mathrm{s} \right)$\\ $\sigma \left( n_\mathrm{s} \right)$\\ $\sigma \left( \tau_{\mathrm{reio}} \right)$\\ $\sigma \left({C_s}\right)$\\ $\sigma \left({k_s}\right)$\\ $\sigma \left( {\phi_s} \right)$\end{tabular} & \begin{tabular}[c]{@{}c@{}}3.94772
\\ 0.00032
\\ 0.12269
\\ 1.19263
\\ 0.00722
\\ 0.00252
\\ 0.00517
\\ 0.01335
\\ 0.44388
\end{tabular} & \begin{tabular}[c]{@{}c@{}}4.02523
\\ 0.00018
\\ 0.06976
\\ 0.83721
\\ 0.00491
\\ 0.00208
\\ 0.00367
\\ 0.00899
\\ 0.29314
\end{tabular} & \begin{tabular}[c]{@{}c@{}}3.69776
\\ 0.00011
\\ 0.04591
\\ 0.64362
\\ 0.00244
\\ 0.00172
\\ 0.00155
\\ 0.00111
\\ 0.06309
\end{tabular} & \begin{tabular}[c]{@{}c@{}}3.89263
\\ 0.00017
\\ 0.06561
\\ 0.82340
\\ 0.00505
\\ 0.00212
\\ 0.00359
\\ 0.00884
\\ 0.28821
\end{tabular} & \begin{tabular}[c]{@{}c@{}}3.64191
\\ 0.00009
\\ 0.03951
\\ 0.61282
\\ 0.00234
\\ 0.00163
\\ 0.00147
\\ 0.00109
\\ 0.05986
\end{tabular} & \begin{tabular}[c]{@{}c@{}}3.23889
\\ 0.00010
\\ 0.04220
\\ 0.64201
\\ 0.00257
\\ 0.00152
\\ 0.00139
\\ 0.00149
\\ 0.07615\end{tabular} \\ \hline
\end{tabular}\par
\bigskip
\parbox{16.52cm}{\captionof{table}{{\fontfamily{bch}The possible 1-$\sigma$ sensitivity of \textbf{Planck+CMB-S4} for sharp feature signal in synergy with SKA1(CS), SKA2(CS), SKA1(IM1), SKA1(IM2), SKA1(IM1+IM2), and Euclid(CS+GC)+SKA1(IM2).}}} \label{table:Planck+CMB-S4-Sharp}
\end{minipage}

\setlength{\tabcolsep}{4.85pt} % Default value: 6pt
\renewcommand{\arraystretch}{1.5} % Default value: 1
\newcolumntype{C}[1]{>{\Centering}m{#1}}
\renewcommand\tabularxcolumn[1]{C{#1}}
\begin{minipage}{\linewidth}
\centering
\captionsetup{font=footnotesize}
\begin{tabular}{|c|c|c|c|c|c|c|c|}
\hline
\textbf{Models}            & \textbf{Parameters}                                                       & \textbf{\begin{tabular}[c]{@{}c@{}}SKA1\\ (CS)\end{tabular}} & \textbf{\begin{tabular}[c]{@{}c@{}}SKA2\\ (CS)\end{tabular}} & \textbf{\begin{tabular}[c]{@{}c@{}}SKA1\\ (IM1)\end{tabular}}            & \textbf{\begin{tabular}[c]{@{}c@{}}SKA1\\ (IM2)\end{tabular}} & \textbf{\begin{tabular}[c]{@{}c@{}}SKA1\\ (IM1+\\ IM2)\end{tabular}}           & \textbf{\begin{tabular}[c]{@{}c@{}}Euclid\\ (CS+GC)+\\ SKA1 (IM2)\end{tabular}}       \\ \hline
\textbf{\begin{tabular}[c]{@{}c@{}}Sharp Feature\\ ${k_s}=0.004$\end{tabular}}      & \begin{tabular}[c]{@{}c@{}}$10^{5}\times\sigma \left(\omega_{\mathrm{b}}\right)$\\ $\sigma \left(\omega_{\mathrm{cdm}}\right)$\\ $\sigma \left( H_0 \right)$\\ $10^{11}\times\sigma \left( A_\mathrm{s} \right)$\\ $\sigma \left( n_\mathrm{s} \right)$\\ $\sigma \left( \tau_{\mathrm{reio}} \right)$\\ $\sigma \left({C_s}\right)$\\ $10^{5}\times\sigma \left({k_s}\right)$\\ $\sigma \left( {\phi_s} \right)$\end{tabular} & \begin{tabular}[c]{@{}c@{}}3.12730
\\ 0.00022
\\ 0.08955
\\ 0.59265
\\ 0.00143
\\ 0.00149
\\ 0.00317
\\ 0.94717
\\ 0.14527
\end{tabular} & \begin{tabular}[c]{@{}c@{}}3.09902
\\ 0.00018
\\ 0.06893
\\ 0.54157
\\ 0.00142
\\ 0.00133
\\ 0.00314
\\ 0.55740
\\ 0.12532
\end{tabular} & \begin{tabular}[c]{@{}c@{}}2.93743
\\ 0.00009
\\ 0.03120
\\ 0.47069
\\ 0.00127
\\ 0.00114
\\ 0.00121
\\ 0.19743
\\ 0.05623
\end{tabular} & \begin{tabular}[c]{@{}c@{}}3.07855
\\ 0.00011
\\ 0.04204
\\ 0.52285
\\ 0.00137
\\ 0.00126
\\ 0.00217
\\ 0.46638
\\ 0.11160
\end{tabular} & \begin{tabular}[c]{@{}c@{}}2.88394
\\ 0.00008
\\ 0.02730
\\ 0.45332
\\ 0.00124
\\ 0.00111
\\ 0.00110
\\ 0.18141
\\ 0.05355
\end{tabular} & \begin{tabular}[c]{@{}c@{}}2.85332
\\ 0.00009
\\ 0.03443
\\ 0.45598
\\ 0.00121
\\ 0.00114
\\ 0.00104
\\ 0.24620
\\ 0.05932\end{tabular} \\ \hline
\textbf{\begin{tabular}[c]{@{}c@{}}Sharp Feature\\ ${k_s}=0.03$\end{tabular}}          & \begin{tabular}[c]{@{}c@{}}$10^{5}\times\sigma \left(\omega_{\mathrm{b}}\right)$\\ $\sigma \left(\omega_{\mathrm{cdm}}\right)$\\ $\sigma \left( H_0 \right)$\\ $10^{11}\times\sigma \left( A_\mathrm{s} \right)$\\ $\sigma \left( n_\mathrm{s} \right)$\\ $\sigma \left( \tau_{\mathrm{reio}} \right)$\\ $\sigma \left({C_s}\right)$\\ $\sigma \left({k_s}\right)$\\ $\sigma \left( {\phi_s} \right)$\end{tabular} & \begin{tabular}[c]{@{}c@{}}3.07202
\\ 0.00022
\\ 0.08690
\\ 0.57917
\\ 0.00141
\\ 0.00150
\\ 0.00118
\\ 0.00033
\\ 0.10019
\end{tabular} & \begin{tabular}[c]{@{}c@{}}3.06946
\\ 0.00015
\\ 0.06103
\\ 0.52303
\\ 0.00137
\\ 0.00134
\\ 0.00118
\\ 0.00029
\\ 0.08884
\end{tabular} & \begin{tabular}[c]{@{}c@{}}3.05657
\\ 0.00009
\\ 0.03993
\\ 0.49446
\\ 0.00129
\\ 0.00121
\\ 0.00091
\\ 0.00009
\\ 0.04117
\end{tabular} & \begin{tabular}[c]{@{}c@{}}3.15320
\\ 0.00013
\\ 0.05253
\\ 0.54455
\\ 0.00139
\\ 0.00135
\\ 0.00110
\\ 0.00030
\\ 0.08773
\end{tabular} & \begin{tabular}[c]{@{}c@{}}3.02840
\\ 0.00008
\\ 0.03360
\\ 0.47956
\\ 0.00126
\\ 0.00118
\\ 0.00087
\\ 0.00008
\\ 0.03834
\end{tabular} & \begin{tabular}[c]{@{}c@{}}2.87056
\\ 0.00009
\\ 0.03597
\\ 0.46886
\\ 0.00118
\\ 0.00121
\\ 0.00082
\\ 0.00009
\\ 0.03778\end{tabular} \\ \hline
\textbf{\begin{tabular}[c]{@{}c@{}}Sharp Feature\\ ${k_s}=0.1$\end{tabular}}      & \begin{tabular}[c]{@{}c@{}}$10^{5}\times\sigma \left(\omega_{\mathrm{b}}\right)$\\ $\sigma \left(\omega_{\mathrm{cdm}}\right)$\\ $\sigma \left( H_0 \right)$\\ $10^{11}\times\sigma \left( A_\mathrm{s} \right)$\\ $\sigma \left( n_\mathrm{s} \right)$\\ $\sigma \left( \tau_{\mathrm{reio}} \right)$\\ $\sigma \left({C_s}\right)$\\ $\sigma \left({k_s}\right)$\\ $\sigma \left( {\phi_s} \right)$\end{tabular} & \begin{tabular}[c]{@{}c@{}}3.73507
\\ 0.00030
\\ 0.11536
\\ 1.06439
\\ 0.00578
\\ 0.00169
\\ 0.00419
\\ 0.01142
\\ 0.37554
\end{tabular} & \begin{tabular}[c]{@{}c@{}}3.83519
\\ 0.00018
\\ 0.06677
\\ 0.77615
\\ 0.00425
\\ 0.00154
\\ 0.00321
\\ 0.00817
\\ 0.26522
\end{tabular} & \begin{tabular}[c]{@{}c@{}}3.52871
\\ 0.00009
\\ 0.04074
\\ 0.52153
\\ 0.00222
\\ 0.00127
\\ 0.00146
\\ 0.00107
\\ 0.06152
\end{tabular} & \begin{tabular}[c]{@{}c@{}}3.73145
\\ 0.00017
\\ 0.06392
\\ 0.76037
\\ 0.00429
\\ 0.00154
\\ 0.00307
\\ 0.00794
\\ 0.25696
\end{tabular} & \begin{tabular}[c]{@{}c@{}}3.48153
\\ 0.00008
\\ 0.03566
\\ 0.50530
\\ 0.00215
\\ 0.00123
\\ 0.00139
\\ 0.00105
\\ 0.05851
\end{tabular} & \begin{tabular}[c]{@{}c@{}}3.22707
\\ 0.00009
\\ 0.03948
\\ 0.54315
\\ 0.00240
\\ 0.00120
\\ 0.00132
\\ 0.00150
\\ 0.07550
\end{tabular} \\ \hline
\end{tabular}\par
\bigskip
\parbox{16.5cm}{\captionof{table}{{\fontfamily{bch}The possible 1-$\sigma$ sensitivity of \textbf{LiteBIRD+CMB-S4} for sharp feature signal in synergy with SKA1(CS), SKA2(CS), SKA1(IM1), SKA1(IM2), SKA1(IM1+IM2), and Euclid(CS+GC)+SKA1(IM2).}}} \label{table:LiteBIRD+CMB-S4-Sharp}
\end{minipage}

\setlength{\tabcolsep}{4.85pt} % Default value: 6pt
\renewcommand{\arraystretch}{1.5} % Default value: 1
\newcolumntype{C}[1]{>{\Centering}m{#1}}
\renewcommand\tabularxcolumn[1]{C{#1}}
\begin{minipage}{\linewidth}
\centering
\captionsetup{font=footnotesize}
\begin{tabular}{|c|c|c|c|c|c|c|c|}
\hline
\textbf{Models}            & \textbf{Parameters}                                                       & \textbf{\begin{tabular}[c]{@{}c@{}}SKA1\\ (CS)\end{tabular}} & \textbf{\begin{tabular}[c]{@{}c@{}}SKA2\\ (CS)\end{tabular}} & \textbf{\begin{tabular}[c]{@{}c@{}}SKA1\\ (IM1)\end{tabular}}            & \textbf{\begin{tabular}[c]{@{}c@{}}SKA1\\ (IM2)\end{tabular}} & \textbf{\begin{tabular}[c]{@{}c@{}}SKA1\\ (IM1+\\ IM2)\end{tabular}}           & \textbf{\begin{tabular}[c]{@{}c@{}}Euclid\\ (CS+GC)+\\ SKA1 (IM2)\end{tabular}}       \\ \hline
\textbf{\begin{tabular}[c]{@{}c@{}}Sharp Feature\\ ${k_s}=0.004$\end{tabular}}      & \begin{tabular}[c]{@{}c@{}}$10^{5}\times\sigma \left(\omega_{\mathrm{b}}\right)$\\ $\sigma \left(\omega_{\mathrm{cdm}}\right)$\\ $\sigma \left( H_0 \right)$\\ $10^{11}\times\sigma \left( A_\mathrm{s} \right)$\\ $\sigma \left( n_\mathrm{s} \right)$\\ $\sigma \left( \tau_{\mathrm{reio}} \right)$\\ $\sigma \left({C_s}\right)$\\ $10^{5}\times\sigma \left({k_s}\right)$\\ $\sigma \left( {\phi_s} \right)$\end{tabular} & \begin{tabular}[c]{@{}c@{}}2.79985
\\ 0.00017
\\ 0.06668
\\ 0.50799
\\ 0.00131
\\ 0.00121
\\ 0.00262
\\ 0.88237
\\ 0.12880
\end{tabular} & \begin{tabular}[c]{@{}c@{}}2.75327
\\ 0.00016
\\ 0.06201
\\ 0.49894
\\ 0.00128
\\ 0.00120
\\ 0.00260
\\ 0.55283
\\ 0.10928
\end{tabular} & \begin{tabular}[c]{@{}c@{}}2.63281
\\ 0.00008
\\ 0.02965
\\ 0.43501
\\ 0.00117
\\ 0.00105
\\ 0.00117
\\ 0.19456
\\ 0.05395
\end{tabular} & \begin{tabular}[c]{@{}c@{}}2.73555
\\ 0.00010
\\ 0.03883
\\ 0.47969
\\ 0.00124
\\ 0.00114
\\ 0.00197
\\ 0.45401
\\ 0.10034
\end{tabular} & \begin{tabular}[c]{@{}c@{}}2.59149
\\ 0.00007
\\ 0.02577
\\ 0.42061
\\ 0.00114
\\ 0.00102
\\ 0.00107
\\ 0.17826
\\ 0.05155
\end{tabular} & \begin{tabular}[c]{@{}c@{}}2.55698
\\ 0.00008
\\ 0.03239
\\ 0.42031
\\ 0.00112
\\ 0.00104
\\ 0.00102
\\ 0.24259
\\ 0.05703\end{tabular} \\ \hline
\textbf{\begin{tabular}[c]{@{}c@{}}Sharp Feature\\ ${k_s}=0.03$\end{tabular}}          & \begin{tabular}[c]{@{}c@{}}$10^{5}\times\sigma \left(\omega_{\mathrm{b}}\right)$\\ $\sigma \left(\omega_{\mathrm{cdm}}\right)$\\ $\sigma \left( H_0 \right)$\\ $10^{11}\times\sigma \left( A_\mathrm{s} \right)$\\ $\sigma \left( n_\mathrm{s} \right)$\\ $\sigma \left( \tau_{\mathrm{reio}} \right)$\\ $\sigma \left({C_s}\right)$\\ $\sigma \left({k_s}\right)$\\ $\sigma \left( {\phi_s} \right)$\end{tabular} & \begin{tabular}[c]{@{}c@{}}2.73437
\\ 0.00017
\\ 0.06649
\\ 0.48242
\\ 0.00127
\\ 0.00121
\\ 0.00097
\\ 0.00028
\\ 0.08221
\end{tabular} & \begin{tabular}[c]{@{}c@{}}2.69912
\\ 0.00014
\\ 0.05386
\\ 0.47101
\\ 0.00124
\\ 0.00119
\\ 0.00097
\\ 0.00026
\\ 0.07689
\end{tabular} & \begin{tabular}[c]{@{}c@{}}2.71925
\\ 0.00009
\\ 0.03706
\\ 0.45241
\\ 0.00118
\\ 0.00110
\\ 0.00080
\\ 0.00009
\\ 0.03749
\end{tabular} & \begin{tabular}[c]{@{}c@{}}2.78809
\\ 0.00012
\\ 0.04827
\\ 0.49601
\\ 0.00125
\\ 0.00121
\\ 0.00092
\\ 0.00027
\\ 0.07707
\end{tabular} & \begin{tabular}[c]{@{}c@{}}2.69725
\\ 0.00007
\\ 0.03093
\\ 0.44077
\\ 0.00115
\\ 0.00108
\\ 0.00077
\\ 0.00008
\\ 0.03510
\end{tabular} & \begin{tabular}[c]{@{}c@{}}2.54784
\\ 0.00008
\\ 0.03320
\\ 0.42524
\\ 0.00108
\\ 0.00110
\\ 0.00074
\\ 0.00009
\\ 0.03478
\end{tabular} \\ \hline
\textbf{\begin{tabular}[c]{@{}c@{}}Sharp Feature\\ ${k_s}=0.1$\end{tabular}}      & \begin{tabular}[c]{@{}c@{}}$10^{5}\times\sigma \left(\omega_{\mathrm{b}}\right)$\\ $\sigma \left(\omega_{\mathrm{cdm}}\right)$\\ $\sigma \left( H_0 \right)$\\ $10^{11}\times\sigma \left( A_\mathrm{s} \right)$\\ $\sigma \left( n_\mathrm{s} \right)$\\ $\sigma \left( \tau_{\mathrm{reio}} \right)$\\ $\sigma \left({C_s}\right)$\\ $\sigma \left({k_s}\right)$\\ $\sigma \left( {\phi_s} \right)$\end{tabular} & \begin{tabular}[c]{@{}c@{}}3.70399
\\ 0.00027
\\ 0.09379
\\ 0.97356
\\ 0.00633
\\ 0.00178
\\ 0.00483
\\ 0.01217
\\ 0.40076
\end{tabular} & \begin{tabular}[c]{@{}c@{}}3.45556
\\ 0.00017
\\ 0.06073
\\ 0.74053
\\ 0.00429
\\ 0.00147
\\ 0.00324
\\ 0.00808
\\ 0.26269
\end{tabular} & \begin{tabular}[c]{@{}c@{}}3.15419
\\ 0.00009
\\ 0.03742
\\ 0.47954
\\ 0.00212
\\ 0.00117
\\ 0.00134
\\ 0.00107
\\ 0.05839
\end{tabular} & \begin{tabular}[c]{@{}c@{}}3.35697
\\ 0.00016
\\ 0.05945
\\ 0.72132
\\ 0.00430
\\ 0.00148
\\ 0.00311
\\ 0.00777
\\ 0.25290
\end{tabular} & \begin{tabular}[c]{@{}c@{}}3.11683
\\ 0.00007
\\ 0.03268
\\ 0.46639
\\ 0.00205
\\ 0.00114
\\ 0.00128
\\ 0.00105
\\ 0.05565
\end{tabular} & \begin{tabular}[c]{@{}c@{}}2.83031
\\ 0.00009
\\ 0.03594
\\ 0.49860
\\ 0.00227
\\ 0.00111
\\ 0.00127
\\ 0.00150
\\ 0.07177
\end{tabular} \\ \hline
\end{tabular}\par
\bigskip
\parbox{16.52cm}{\captionof{table}{{\fontfamily{bch}The possible 1-$\sigma$ sensitivity of \textbf{CORE-M5+CMB-S4} for sharp feature signal in synergy with SKA1(CS), SKA2(CS), SKA1(IM1), SKA1(IM2), SKA1(IM1+IM2), and Euclid(CS+GC)+SKA1(IM2).}}} \label{table:CORE-M5+CMB-S4-Sharp}
\end{minipage}

\setlength{\tabcolsep}{3.25pt} % Default value: 6pt
\renewcommand{\arraystretch}{1.5} % Default value: 1
\newcolumntype{C}[1]{>{\Centering}m{#1}}
\renewcommand\tabularxcolumn[1]{C{#1}}
\begin{minipage}{\linewidth}
\centering
\captionsetup{font=footnotesize}
\begin{tabular}{|c|c|c|c|c|c|c|c|}
\hline
\textbf{Models}            & \textbf{Parameters}                                                       & \textbf{\begin{tabular}[c]{@{}c@{}}SKA1\\ (CS)\end{tabular}} & \textbf{\begin{tabular}[c]{@{}c@{}}SKA2\\ (CS)\end{tabular}} & \textbf{\begin{tabular}[c]{@{}c@{}}SKA1\\ (IM1)\end{tabular}}            & \textbf{\begin{tabular}[c]{@{}c@{}}SKA1\\ (IM2)\end{tabular}} & \textbf{\begin{tabular}[c]{@{}c@{}}SKA1\\ (IM1+\\ IM2)\end{tabular}}           & \textbf{\begin{tabular}[c]{@{}c@{}}Euclid\\ (CS+GC)+\\ SKA1 (IM2)\end{tabular}}       \\ \hline
\textbf{\begin{tabular}[c]{@{}c@{}}Resonance Feature\\ ${k_r}=5$\end{tabular}}      & \begin{tabular}[c]{@{}c@{}}$10^{5}\times\sigma \left(\omega_{\mathrm{b}}\right)$\\ $\sigma \left(\omega_{\mathrm{cdm}}\right)$\\ $\sigma \left( H_0 \right)$\\ $10^{11}\times\sigma \left( A_\mathrm{s} \right)$\\ $\sigma \left( n_\mathrm{s} \right)$\\ $\sigma \left( \tau_{\mathrm{reio}} \right)$\\ $\sigma \left({C_r}\right)$\\ $\sigma \left({k_r}\right)$\\ $\sigma \left( {\phi_r} \right)$\end{tabular} & \begin{tabular}[c]{@{}c@{}}3.80799
\\ 0.00024
\\ 0.09962
\\ 0.89633
\\ 0.00175
\\ 0.00234
\\ 0.00119
\\ 0.08103
\\ 0.11813
\end{tabular} & \begin{tabular}[c]{@{}c@{}}3.74756
\\ 0.00016
\\ 0.06625
\\ 0.71608
\\ 0.00170
\\ 0.00185
\\ 0.00118
\\ 0.07821
\\ 0.11416\end{tabular} & \begin{tabular}[c]{@{}c@{}}3.35984
\\ 0.00011
\\ 0.04651
\\ 0.61495
\\ 0.00146
\\ 0.00158
\\ 0.00089
\\ 0.03805
\\ 0.05940
\end{tabular} & \begin{tabular}[c]{@{}c@{}}3.55346
\\ 0.00014
\\ 0.05916
\\ 0.70968
\\ 0.00161
\\ 0.00180
\\ 0.00111
\\ 0.07388
\\ 0.10925
\end{tabular} & \begin{tabular}[c]{@{}c@{}}3.30718
\\ 0.00009
\\ 0.03948
\\ 0.58741
\\ 0.00142
\\ 0.00151
\\ 0.00084
\\ 0.03312
\\ 0.05185\end{tabular} & \begin{tabular}[c]{@{}c@{}}3.39728
\\ 0.00010
\\ 0.04299
\\ 0.59531
\\ 0.00149
\\ 0.00152
\\ 0.00080
\\ 0.04256
\\ 0.06504
\end{tabular} \\ \hline
\textbf{\begin{tabular}[c]{@{}c@{}}Resonance Feature\\ ${k_r}=30$\end{tabular}}          & \begin{tabular}[c]{@{}c@{}}$10^{5}\times\sigma \left(\omega_{\mathrm{b}}\right)$\\ $\sigma \left(\omega_{\mathrm{cdm}}\right)$\\ $\sigma \left( H_0 \right)$\\ $10^{11}\times\sigma \left( A_\mathrm{s} \right)$\\ $\sigma \left( n_\mathrm{s} \right)$\\ $\sigma \left( \tau_{\mathrm{reio}} \right)$\\ $\sigma \left({C_r}\right)$\\ $\sigma \left({k_r}\right)$\\ $\sigma \left( {\phi_r} \right)$\end{tabular} & \begin{tabular}[c]{@{}c@{}}3.46382
\\ 0.00024
\\ 0.10094
\\ 0.89348
\\ 0.00163
\\ 0.00232
\\ 0.00228
\\ 0.17535
\\ 0.26769
\end{tabular} & \begin{tabular}[c]{@{}c@{}}3.39077
\\ 0.00017
\\ 0.07068
\\ 0.72046
\\ 0.00159
\\ 0.00184
\\ 0.00227
\\ 0.14768
\\ 0.22506\end{tabular} & \begin{tabular}[c]{@{}c@{}}3.24465
\\ 0.00011
\\ 0.04210
\\ 0.63707
\\ 0.00147
\\ 0.00162
\\ 0.00117
\\ 0.04436
\\ 0.07146
\end{tabular} & \begin{tabular}[c]{@{}c@{}}3.41539
\\ 0.00015
\\ 0.05906
\\ 0.77648
\\ 0.00160
\\ 0.00196
\\ 0.00179
\\ 0.13698
\\ 0.21053
\end{tabular} & \begin{tabular}[c]{@{}c@{}}3.20295
\\ 0.00009
\\ 0.03571
\\ 0.60720
\\ 0.00143
\\ 0.00156
\\ 0.00108
\\ 0.03975
\\ 0.06323\end{tabular} & \begin{tabular}[c]{@{}c@{}}3.13722
\\ 0.00010
\\ 0.04020
\\ 0.59441
\\ 0.00138
\\ 0.00153
\\ 0.00102
\\ 0.05213
\\ 0.08108
\end{tabular} \\ \hline
\textbf{\begin{tabular}[c]{@{}c@{}}Resonance Feature\\ ${k_r}=100$\end{tabular}}      & \begin{tabular}[c]{@{}c@{}}$10^{5}\times\sigma \left(\omega_{\mathrm{b}}\right)$\\ $\sigma \left(\omega_{\mathrm{cdm}}\right)$\\ $\sigma \left( H_0 \right)$\\ $10^{11}\times\sigma \left( A_\mathrm{s} \right)$\\ $\sigma \left( n_\mathrm{s} \right)$\\ $\sigma \left( \tau_{\mathrm{reio}} \right)$\\ $\sigma \left({C_r}\right)$\\ $\sigma \left({k_r}\right)$\\ $\sigma \left( {\phi_r} \right)$\end{tabular} & \begin{tabular}[c]{@{}c@{}}3.54018
\\ 0.00024
\\ 0.09863
\\ 0.88304
\\ 0.00172
\\ 0.00228
\\ 0.00583
\\ 0.34834
\\ 0.56705
\end{tabular} & \begin{tabular}[c]{@{}c@{}}3.47983
\\ 0.00017
\\ 0.06782
\\ 0.70850
\\ 0.00169
\\ 0.00182
\\ 0.00576
\\ 0.23494
\\ 0.35089\end{tabular} & \begin{tabular}[c]{@{}c@{}}3.13697
\\ 0.00007
\\ 0.02536
\\ 0.45981
\\ 0.00144
\\ 0.00119
\\ 0.00144
\\ 0.05694
\\ 0.09403
\end{tabular} & \begin{tabular}[c]{@{}c@{}}3.23768
\\ 0.00009
\\ 0.03419
\\ 0.58297
\\ 0.00156
\\ 0.00147
\\ 0.00270
\\ 0.16329
\\ 0.25819
\end{tabular} & \begin{tabular}[c]{@{}c@{}}3.14395
\\ 0.00007
\\ 0.02359
\\ 0.42472
\\ 0.00141
\\ 0.00113
\\ 0.00130
\\ 0.05036
\\ 0.08321\end{tabular} & \begin{tabular}[c]{@{}c@{}}3.22349
\\ 0.00007
\\ 0.02449
\\ 0.47144
\\ 0.00143
\\ 0.00119
\\ 0.00123
\\ 0.06479
\\ 0.09791
\end{tabular} \\ \hline
\end{tabular}\par
\bigskip
\parbox{16.53cm}{\captionof{table}{{\fontfamily{bch}The possible 1-$\sigma$ sensitivity of \textbf{CMB-S4} for resonance feature signal in synergy with SKA1(CS), SKA2(CS), SKA1(IM1), SKA1(IM2), SKA1(IM1+IM2), and Euclid(CS+GC)+SKA1(IM2).}}} \label{table:CMB-S4-Resonance}
\end{minipage}

\setlength{\tabcolsep}{3.25pt} % Default value: 6pt
\renewcommand{\arraystretch}{1.5} % Default value: 1
\newcolumntype{C}[1]{>{\Centering}m{#1}}
\renewcommand\tabularxcolumn[1]{C{#1}}
\begin{minipage}{\linewidth}
\centering
\captionsetup{font=footnotesize}
\begin{tabular}{|c|c|c|c|c|c|c|c|}
\hline
\textbf{Models}            & \textbf{Parameters}                                                       & \textbf{\begin{tabular}[c]{@{}c@{}}SKA1\\ (CS)\end{tabular}} & \textbf{\begin{tabular}[c]{@{}c@{}}SKA2\\ (CS)\end{tabular}} & \textbf{\begin{tabular}[c]{@{}c@{}}SKA1\\ (IM1)\end{tabular}}            & \textbf{\begin{tabular}[c]{@{}c@{}}SKA1\\ (IM2)\end{tabular}} & \textbf{\begin{tabular}[c]{@{}c@{}}SKA1\\ (IM1+\\ IM2)\end{tabular}}           & \textbf{\begin{tabular}[c]{@{}c@{}}Euclid\\ (CS+GC)+\\ SKA1 (IM2)\end{tabular}}       \\ \hline
\textbf{\begin{tabular}[c]{@{}c@{}}Resonance Feature\\ ${k_r}=5$\end{tabular}}      & \begin{tabular}[c]{@{}c@{}}$10^{5}\times\sigma \left(\omega_{\mathrm{b}}\right)$\\ $\sigma \left(\omega_{\mathrm{cdm}}\right)$\\ $\sigma \left( H_0 \right)$\\ $10^{11}\times\sigma \left( A_\mathrm{s} \right)$\\ $\sigma \left( n_\mathrm{s} \right)$\\ $\sigma \left( \tau_{\mathrm{reio}} \right)$\\ $\sigma \left({C_r}\right)$\\ $\sigma \left({k_r}\right)$\\ $\sigma \left( {\phi_r} \right)$\end{tabular} & \begin{tabular}[c]{@{}c@{}}4.06178
\\ 0.00017
\\ 0.07318
\\ 0.54325
\\ 0.00145
\\ 0.00132
\\ 0.00096
\\ 0.07059
\\ 0.11234
\end{tabular} & \begin{tabular}[c]{@{}c@{}}3.94988
\\ 0.00012
\\ 0.05393
\\ 0.53006
\\ 0.00141
\\ 0.00129
\\ 0.00096
\\ 0.06895
\\ 0.10987\end{tabular} & \begin{tabular}[c]{@{}c@{}}3.76041
\\ 0.00009
\\ 0.04311
\\ 0.49019
\\ 0.00129
\\ 0.00120
\\ 0.00078
\\ 0.03624
\\ 0.05920
\end{tabular} & \begin{tabular}[c]{@{}c@{}}3.90061
\\ 0.00012
\\ 0.05513
\\ 0.53671
\\ 0.00139
\\ 0.00130
\\ 0.00091
\\ 0.06601
\\ 0.10554
\end{tabular} & \begin{tabular}[c]{@{}c@{}}3.70450
\\ 0.00008
\\ 0.03701
\\ 0.47663
\\ 0.00126
\\ 0.00117
\\ 0.00074
\\ 0.03169
\\ 0.05171\end{tabular} & \begin{tabular}[c]{@{}c@{}}3.66881
\\ 0.00009
\\ 0.04050
\\ 0.48372
\\ 0.00128
\\ 0.00119
\\ 0.00071
\\ 0.04058
\\ 0.06482
\end{tabular} \\ \hline
\textbf{\begin{tabular}[c]{@{}c@{}}Resonance Feature\\ ${k_r}=30$\end{tabular}}          & \begin{tabular}[c]{@{}c@{}}$10^{5}\times\sigma \left(\omega_{\mathrm{b}}\right)$\\ $\sigma \left(\omega_{\mathrm{cdm}}\right)$\\ $\sigma \left( H_0 \right)$\\ $10^{11}\times\sigma \left( A_\mathrm{s} \right)$\\ $\sigma \left( n_\mathrm{s} \right)$\\ $\sigma \left( \tau_{\mathrm{reio}} \right)$\\ $\sigma \left({C_r}\right)$\\ $\sigma \left({k_r}\right)$\\ $\sigma \left( {\phi_r} \right)$\end{tabular} & \begin{tabular}[c]{@{}c@{}}4.02926
\\ 0.00017
\\ 0.07536
\\ 0.54759
\\ 0.00142
\\ 0.00132
\\ 0.00211
\\ 0.15830
\\ 0.25738
\end{tabular} & \begin{tabular}[c]{@{}c@{}}3.91496
\\ 0.00014
\\ 0.06203
\\ 0.53901
\\ 0.00137
\\ 0.00130
\\ 0.00210
\\ 0.13881
\\ 0.22552\end{tabular} & \begin{tabular}[c]{@{}c@{}}3.65666
\\ 0.00009
\\ 0.03949
\\ 0.48523
\\ 0.00129
\\ 0.00117
\\ 0.00114
\\ 0.04365
\\ 0.07138
\end{tabular} & \begin{tabular}[c]{@{}c@{}}3.89467
\\ 0.00012
\\ 0.05344
\\ 0.53933
\\ 0.00138
\\ 0.00128
\\ 0.00171
\\ 0.12803
\\ 0.20562
\end{tabular} & \begin{tabular}[c]{@{}c@{}}3.61260
\\ 0.00008
\\ 0.03386
\\ 0.47178
\\ 0.00126
\\ 0.00114
\\ 0.00106
\\ 0.03910
\\ 0.06308\end{tabular} & \begin{tabular}[c]{@{}c@{}}3.58617
\\ 0.00009
\\ 0.03794
\\ 0.47539
\\ 0.00123
\\ 0.00117
\\ 0.00101
\\ 0.05129
\\ 0.08078
\end{tabular} \\ \hline
\textbf{\begin{tabular}[c]{@{}c@{}}Resonance Feature\\ ${k_r}=100$\end{tabular}}      & \begin{tabular}[c]{@{}c@{}}$10^{5}\times\sigma \left(\omega_{\mathrm{b}}\right)$\\ $\sigma \left(\omega_{\mathrm{cdm}}\right)$\\ $\sigma \left( H_0 \right)$\\ $10^{11}\times\sigma \left( A_\mathrm{s} \right)$\\ $\sigma \left( n_\mathrm{s} \right)$\\ $\sigma \left( \tau_{\mathrm{reio}} \right)$\\ $\sigma \left({C_r}\right)$\\ $\sigma \left({k_r}\right)$\\ $\sigma \left( {\phi_r} \right)$\end{tabular} & \begin{tabular}[c]{@{}c@{}}3.89769
\\ 0.00017
\\ 0.07353
\\ 0.54193
\\ 0.00143
\\ 0.00131
\\ 0.00502
\\ 0.32976
\\ 0.58959
\end{tabular} & \begin{tabular}[c]{@{}c@{}}3.76804
\\ 0.00013
\\ 0.05554
\\ 0.52731
\\ 0.00139
\\ 0.00128
\\ 0.00497
\\ 0.21674
\\ 0.36293\end{tabular} & \begin{tabular}[c]{@{}c@{}}3.61245
\\ 0.00007
\\ 0.02692
\\ 0.40973
\\ 0.00127
\\ 0.00100
\\ 0.00142
\\ 0.05689
\\ 0.09580
\end{tabular} & \begin{tabular}[c]{@{}c@{}}3.72614
\\ 0.00009
\\ 0.03586
\\ 0.49997
\\ 0.00135
\\ 0.00119
\\ 0.00261
\\ 0.16102
\\ 0.26579
\end{tabular} & \begin{tabular}[c]{@{}c@{}}3.72782
\\ 0.00006
\\ 0.02593
\\ 0.38545
\\ 0.00127
\\ 0.00096
\\ 0.00129
\\ 0.05031
\\ 0.08557\end{tabular} & \begin{tabular}[c]{@{}c@{}}3.61609
\\ 0.00007
\\ 0.02592
\\ 0.42199
\\ 0.00124
\\ 0.00102
\\ 0.00122
\\ 0.06437
\\ 0.09898
\end{tabular} \\ \hline
\end{tabular}\par
\bigskip
\parbox{16.52cm}{\captionof{table}{{\fontfamily{bch}The possible 1-$\sigma$ sensitivity of \textbf{CORE-M5} for resonance feature signal in synergy with SKA1(CS), SKA2(CS), SKA1(IM1), SKA1(IM2), SKA1(IM1+IM2), and Euclid(CS+GC)+SKA1(IM2).}}} \label{table:CORE-M5-Resonance}
\end{minipage}

\setlength{\tabcolsep}{3.25pt} % Default value: 6pt
\renewcommand{\arraystretch}{1.5} % Default value: 1
\newcolumntype{C}[1]{>{\Centering}m{#1}}
\renewcommand\tabularxcolumn[1]{C{#1}}
\begin{minipage}{\linewidth}
\centering
\captionsetup{font=footnotesize}
\begin{tabular}{|c|c|c|c|c|c|c|c|}
\hline
\textbf{Models}            & \textbf{Parameters}                                                       & \textbf{\begin{tabular}[c]{@{}c@{}}SKA1\\ (CS)\end{tabular}} & \textbf{\begin{tabular}[c]{@{}c@{}}SKA2\\ (CS)\end{tabular}} & \textbf{\begin{tabular}[c]{@{}c@{}}SKA1\\ (IM1)\end{tabular}}            & \textbf{\begin{tabular}[c]{@{}c@{}}SKA1\\ (IM2)\end{tabular}} & \textbf{\begin{tabular}[c]{@{}c@{}}SKA1\\ (IM1+\\ IM2)\end{tabular}}           & \textbf{\begin{tabular}[c]{@{}c@{}}Euclid\\ (CS+GC)+\\ SKA1 (IM2)\end{tabular}}       \\ \hline
\textbf{\begin{tabular}[c]{@{}c@{}}Resonance Feature\\ ${k_r}=5$\end{tabular}}      & \begin{tabular}[c]{@{}c@{}}$10^{5}\times\sigma \left(\omega_{\mathrm{b}}\right)$\\ $\sigma \left(\omega_{\mathrm{cdm}}\right)$\\ $\sigma \left( H_0 \right)$\\ $10^{11}\times\sigma \left( A_\mathrm{s} \right)$\\ $\sigma \left( n_\mathrm{s} \right)$\\ $\sigma \left( \tau_{\mathrm{reio}} \right)$\\ $\sigma \left({C_r}\right)$\\ $\sigma \left({k_r}\right)$\\ $\sigma \left( {\phi_r} \right)$\end{tabular} & \begin{tabular}[c]{@{}c@{}}3.15644
\\ 0.00015
\\ 0.05857
\\ 0.48764
\\ 0.00136
\\ 0.00119
\\ 0.00091
\\ 0.06228
\\ 0.09196
\end{tabular} & \begin{tabular}[c]{@{}c@{}}2.96168
\\ 0.00011
\\ 0.04749
\\ 0.47480
\\ 0.00126
\\ 0.00118
\\ 0.00091
\\ 0.06091
\\ 0.09006\end{tabular} & \begin{tabular}[c]{@{}c@{}}2.85708
\\ 0.00009
\\ 0.03814
\\ 0.44193
\\ 0.00117
\\ 0.00109
\\ 0.00075
\\ 0.03507
\\ 0.05424
\end{tabular} & \begin{tabular}[c]{@{}c@{}}2.95704
\\ 0.00011
\\ 0.04747
\\ 0.47496
\\ 0.00125
\\ 0.00116
\\ 0.00087
\\ 0.05875
\\ 0.08744
\end{tabular} & \begin{tabular}[c]{@{}c@{}}2.81261
\\ 0.00007
\\ 0.03208
\\ 0.42498
\\ 0.00115
\\ 0.00105
\\ 0.00072
\\ 0.03108
\\ 0.04826\end{tabular} & \begin{tabular}[c]{@{}c@{}}2.92777
\\ 0.00009
\\ 0.03598
\\ 0.43262
\\ 0.00122
\\ 0.00106
\\ 0.00069
\\ 0.03884
\\ 0.05904
\end{tabular} \\ \hline
\textbf{\begin{tabular}[c]{@{}c@{}}Resonance Feature\\ ${k_r}=30$\end{tabular}}          & \begin{tabular}[c]{@{}c@{}}$10^{5}\times\sigma \left(\omega_{\mathrm{b}}\right)$\\ $\sigma \left(\omega_{\mathrm{cdm}}\right)$\\ $\sigma \left( H_0 \right)$\\ $10^{11}\times\sigma \left( A_\mathrm{s} \right)$\\ $\sigma \left( n_\mathrm{s} \right)$\\ $\sigma \left( \tau_{\mathrm{reio}} \right)$\\ $\sigma \left({C_r}\right)$\\ $\sigma \left({k_r}\right)$\\ $\sigma \left( {\phi_r} \right)$\end{tabular} & \begin{tabular}[c]{@{}c@{}}2.90430
\\ 0.00015
\\ 0.05952
\\ 0.48725
\\ 0.00127
\\ 0.00119
\\ 0.00179
\\ 0.14091
\\ 0.21882
\end{tabular} & \begin{tabular}[c]{@{}c@{}}2.86878
\\ 0.00013
\\ 0.05264
\\ 0.47992
\\ 0.00125
\\ 0.00118
\\ 0.00179
\\ 0.12091
\\ 0.18795\end{tabular} & \begin{tabular}[c]{@{}c@{}}2.74777
\\ 0.00009
\\ 0.03523
\\ 0.43437
\\ 0.00117
\\ 0.00105
\\ 0.00108
\\ 0.04295
\\ 0.06889
\end{tabular} & \begin{tabular}[c]{@{}c@{}}2.85803
\\ 0.00011
\\ 0.04608
\\ 0.47322
\\ 0.00124
\\ 0.00114
\\ 0.00153
\\ 0.11950
\\ 0.18467
\end{tabular} & \begin{tabular}[c]{@{}c@{}}2.71068
\\ 0.00007
\\ 0.02942
\\ 0.41908
\\ 0.00115
\\ 0.00102
\\ 0.00101
\\ 0.03881
\\ 0.06141\end{tabular} & \begin{tabular}[c]{@{}c@{}}2.67901
\\ 0.00008
\\ 0.03358
\\ 0.42124
\\ 0.00112
\\ 0.00103
\\ 0.00097
\\ 0.05057
\\ 0.07877\end{tabular} \\ \hline
\textbf{\begin{tabular}[c]{@{}c@{}}Resonance Feature\\ ${k_r}=100$\end{tabular}}      & \begin{tabular}[c]{@{}c@{}}$10^{5}\times\sigma \left(\omega_{\mathrm{b}}\right)$\\ $\sigma \left(\omega_{\mathrm{cdm}}\right)$\\ $\sigma \left( H_0 \right)$\\ $10^{11}\times\sigma \left( A_\mathrm{s} \right)$\\ $\sigma \left( n_\mathrm{s} \right)$\\ $\sigma \left( \tau_{\mathrm{reio}} \right)$\\ $\sigma \left({C_r}\right)$\\ $\sigma \left({k_r}\right)$\\ $\sigma \left( {\phi_r} \right)$\end{tabular} & \begin{tabular}[c]{@{}c@{}}2.97572
\\ 0.00015
\\ 0.05841
\\ 0.48557
\\ 0.00135
\\ 0.00118
\\ 0.00455
\\ 0.28335
\\ 0.47615
\end{tabular} & \begin{tabular}[c]{@{}c@{}}2.75848
\\ 0.00012
\\ 0.04872
\\ 0.47245
\\ 0.00124
\\ 0.00117
\\ 0.00451
\\ 0.20343
\\ 0.32107\end{tabular} & \begin{tabular}[c]{@{}c@{}}2.67926
\\ 0.00006
\\ 0.02312
\\ 0.36722
\\ 0.00116
\\ 0.00090
\\ 0.00141
\\ 0.05622
\\ 0.09215
\end{tabular} & \begin{tabular}[c]{@{}c@{}}2.7365
\\ 0.00008
\\ 0.03073
\\ 0.42414
\\ 0.00122
\\ 0.00102
\\ 0.00253
\\ 0.15321
\\ 0.24453
\end{tabular} & \begin{tabular}[c]{@{}c@{}}2.68875
\\ 0.00006
\\ 0.02131
\\ 0.34769
\\ 0.00114
\\ 0.00087
\\ 0.00128
\\ 0.04958
\\ 0.08104\end{tabular} & \begin{tabular}[c]{@{}c@{}}2.82027
\\ 0.00006
\\ 0.02270
\\ 0.37424
\\ 0.00119
\\ 0.00090
\\ 0.00121
\\ 0.06330
\\ 0.09573
\end{tabular} \\ \hline
\end{tabular}\par
\bigskip
\parbox{16.53cm}{\captionof{table}{{\fontfamily{bch}The possible 1-$\sigma$ sensitivity of \textbf{PICO} for resonance feature signal in synergy with SKA1(CS), SKA2(CS), SKA1(IM1), SKA1(IM2), SKA1(IM1+IM2), and Euclid(CS+GC)+SKA1(IM2).}}} \label{table:PICO-Resonance}
\end{minipage}

\setlength{\tabcolsep}{3.25pt} % Default value: 6pt
\renewcommand{\arraystretch}{1.5} % Default value: 1
\newcolumntype{C}[1]{>{\Centering}m{#1}}
\renewcommand\tabularxcolumn[1]{C{#1}}
\begin{minipage}{\linewidth}
\centering
\captionsetup{font=footnotesize}
\begin{tabular}{|c|c|c|c|c|c|c|c|}
\hline
\textbf{Models}            & \textbf{Parameters}                                                       & \textbf{\begin{tabular}[c]{@{}c@{}}SKA1\\ (CS)\end{tabular}} & \textbf{\begin{tabular}[c]{@{}c@{}}SKA2\\ (CS)\end{tabular}} & \textbf{\begin{tabular}[c]{@{}c@{}}SKA1\\ (IM1)\end{tabular}}            & \textbf{\begin{tabular}[c]{@{}c@{}}SKA1\\ (IM2)\end{tabular}} & \textbf{\begin{tabular}[c]{@{}c@{}}SKA1\\ (IM1+\\ IM2)\end{tabular}}           & \textbf{\begin{tabular}[c]{@{}c@{}}Euclid\\ (CS+GC)+\\ SKA1 (IM2)\end{tabular}}       \\ \hline
\textbf{\begin{tabular}[c]{@{}c@{}}Resonance Feature\\ ${k_r}=5$\end{tabular}}      & \begin{tabular}[c]{@{}c@{}}$10^{5}\times\sigma \left(\omega_{\mathrm{b}}\right)$\\ $\sigma \left(\omega_{\mathrm{cdm}}\right)$\\ $\sigma \left( H_0 \right)$\\ $10^{11}\times\sigma \left( A_\mathrm{s} \right)$\\ $\sigma \left( n_\mathrm{s} \right)$\\ $\sigma \left( \tau_{\mathrm{reio}} \right)$\\ $\sigma \left({C_r}\right)$\\ $\sigma \left({k_r}\right)$\\ $\sigma \left( {\phi_r} \right)$\end{tabular} & \begin{tabular}[c]{@{}c@{}}3.68409
\\ 0.00022
\\ 0.09159
\\ 0.78965
\\ 0.00163
\\ 0.00204
\\ 0.00114
\\ 0.07590
\\ 0.11280
\end{tabular} & \begin{tabular}[c]{@{}c@{}}3.64553
\\ 0.00015
\\ 0.06201
\\ 0.65650
\\ 0.00159
\\ 0.00167
\\ 0.00113
\\ 0.07359
\\ 0.10951\end{tabular} & \begin{tabular}[c]{@{}c@{}}3.31910
\\ 0.00011
\\ 0.04639
\\ 0.61237
\\ 0.00139
\\ 0.00157
\\ 0.00087
\\ 0.03748
\\ 0.05898
\end{tabular} & \begin{tabular}[c]{@{}c@{}}3.49710
\\ 0.00014
\\ 0.05908
\\ 0.70855
\\ 0.00152
\\ 0.00179
\\ 0.00107
\\ 0.06993
\\ 0.10512
\end{tabular} & \begin{tabular}[c]{@{}c@{}}3.26886
\\ 0.00009
\\ 0.03933
\\ 0.58497
\\ 0.00135
\\ 0.00150
\\ 0.00082
\\ 0.03270
\\ 0.05156\end{tabular} & \begin{tabular}[c]{@{}c@{}}3.25700
\\ 0.00010
\\ 0.04246
\\ 0.57807
\\ 0.00138
\\ 0.00147
\\ 0.00078
\\ 0.04171
\\ 0.06425
\end{tabular} \\ \hline
\textbf{\begin{tabular}[c]{@{}c@{}}Resonance Feature\\ ${k_r}=30$\end{tabular}}          & \begin{tabular}[c]{@{}c@{}}$10^{5}\times\sigma \left(\omega_{\mathrm{b}}\right)$\\ $\sigma \left(\omega_{\mathrm{cdm}}\right)$\\ $\sigma \left( H_0 \right)$\\ $10^{11}\times\sigma \left( A_\mathrm{s} \right)$\\ $\sigma \left( n_\mathrm{s} \right)$\\ $\sigma \left( \tau_{\mathrm{reio}} \right)$\\ $\sigma \left({C_r}\right)$\\ $\sigma \left({k_r}\right)$\\ $\sigma \left( {\phi_r} \right)$\end{tabular} & \begin{tabular}[c]{@{}c@{}}3.33974
\\ 0.00022
\\ 0.09284
\\ 0.78550
\\ 0.00153
\\ 0.00202
\\ 0.00222
\\ 0.16953
\\ 0.26178
\end{tabular} & \begin{tabular}[c]{@{}c@{}}3.28643
\\ 0.00016
\\ 0.06684
\\ 0.66022
\\ 0.00150
\\ 0.00167
\\ 0.00222
\\ 0.14359
\\ 0.22161\end{tabular} & \begin{tabular}[c]{@{}c@{}}3.16120
\\ 0.00010
\\ 0.04097
\\ 0.59829
\\ 0.00139
\\ 0.00150
\\ 0.00116
\\ 0.04419
\\ 0.07133
\end{tabular} & \begin{tabular}[c]{@{}c@{}}3.31127
\\ 0.00014
\\ 0.05641
\\ 0.70790
\\ 0.00149
\\ 0.00176
\\ 0.00177
\\ 0.13416
\\ 0.20761
\end{tabular} & \begin{tabular}[c]{@{}c@{}}3.12406
\\ 0.00009
\\ 0.03485
\\ 0.57358
\\ 0.00135
\\ 0.00145
\\ 0.00107
\\ 0.03962
\\ 0.06310\end{tabular} & \begin{tabular}[c]{@{}c@{}}3.06367
\\ 0.00009
\\ 0.03919
\\ 0.56233
\\ 0.00131
\\ 0.00143
\\ 0.00102
\\ 0.05185
\\ 0.08083
\end{tabular} \\ \hline
\textbf{\begin{tabular}[c]{@{}c@{}}Resonance Feature\\ ${k_r}=100$\end{tabular}}      & \begin{tabular}[c]{@{}c@{}}$10^{5}\times\sigma \left(\omega_{\mathrm{b}}\right)$\\ $\sigma \left(\omega_{\mathrm{cdm}}\right)$\\ $\sigma \left( H_0 \right)$\\ $10^{11}\times\sigma \left( A_\mathrm{s} \right)$\\ $\sigma \left( n_\mathrm{s} \right)$\\ $\sigma \left( \tau_{\mathrm{reio}} \right)$\\ $\sigma \left({C_r}\right)$\\ $\sigma \left({k_r}\right)$\\ $\sigma \left( {\phi_r} \right)$\end{tabular} & \begin{tabular}[c]{@{}c@{}}3.40735
\\ 0.00022
\\ 0.09076
\\ 0.77844
\\ 0.00161
\\ 0.00199
\\ 0.00568
\\ 0.33097
\\ 0.55054
\end{tabular} & \begin{tabular}[c]{@{}c@{}}3.36601
\\ 0.00015
\\ 0.06342
\\ 0.64907
\\ 0.00158
\\ 0.00165
\\ 0.00561
\\ 0.22696
\\ 0.34726\end{tabular} & \begin{tabular}[c]{@{}c@{}}3.12658
\\ 0.00007
\\ 0.02543
\\ 0.45591
\\ 0.00137
\\ 0.00117
\\ 0.00144
\\ 0.05690
\\ 0.09411
\end{tabular} & \begin{tabular}[c]{@{}c@{}}3.22407
\\ 0.00009
\\ 0.03460
\\ 0.58941
\\ 0.00148
\\ 0.00145
\\ 0.00269
\\ 0.16148
\\ 0.25719
\end{tabular} & \begin{tabular}[c]{@{}c@{}}3.09377
\\ 0.00007
\\ 0.02331
\\ 0.42192
\\ 0.00134
\\ 0.00111
\\ 0.00130
\\ 0.05029
\\ 0.08285\end{tabular} & \begin{tabular}[c]{@{}c@{}}3.13620
\\ 0.00007
\\ 0.02447
\\ 0.46182
\\ 0.00134
\\ 0.00115
\\ 0.00123
\\ 0.06459
\\ 0.09784
\end{tabular} \\ \hline
\end{tabular}\par
\bigskip
\parbox{16.52cm}{\captionof{table}{{\fontfamily{bch}The possible 1-$\sigma$ sensitivity of \textbf{Planck+CMB-S4} for resonance feature signal in synergy with SKA1(CS), SKA2(CS), SKA1(IM1), SKA1(IM2), SKA1(IM1+IM2), and Euclid(CS+GC)+SKA1(IM2).}}} \label{table:Planck+CMB-S4-Resonance}
\end{minipage}

\setlength{\tabcolsep}{3.25pt} % Default value: 6pt
\renewcommand{\arraystretch}{1.5} % Default value: 1
\newcolumntype{C}[1]{>{\Centering}m{#1}}
\renewcommand\tabularxcolumn[1]{C{#1}}
\begin{minipage}{\linewidth}
\centering
\captionsetup{font=footnotesize}
\begin{tabular}{|c|c|c|c|c|c|c|c|}
\hline
\textbf{Models}            & \textbf{Parameters}                                                       & \textbf{\begin{tabular}[c]{@{}c@{}}SKA1\\ (CS)\end{tabular}} & \textbf{\begin{tabular}[c]{@{}c@{}}SKA2\\ (CS)\end{tabular}} & \textbf{\begin{tabular}[c]{@{}c@{}}SKA1\\ (IM1)\end{tabular}}            & \textbf{\begin{tabular}[c]{@{}c@{}}SKA1\\ (IM2)\end{tabular}} & \textbf{\begin{tabular}[c]{@{}c@{}}SKA1\\ (IM1+\\ IM2)\end{tabular}}           & \textbf{\begin{tabular}[c]{@{}c@{}}Euclid\\ (CS+GC)+\\ SKA1 (IM2)\end{tabular}}       \\ \hline
\textbf{\begin{tabular}[c]{@{}c@{}}Resonance Feature\\ ${k_r}=5$\end{tabular}}      & \begin{tabular}[c]{@{}c@{}}$10^{5}\times\sigma \left(\omega_{\mathrm{b}}\right)$\\ $\sigma \left(\omega_{\mathrm{cdm}}\right)$\\ $\sigma \left( H_0 \right)$\\ $10^{11}\times\sigma \left( A_\mathrm{s} \right)$\\ $\sigma \left( n_\mathrm{s} \right)$\\ $\sigma \left( \tau_{\mathrm{reio}} \right)$\\ $\sigma \left({C_r}\right)$\\ $\sigma \left({k_r}\right)$\\ $\sigma \left( {\phi_r} \right)$\end{tabular} & \begin{tabular}[c]{@{}c@{}}3.52351
\\ 0.00019
\\ 0.07973
\\ 0.59983
\\ 0.00150
\\ 0.00152
\\ 0.00116
\\ 0.07034
\\ 0.10633
\end{tabular} & \begin{tabular}[c]{@{}c@{}}3.52600
\\ 0.00013
\\ 0.05418
\\ 0.53763
\\ 0.00147
\\ 0.00134
\\ 0.00115
\\ 0.06859
\\ 0.10385\end{tabular} & \begin{tabular}[c]{@{}c@{}}3.21419
\\ 0.00010
\\ 0.04210
\\ 0.49504
\\ 0.00129
\\ 0.00123
\\ 0.00087
\\ 0.03646
\\ 0.05800
\end{tabular} & \begin{tabular}[c]{@{}c@{}}3.36541
\\ 0.00013
\\ 0.05323
\\ 0.53944
\\ 0.00140
\\ 0.00133
\\ 0.00108
\\ 0.06553
\\ 0.10014
\end{tabular} & \begin{tabular}[c]{@{}c@{}}3.17217
\\ 0.00008
\\ 0.03606
\\ 0.48082
\\ 0.00126
\\ 0.00119
\\ 0.00083
\\ 0.03198
\\ 0.05097\end{tabular} & \begin{tabular}[c]{@{}c@{}}3.24568
\\ 0.00009
\\ 0.03952
\\ 0.48535
\\ 0.00133
\\ 0.00120
\\ 0.00079
\\ 0.04071
\\ 0.06332
\end{tabular} \\ \hline
\textbf{\begin{tabular}[c]{@{}c@{}}Resonance Feature\\ ${k_r}=30$\end{tabular}}          & \begin{tabular}[c]{@{}c@{}}$10^{5}\times\sigma \left(\omega_{\mathrm{b}}\right)$\\ $\sigma \left(\omega_{\mathrm{cdm}}\right)$\\ $\sigma \left( H_0 \right)$\\ $10^{11}\times\sigma \left( A_\mathrm{s} \right)$\\ $\sigma \left( n_\mathrm{s} \right)$\\ $\sigma \left( \tau_{\mathrm{reio}} \right)$\\ $\sigma \left({C_r}\right)$\\ $\sigma \left({k_r}\right)$\\ $\sigma \left( {\phi_r} \right)$\end{tabular} & \begin{tabular}[c]{@{}c@{}}3.23587
\\ 0.00019
\\ 0.08132
\\ 0.59760
\\ 0.00142
\\ 0.00151
\\ 0.00222
\\ 0.16235
\\ 0.25254
\end{tabular} & \begin{tabular}[c]{@{}c@{}}3.21655
\\ 0.00015
\\ 0.06059
\\ 0.53971
\\ 0.00140
\\ 0.00134
\\ 0.00222
\\ 0.13848
\\ 0.21534\end{tabular} & \begin{tabular}[c]{@{}c@{}}3.07530
\\ 0.00009
\\ 0.03820
\\ 0.48760
\\ 0.00129
\\ 0.00119
\\ 0.00116
\\ 0.04397
\\ 0.07096
\end{tabular} & \begin{tabular}[c]{@{}c@{}}3.20257
\\ 0.00012
\\ 0.05095
\\ 0.53318
\\ 0.00138
\\ 0.00130
\\ 0.00176
\\ 0.13074
\\ 0.20303
\end{tabular} & \begin{tabular}[c]{@{}c@{}}3.04428
\\ 0.00008
\\ 0.03263
\\ 0.47393
\\ 0.00126
\\ 0.00116
\\ 0.00107
\\ 0.03948
\\ 0.06280\end{tabular} & \begin{tabular}[c]{@{}c@{}}2.99212
\\ 0.00009
\\ 0.03675
\\ 0.47350
\\ 0.00123
\\ 0.00117
\\ 0.00102
\\ 0.05155
\\ 0.08052
\end{tabular} \\ \hline
\textbf{\begin{tabular}[c]{@{}c@{}}Resonance Feature\\ ${k_r}=100$\end{tabular}}      & \begin{tabular}[c]{@{}c@{}}$10^{5}\times\sigma \left(\omega_{\mathrm{b}}\right)$\\ $\sigma \left(\omega_{\mathrm{cdm}}\right)$\\ $\sigma \left( H_0 \right)$\\ $10^{11}\times\sigma \left( A_\mathrm{s} \right)$\\ $\sigma \left( n_\mathrm{s} \right)$\\ $\sigma \left( \tau_{\mathrm{reio}} \right)$\\ $\sigma \left({C_r}\right)$\\ $\sigma \left({k_r}\right)$\\ $\sigma \left( {\phi_r} \right)$\end{tabular} & \begin{tabular}[c]{@{}c@{}}3.28362
\\ 0.00019
\\ 0.07940
\\ 0.59578
\\ 0.00148
\\ 0.00150
\\ 0.00566
\\ 0.31397
\\ 0.53226
\end{tabular} & \begin{tabular}[c]{@{}c@{}}3.27243
\\ 0.00013
\\ 0.05580
\\ 0.53362
\\ 0.00147
\\ 0.00133
\\ 0.00559
\\ 0.22006
\\ 0.34293\end{tabular} & \begin{tabular}[c]{@{}c@{}}3.01355
\\ 0.00007
\\ 0.02492
\\ 0.40518
\\ 0.00128
\\ 0.00101
\\ 0.00144
\\ 0.05672
\\ 0.09377
\end{tabular} & \begin{tabular}[c]{@{}c@{}}3.08594
\\ 0.00009
\\ 0.03344
\\ 0.48168
\\ 0.00136
\\ 0.00117
\\ 0.00269
\\ 0.15878
\\ 0.25441
\end{tabular} & \begin{tabular}[c]{@{}c@{}}3.02311
\\ 0.00007
\\ 0.02318
\\ 0.38011
\\ 0.00126
\\ 0.00097
\\ 0.00130
\\ 0.05011
\\ 0.08280\end{tabular} & \begin{tabular}[c]{@{}c@{}}3.08571
\\ 0.00007
\\ 0.02409
\\ 0.41400
\\ 0.00128
\\ 0.00101
\\ 0.00122
\\ 0.06419
\\ 0.09734
\end{tabular} \\ \hline
\end{tabular}\par
\bigskip
\parbox{16.52cm}{\captionof{table}{{\fontfamily{bch}The possible 1-$\sigma$ sensitivity of \textbf{LiteBIRD+CMB-S4} for resonance feature signal in synergy with SKA1(CS), SKA2(CS), SKA1(IM1), SKA1(IM2), SKA1(IM1+IM2), and Euclid(CS+GC)+SKA1(IM2).}}} \label{table:LiteBIRD+CMB-S4-Resonance}
\end{minipage}

\setlength{\tabcolsep}{3.25pt} % Default value: 6pt
\renewcommand{\arraystretch}{1.5} % Default value: 1
\newcolumntype{C}[1]{>{\Centering}m{#1}}
\renewcommand\tabularxcolumn[1]{C{#1}}
\begin{minipage}{\linewidth}
\centering
\captionsetup{font=footnotesize}
\begin{tabular}{|c|c|c|c|c|c|c|c|}
\hline
\textbf{Models}            & \textbf{Parameters}                                                       & \textbf{\begin{tabular}[c]{@{}c@{}}SKA1\\ (CS)\end{tabular}} & \textbf{\begin{tabular}[c]{@{}c@{}}SKA2\\ (CS)\end{tabular}} & \textbf{\begin{tabular}[c]{@{}c@{}}SKA1\\ (IM1)\end{tabular}}            & \textbf{\begin{tabular}[c]{@{}c@{}}SKA1\\ (IM2)\end{tabular}} & \textbf{\begin{tabular}[c]{@{}c@{}}SKA1\\ (IM1+\\ IM2)\end{tabular}}           & \textbf{\begin{tabular}[c]{@{}c@{}}Euclid\\ (CS+GC)+\\ SKA1 (IM2)\end{tabular}}       \\ \hline
\textbf{\begin{tabular}[c]{@{}c@{}}Resonance Feature\\ ${k_r}=5$\end{tabular}}      & \begin{tabular}[c]{@{}c@{}}$10^{5}\times\sigma \left(\omega_{\mathrm{b}}\right)$\\ $\sigma \left(\omega_{\mathrm{cdm}}\right)$\\ $\sigma \left( H_0 \right)$\\ $10^{11}\times\sigma \left( A_\mathrm{s} \right)$\\ $\sigma \left( n_\mathrm{s} \right)$\\ $\sigma \left( \tau_{\mathrm{reio}} \right)$\\ $\sigma \left({C_r}\right)$\\ $\sigma \left({k_r}\right)$\\ $\sigma \left( {\phi_r} \right)$\end{tabular} & \begin{tabular}[c]{@{}c@{}}3.16131
\\ 0.00015
\\ 0.06039
\\ 0.49997
\\ 0.00137
\\ 0.00122
\\ 0.00093
\\ 0.06394
\\ 0.09592
\end{tabular} & \begin{tabular}[c]{@{}c@{}}3.12272
\\ 0.00011
\\ 0.04816
\\ 0.48683
\\ 0.00134
\\ 0.00120
\\ 0.00092
\\ 0.06251
\\ 0.09394\end{tabular} & \begin{tabular}[c]{@{}c@{}}2.87038
\\ 0.00009
\\ 0.03869
\\ 0.45257
\\ 0.00119
\\ 0.00112
\\ 0.00076
\\ 0.03534
\\ 0.05532
\end{tabular} & \begin{tabular}[c]{@{}c@{}}2.97878
\\ 0.00012
\\ 0.04820
\\ 0.48830
\\ 0.00127
\\ 0.00119
\\ 0.00088
\\ 0.06023
\\ 0.09101
\end{tabular} & \begin{tabular}[c]{@{}c@{}}2.83766
\\ 0.00007
\\ 0.03308
\\ 0.44151
\\ 0.00117
\\ 0.00109
\\ 0.00073
\\ 0.03123
\\ 0.04900\end{tabular} & \begin{tabular}[c]{@{}c@{}}2.91845
\\ 0.00009
\\ 0.03638
\\ 0.44418
\\ 0.00123
\\ 0.00109
\\ 0.00070
\\ 0.03928
\\ 0.06043\end{tabular} \\ \hline
\textbf{\begin{tabular}[c]{@{}c@{}}Resonance Feature\\ ${k_r}=30$\end{tabular}}          & \begin{tabular}[c]{@{}c@{}}$10^{5}\times\sigma \left(\omega_{\mathrm{b}}\right)$\\ $\sigma \left(\omega_{\mathrm{cdm}}\right)$\\ $\sigma \left( H_0 \right)$\\ $10^{11}\times\sigma \left( A_\mathrm{s} \right)$\\ $\sigma \left( n_\mathrm{s} \right)$\\ $\sigma \left( \tau_{\mathrm{reio}} \right)$\\ $\sigma \left({C_r}\right)$\\ $\sigma \left({k_r}\right)$\\ $\sigma \left( {\phi_r} \right)$\end{tabular} & \begin{tabular}[c]{@{}c@{}}2.91829
\\ 0.00015
\\ 0.06132
\\ 0.49993
\\ 0.00130
\\ 0.00121
\\ 0.00185
\\ 0.14155
\\ 0.22125
\end{tabular} & \begin{tabular}[c]{@{}c@{}}2.88014
\\ 0.00013
\\ 0.05333
\\ 0.49080
\\ 0.00126
\\ 0.00120
\\ 0.00185
\\ 0.12261
\\ 0.19185\end{tabular} & \begin{tabular}[c]{@{}c@{}}2.75542
\\ 0.00009
\\ 0.03557
\\ 0.44615
\\ 0.00119
\\ 0.00108
\\ 0.00110
\\ 0.04307
\\ 0.06937
\end{tabular} & \begin{tabular}[c]{@{}c@{}}2.86848
\\ 0.00011
\\ 0.04681
\\ 0.48732
\\ 0.00125
\\ 0.00117
\\ 0.00156
\\ 0.11911
\\ 0.18621
\end{tabular} & \begin{tabular}[c]{@{}c@{}}2.71936
\\ 0.00007
\\ 0.02986
\\ 0.43120
\\ 0.00116
\\ 0.00105
\\ 0.00102
\\ 0.03886
\\ 0.06172\end{tabular} & \begin{tabular}[c]{@{}c@{}}2.68033
\\ 0.00008
\\ 0.03401
\\ 0.43351
\\ 0.00113
\\ 0.00106
\\ 0.00097
\\ 0.05060
\\ 0.07903
\end{tabular} \\ \hline
\textbf{\begin{tabular}[c]{@{}c@{}}Resonance Feature\\ ${k_r}=100$\end{tabular}}      & \begin{tabular}[c]{@{}c@{}}$10^{5}\times\sigma \left(\omega_{\mathrm{b}}\right)$\\ $\sigma \left(\omega_{\mathrm{cdm}}\right)$\\ $\sigma \left( H_0 \right)$\\ $10^{11}\times\sigma \left( A_\mathrm{s} \right)$\\ $\sigma \left( n_\mathrm{s} \right)$\\ $\sigma \left( \tau_{\mathrm{reio}} \right)$\\ $\sigma \left({C_r}\right)$\\ $\sigma \left({k_r}\right)$\\ $\sigma \left( {\phi_r} \right)$\end{tabular} & \begin{tabular}[c]{@{}c@{}}2.95699
\\ 0.00015
\\ 0.06030
\\ 0.49728
\\ 0.00136
\\ 0.00120
\\ 0.00464
\\ 0.28708
\\ 0.48681
\end{tabular} & \begin{tabular}[c]{@{}c@{}}2.91117
\\ 0.00012
\\ 0.04945
\\ 0.48410
\\ 0.00132
\\ 0.00119
\\ 0.00460
\\ 0.20531
\\ 0.32654\end{tabular} & \begin{tabular}[c]{@{}c@{}}2.68131
\\ 0.00006
\\ 0.02331
\\ 0.37691
\\ 0.00117
\\ 0.00093
\\ 0.00141
\\ 0.05635
\\ 0.09256
\end{tabular} & \begin{tabular}[c]{@{}c@{}}2.74026
\\ 0.00008
\\ 0.03120
\\ 0.44087
\\ 0.00124
\\ 0.00106
\\ 0.00255
\\ 0.15428
\\ 0.24749
\end{tabular} & \begin{tabular}[c]{@{}c@{}}2.69282
\\ 0.00006
\\ 0.02151
\\ 0.35596
\\ 0.00116
\\ 0.00089
\\ 0.00128
\\ 0.04974
\\ 0.08150\end{tabular} & \begin{tabular}[c]{@{}c@{}}2.76160
\\ 0.00006
\\ 0.02249
\\ 0.38411
\\ 0.00119
\\ 0.00093
\\ 0.00121
\\ 0.06353
\\ 0.09609
\end{tabular} \\ \hline
\end{tabular}\par
\bigskip
\parbox{16.53cm}{\captionof{table}{{\fontfamily{bch}The possible 1-$\sigma$ sensitivity of \textbf{CORE-M5+CMB-S4} for resonance feature signal in synergy with SKA1(CS), SKA2(CS), SKA1(IM1), SKA1(IM2), SKA1(IM1+IM2), and Euclid(CS+GC)+SKA1(IM2).}}} \label{table:CORE-M5+CMB-S4-Resonance}
\end{minipage}
}

{\fontfamily{qpl}

\section{\textbf{21cm intensity mapping power spectrum and likelihood}}\label{pwrspec:likhd}
Here we present a succinct review on the modelling of 21 cm intensity mapping power spectrum and its likelihood, where only the required information relevant to the current article has been provided. For a comprehensive review, one can refer to the article~\cite{Sprenger:2018tdb}.

%A neutral hydrogen~(HI) atom has two states with marginally different energy levels. 
A neutral hydrogen~(HI) atom possesses two states that are marginally different in energy. In one state, the proton and electron spins are aligned in the same direction, which is called a singlet state~($ {\cal{n}}_{0} $), and another state is where their spins are opposite in direction, called a triplet state~($ {\cal{n}}_{1} $). The abundance of these two states can be associated with a temperature called a spin temperature~($ {\mathscr{T}}_\text{spin} $) using the Boltzmann factor~\cite{Wouthuysen,Field}. The statistical weights for the 21cm hyperfine transitions are ${\cal{g}}_{0} = 1$ and ${\cal{g}}_{1} = 3$; and as ${\mathscr{T}}_\text{spin} \gg {\cal{E}}_{10}$, the spin temperature we get on simplification is as follows:
\be \label{spintemp}
\frac{{\cal{n}}_{1}}{{\cal{n}}_{0}}=\frac{{\cal{g}}_{1}}{{\cal{g}}_{0}} \exp \left(-\frac{{\cal{E}}_{10}}{{\mathscr{T}}_\text{spin}} \right)= 3\exp \left( -\frac{{\cal{E}}_{10}}{{\mathscr{T}}_\text{spin}} \right) \approx 3 \left(1 -\frac{{\cal{E}}_{10}}{{\mathscr{T}}_\text{spin}} \right).
\ee
A 21cm hyperfine transition produces or absorbs a photon with a wavelength~$(\lambda)=21~\text{cm}$~(frequency~$(\nu_{0}) = 1420.4057\,\text{MHz}$) as the energy difference~$({\cal{E}}_{10})$ is $\sim0.068$~$\text{K}$.

The emission and absorption in neutral hydrogen are described using spin temperature~(${\mathscr{T}}_\text{spin}$) and optical depth~($\tau$):
\be
{\mathscr{T}}_\text{bright} = {\mathscr{T}}_\text{spin}(1-e^{-\tau})+{\mathscr{T}}_\text{cmb}e^{-\tau} \ .
\ee
Since, we can not measure the spin temperature~$({\mathscr{T}}_\text{spin})$ directly, the 21cm intensity mapping surveys measure the differential brightness temperature~$ (\Delta {\mathscr{T}}_\text{bright}) $ of 21cm fluctuations against the background of CMB photons. The differential brightness temperature is defined below, where the higher order terms of $ \tau $ are ignored as the optical depth is small due to the low probability of a 21cm transition:
\begin{equation}
\Delta {\mathscr{T}}_\text{bright} \equiv \frac{{\mathscr{T}}_\text{bright}-{\mathscr{T}}_\text{cmb}}{1+z}= \frac{{\mathscr{T}}_\text{spin}-{\mathscr{T}}_\text{cmb}}{1+z}\left(1-e^{-\tau}\right) \approx \frac{{\mathscr{T}}_\text{spin}-{\mathscr{T}}_\text{cmb}}{1+z}\tau \ .
\end{equation}
The absorption coefficient $\alpha$ is evaluated from the equation of radiative transfer for computing $\tau$:
\be
\frac{\dd {\cal{I}}}{\dd {\cal{s}}} = -\alpha {\cal{I}} + {\cal{j}} \ ,
\ee
where ${\cal{I}}$ and ${\cal{s}}$ are the specific intensity and the radial distance (in physical units), respectively. The specific intensity~(${\cal{I}}$) is by definition the energy flux per frequency and solid angle.

Below we have the radial derivative term as:
\be
\frac{\dd {\cal{I}}}{\dd {\cal{s}}} = {\cal{E}}_{10}\frac{\phi(\nu)}{4\pi}\frac{\dd {\cal{n}}_0}{\dd t} \ .
\ee
Since, we get a photon of energy ${\cal{E}}$ each time an atom falls from the excited state 1 into the ground state 0, thus the derivative of the energy flux with respect to the radial distance is directly proportional to the derivative of the number density of atoms in the ground state with respect to time. The  derivative with respect to solid angle yields a factor of $1/4\pi$, if isotropy is assumed. $\phi(\nu)$ refers to the line profile, which describes a small band of frequencies associated with the measurement of a single frequency, and is normalized to $\int \phi(\nu) \dd \nu = 1$.

The number density can be expressed in terms of Einstein coefficients as follows:
\be\label{numdenhy}
\frac{\dd {\cal{n}}_0}{\dd t} ={\cal{n}}_1 {\cal{A}}_{10} + {\cal{n}}_1 {\cal{B}}_{10} {\cal{I}} - {\cal{n}}_0 {\cal{B}}_{01} {\cal{I}}  \ .
\ee
For Einstein coefficients we have the following relations: ${\cal{A}}_{10} = 4\pi\nu_0^3 {\cal{B}}_{10}$ and ${\cal{g}}_0 {\cal{B}}_{01} = {\cal{g}}_1 {\cal{B}}_{10}$, where natural units have been used ($\hslash = c = k_B = 1$).

Using equation~(\ref{spintemp}), equation~(\ref{numdenhy}) and ${\cal{n}}_{\text{HI}} \equiv {\cal{n}}_0+{\cal{n}}_1 \simeq \frac{4}{3} {\cal{n}}_1 \simeq 4{\cal{n}}_0 $ one gets,
\be
\frac{\dd {\cal{n}}_0}{\dd t} = \frac{3}{4} {\cal{n}}_{\text{HI}} {\cal{A}}_{10}-\frac{3}{4}\frac{{\cal{A}}_{10}}{4\pi\nu_0^3} \frac{{\cal{E}}_{10}}{{\cal{T}}_\text{spin}}{\cal{n}}_{\text{HI}} {\cal{I}}  \ ,
\ee
which provides an expression of $\alpha$, 
\be
\alpha = \frac{3}{16}\frac{{\cal{A}}_{10}}{{\cal{T}}_\text{spin}}\frac{\phi(\nu)}{\nu_0}{\cal{n}}_{\text{HI}} \ .
\ee
To describe line profile corresponding to a small Doppler shift arising because of a constant velocity dispersion~($\frac{\dd v}{\dd {\cal{s}}}$) across a region of neutral hydrogen of radial range $\delta {\cal{s}}$, a simple model of constant distribution over a given range of $\delta \nu$ has been assumed:
\be
\phi(\nu) = \frac{1}{\delta \nu} = \frac{1}{\frac{\dd v}{\dd {\cal{s}}}\delta {\cal{s}} \cdot\nu_0} \ .
\ee
On averaging over large volumes, one can approximate a constant velocity dispersion to a constant Hubble flow $\frac{\dd v}{\dd {\cal{s}}} = H(z)$.
Now we have the optical depth as,
\be
\tau \equiv \int_{\delta {\cal{s}}} \alpha \dd {\cal{s}} = \frac{3{\cal{A}}_{10}}{16\nu_0^2 {\cal{T}}_\text{spin}}\frac{1}{H(z)}{\cal{n}}_{\text{HI}} \ .
\ee
The number density of neutral hydrogen atom is a combination of its background value and a perturbation in the density field,
\be
{\cal{n}}_{\text{HI}} = \frac{1}{m_{\text{H}}}\frac{3 H_0^2}{8\pi G}\Omega_{\text{HI}}(z)(1+\delta_{\text{HI}})(1+z)^3 \ .
\ee
Putting everything together the final expression of the differential brightness temperature appears as,
\be
\Delta {\mathscr{T}}_\text{bright} = \frac{9{\cal{A}}_{10} H_0}{16m_{\text{H}} \nu_0^2  } \frac{\hslash c^3}{8\pi Gk_B}\left(\frac{H_0 (1+z)^2}{H(z)}\right)\Omega_{\text{HI}}(z)(1+\delta_{\text{HI}}) \left(1-\frac{{\mathscr{T}}_{\text{cmb}}}{{\mathscr{T}}_\text{spin}}\right) \ .
\ee
One can ignore the last term as ${\mathscr{T}}_S \gg {\mathscr{T}}_{\gamma}$ inside the galaxies, so that equation~(\ref{DelTb189}) is obtained. 
\begin{equation}
\label{DelTb189}
\overline{\Delta {\mathscr{T}}}_\text{bright} \simeq 189\left[\frac{H_0(1+z)^2}{H(z)}\right]\Omega_{\text{HI}}(z) \, h \, \text{mK} \ ,
\end{equation}
where $H_{0}$ represents the Hubble constant $H_{0} = 100h\,\text{km}/(\text{s}\,\text{Mpc})$, and $\Omega_{\text{HI}}(z) = \rho_{\text{HI}}(z)/\rho_c$ refers to the fraction of mass density of neutral hydrogen and the critical density of the present-day Universe.

Here we consider the signals coming from the neutral hydrogen within the galaxies at low redshifts. For instance, this above approach to modeling the differential brightness temperature can be found in refs.~\cite{Bull:2014rha,Battye:2012tg,Hall:2012wd}.

In determining the power spectrum of fluctuations in the differential brightness temperature, one can reasonably ignore local fluctuations in the Hubble parameter~($H(z)$), which allows us to write the 21cm power spectrum~(${\cal{P}}_{21}$) proportional to the power spectrum~(${\cal{P}}_{\text{HI}}$) of neutral hydrogen~(HI) density fluctuations:
\begin{equation}
\Delta {\mathscr{T}}_\text{bright} - \overline{\Delta {\mathscr{T}}}_\text{bright} = \overline{\Delta {\mathscr{T}}}_\text{bright} \delta_{\text{HI}} = \overline{\Delta {\mathscr{T}}}_\text{bright} b_{\text{HI}} \delta_\text{m} \ .
\end{equation}
Hence, the power spectrum for these deviations can be written as:
\begin{equation}
\label{eq:P21_1}
{\mathscr{P}}_{21} = b_{21}^2 {\mathscr{P}}_{\text{m}},
\end{equation}
where $b_{21} \equiv \overline{\Delta {\mathscr{T}}}_\text{bright} b_{\text{HI}}$.
Following the refs.~\cite{Villaescusa-Navarro:2016kbz,Sprenger:2018tdb}, the redshift dependence of $\Omega_{\text{HI}}(z)$ and $b_{\text{HI}}(z)$ are considered as follows:
\be
\label{reddep:omega}
\Omega_{\text{HI}}(z) = \Omega_{\text{HI},0} (1+z)^{{\mathscr M}_{\text{HI}}} \\
\ee
\be
b_{\text{HI}}(z) = 0.904+0.135(1+z)^{1.696} \ ,
\ee
where the assumed fiducial values for $\Omega_{\text{HI},0}$ and ${\mathscr M}_{\text{HI}}$ are $4\times10^{-4}$ and $0.6$, respectively, and allowed to vary in our forecasts. In modeling the bias with further accuracy, we use two nuisance parameters~(${\mathscr N}_1^{\text{IM}},{\mathscr N}_2^{\text{IM}}$) for bias in this analysis, with a mean value of one,
\begin{equation}
\label{bias:HI}
b_{\text{HI}}(z) = {\mathscr N}_1^{\text{IM}} [0.904+0.135(1+z)^{1.696 {\mathscr N}_2^{\text{IM}}}].
\end{equation}
Apart from all these we have to consider the observational effects as well to construct the 21cm power spectrum:
\be
\label{eq:P21}
\begin{split}
{\mathscr{P}}_{21}(k, \mu, z) = T_{\text{AP}}(z)\times T_{\text{RE}}(k, \mu, z) \times T_{\text{RSD}}(k_{\text{t}}, \mu_{\text{t}} ,z)\times
b_{21}^2(z)\times {\mathscr{P}}_{\text{m}}(k_{\text{t}},z) \,.
\end{split}
\ee
where $ {\mathscr{P}}_{\text{m}}(k,z)$ stand for the $\text{matter power spectrum}$. The prefactors are the explicit mentioned and elaborated subsequently.
\begin{equation}\label{Terms}
\textbf{Term}=\left\lbrace
\begin{array}{lll}
\left[ \frac{D_{\text{A}}(z)}{D_{\text{A,t}}(z)} \right]^2
\frac{H_{\text{t}}(z)}{H(z)} & ~~~\textbf{Alcock-Paczinsky Term~($\mathbf{T_{AP}}$)} & \\
e^{-k^2\left[\mu^2\cdot\left(\sigma_{\shortparallel}^2(z)-\sigma_{\perp}^2(z)\right)+\sigma_{\perp}^2(z)\right]} & ~~~\textbf{Resolution Effect Term~($\mathbf{T_{RE}}$)} & \\
\underbrace{\left( 1+\beta(k_{\text{t}},z) \, \mu_{\text{t}}^2 \right)^2}_{\textbf{Kaiser Effect}}\underbrace{ e^{-k_{\text{t}}^2\mu_{\text{t}}^2\sigma_{\text{nl}}^2}}_{\textbf{FoG Effect}} & ~~~\textbf{Redshift Space Distortions Term~($\mathbf{T_{RSD}}$)}  & 
\end{array}
\right..
\end{equation}
\begin{itemize}[itemsep=-.3em]
\item[\ding{109}] The $\textbf{Alcock-Paczinsky Term~($\mathbf{T_{AP}}$)}$ of equation~(\ref{eq:P21}) refers to a geometric correction term~\cite{Alcock:1979mp,Seo:2003pu}, brought in to compensate for the lack of knowledge of true cosmology in modelling the observed galaxy power spectrum from the matter power spectrum. The term associted to this effect is shown in equation~(\ref{Terms}). The parameters which have a '\textit{t}' in the suffix refer to the true cosmology, and that may not agree with fiducial cosmology. The transformation equation of coordinates that relates the true geometry with fiducial geometry is as follows:
\begin{equation}\label{coor}
k_\perp = \frac{D_{\text{A,t}}(z)}{D_{\text{A}}(z)}\,k_{\perp\,,\text{t}}~ ,
\qquad k_\shortparallel=\frac{H(z)}{H_{\text{t}}(z)}\,k_{\shortparallel\,,{\text{t}}}.
\end{equation}
In equation~(\ref{coor}), the parameter $H=\frac{\dot{a}}{a}$ denotes the Hubble parameter, and $D_\text{A} = \frac{r(z)}{(1+z)}$ stands for the angular diameter distance, where $r(z)$ connotes the comoving distance and $ a $ refers to the scale factor. The variables $k_{\shortparallel}$ and $k_{\perp}$ are the parallel and perpendicular components of the  Fourier mode vector in respect of the line-of-sight, respectively.  The $k_{\shortparallel}$ and $k_{\perp}$ are defined as follows:
\begin{equation}\label{mode_vec}
k = \vert\vec{k}\vert =\sqrt{k_{\perp}^2 + k_{\shortparallel}^2} \ ; \ \mu = \frac{\vec{k}\cdot\vec{r}}{kr}=\frac{k_{\shortparallel}}{k} \ ,
\end{equation} 
where $ \vec{k} $ and $ \vec{r} $ connote the Fourier mode and the line-of-sight distance vector, respectively, and $ \mu $ denotes the cosine of the angle between the unit vectors of the mode vector and the line-of-sight distance vector.

From the equations~(\ref{coor}) and~(\ref{mode_vec}), the following conversion relations for the Fourier modes of true and fiducial geometry can be obtained: 
\begin{equation}
\label{k^s}
k_{\text{t}} =k\left[\left(\frac{H_{\text{t}}}{H}\right)^2\mu^2 + \left(\frac{D_{\text{A}}}{D_{\text{A,t}}}\right)^2\left(1-\mu^2\right)\right]^{1/2},
\end{equation}
and
\begin{equation}
\label{mu^s}
\frac{\mu_{\text{t}} k_{\text{t}}}{H_{\text{t}}} = \frac{{\mu}k}{H} \ .
\end{equation}
\item[\ding{109}] The $\textbf{Resolution~Effect~Term~($\mathbf{T_{RE}}$)}$ of equation~(\ref{eq:P21}) is introduced to model the apparent reduction in the power arises on small scales due to instrumental resolving capacity. 
The parameters $\sigma_{\shortparallel}(z)$ and $\sigma_{\perp}(z)$ in equation~(\ref{Terms}) are Gaussian errors related to the coordinates, parallel and perpendicular to the line of sight, respectively. 
\item[\ding{109}] The $\textbf{Redshift~Space~Distortion~Term~($\mathbf{T_{RSD}}$)}$ includes two separate events, specifically the  $\text{Kaiser effect}$~\cite{Kaiser:1987qv}, and the $\text{Finger of God effect~(FoG)}$~\cite{Jackson:1971sky}. The term encompassing these two distinct effects in modelling the galaxy power spectrum is illustrated in equation~(\ref{Terms}). The $\text{Kaiser effect}$ refers to the anisotropy arises in the measurements due to the usual Doppler effect, which contributes to the redshift-space power spectrum on large scales. Similarly, on small scale another effect contributes to the redshift-space power spectrum which arises from the peculiar velocities of the galaxies known as $\text{Finger of God effect}$.  
\end{itemize}
In equation~(\ref{Terms}), the parameter $ \beta(k_{\text{t}},z) $, which refers to the redshift space distortion term, is defined as follows:
\begin{equation}
\label{eq:beta}
\beta(k_{\text{t}},z) = -\frac{1}{2}\frac{1+z}{b_{\text{g}}(z)} \frac{\dd \ln {\mathscr{P}}_{\text{m}}(k_{\text{t}},z)}{\dd z} \ .
\end{equation}
The factor $ b_{\text{g}}(z) $ appearing in equation~(\ref{eq:beta}) is the galaxy bias. In our forecasts, the fiducial value assumed for the parameter $ \sigma_{\text{nl}} $ related to the FoG effect is 7 Mpc, and it has been allowed to vary between 4-10 Mpc.

Finally, the observed 21cm power spectrum can be written as follows by taking into account the noise term:
\begin{equation}
{\mathscr{P}}_{21,\text{obs}}(k,\mu,z) = {\mathscr{P}}_{21}(k,\mu,z)  + {\mathscr{P}}_\text{noise}(z)
\label{eq:P21_PN}
\end{equation}
where ${\mathscr{P}}_\text{noise}(z)$ has been described in subsection~\ref{COS:EXPSPEC}. 

The likelihood~($\mathscr L$) or the chi-square~($\chi^2$) function for the observed 21cm power spectrum appears as follows\footnote{For the detailed scheme of constructing the likelihood or chi-square function, the reader can refer to article~\cite{Sprenger:2018tdb}.}~\cite{Sprenger:2018tdb}:
\begin{equation}
\label{chi2-result}
\chi^2 = -2\ln{\mathscr L} =\sum_{\bar{z}} \frac{{\mathscr{V}}_{\text{bin}}(\bar{z})}{2(2\pi)^3} \int  \frac{\left(\hat{{\mathscr{P}}}_{21,\text{obs}}(\vec{k},\bar{z})-{\mathscr{P}}_{21,\text{obs}}(\vec{k},\bar{z})\right)^2}{{\mathscr{P}}_{21,\text{obs}}^2(\vec{k},\bar{z})} \dd^3\vec{k} \ ,
\end{equation}
where $\hat{{\mathscr{P}}}_{21,\text{obs}}(\vec{k},\bar{z})$ and ${\mathscr{P}}_{21,\text{obs}}(\vec{k},\bar{z})$ are the mock data and the theoretical 21cm power spectrum appearing in equation~(\ref{eq:P21_PN}), respectively.
The survey volume of each redshift bin can be obtained from the following equation:
\begin{equation}
\label{eq:Vr}
 {\mathscr{V}}_{\text{bin}}(\bar{z}) = \frac{4\pi f_{\text{sky}}}{3} \left\lbrace r^3\left(\bar{z}+\frac{\Delta z}{2}\right)-r^3\left(\bar{z}-\frac{\Delta z}{2}\right)\right\rbrace \ .
\end{equation}
The $r(z)$ in equation~(\ref{eq:Vr}) refers to the comoving distance, and defined as follows:
\be
r(z) = c\int_0^z \dfrac{\dd z}{H(z)}~,
\ee
and $f_{\text{sky}}$ and $\bar{z}$ represent the sky-fraction and mean redshift per bin, respectively.}

%%%%%%%%%%%%%%%%%%%%%%%%%%%%%%%%%%%%%%%%%%%%%%%%%%%%%%%%%%%%%%%%%%%%%%%%%%%%%

\section*{Acknowledgments}
DC thanks ISI Kolkata for financial support through the Senior Research Fellowship and gratefully acknowledges the computational facilities of ISI Kolkata. I am grateful to Supratik Pal for his useful comments and careful reading of the manuscript, as well as for his continuous encouragement and stimulating discussions in the making of this article.
%%%%%%%%%%%%%%%%%%%%%%%%%%%%%%%%%%%%%%%%%%%%%%%%%%%%%%%%%%%%%%%%%%%%%%%%%%%%%%%%%%%%%%%%%%%%%%%%%%%%%%%%%%%%%%%%%%%%%%%%%%%%%%%%%%%%%%%%%%%%%%%%%%%%%%%%%%

\end{document}